\documentclass[aps,onecolumn,nofootinbib,superscriptaddress,balancelastpage]{revtex4}

\usepackage[T1]{fontenc} 
\usepackage[utf8]{inputenc}
\usepackage{subcaption}
\usepackage{graphicx}
\usepackage{graphics}
\usepackage[mathscr]{eucal}
\usepackage{amssymb} 
\usepackage{amsmath}
\usepackage{xspace}
\usepackage{listings}
\usepackage{ulem}
\usepackage{xcolor}
\usepackage{bm}         
\usepackage{array}
\usepackage{multirow}   
\usepackage{booktabs}   
\usepackage{mleftright}
\usepackage[export]{adjustbox}

\usepackage{bbold}
\usepackage{amsmath,amscd}
\usepackage{slashed}
\usepackage{placeins}
\usepackage{soul}
\usepackage{braket}
\usepackage{tikz-cd}
\usepackage{empheq}
\usepackage[most]{tcolorbox}
\usetikzlibrary{arrows, shapes,shadows}
\usepackage{mathtools,slashed}
\usepackage{dsfont}
\usepackage[makeroom]{cancel}
\usepackage{multirow}

\usepackage{amssymb}

\usepackage{dcolumn}
\usepackage{physics}
\usepackage{makecell} 

\usepackage{hyperref}
\hypersetup{colorlinks=true, linkcolor=blue, urlcolor=blue, citecolor=blue}

\usepackage{feynmp-auto}

\DeclareMathSymbol{\comma}{\mathpunct}{letters}{"3B}

\definecolor{ForestGreen}{rgb}{0.13, 0.55, 0.13}
\definecolor{Mulberry}{rgb}{0.77, 0.29, 0.55}
\definecolor{bostonuniversityred}{rgb}{0.8, 0.0, 0.0}
\definecolor{amber}{rgb}{1.0, 0.49, 0.0}
\newcommand{\amber}[0]{\color{amber}}
\newcommand{\blue}[0]{\color{blue}}
\newcommand{\green}[0]{\color{ForestGreen}}

\newcommand{\red}[0]{\color{red}}

\newcommand\varpm{\mathbin{\vcenter{\hbox{%
  \oalign{\hfil$\scriptstyle\hspace{-0.1ex}+\hspace{-0.1ex}$\hfil\cr
          \noalign{\kern-.5ex}
          $\scriptscriptstyle({-})$\cr}%
}}}}

\begin{document}

\title{On interplay between flavour anomalies and neutrino properties}

\author{Felipe F. Freitas}
\email{felipefreitas@ua.pt}
\affiliation{Departamento de F\'isica, Universidade de Aveiro and CIDMA, Campus de Santiago, 
3810-183 Aveiro, Portugal}

\author{João Gonçalves}
\email{jpedropino@ua.pt}
\affiliation{Departamento de F\'isica, Universidade de Aveiro and CIDMA, Campus de Santiago, 
3810-183 Aveiro, Portugal}
\affiliation{Department of Astronomy and Theoretical Physics, Lund University, 221 00 Lund, Sweden}

\author{Ant{\'o}nio~P.~Morais}
\email{aapmorais@ua.pt}
\affiliation{Departamento de F\'isica, Universidade de Aveiro and CIDMA, Campus de Santiago, 
3810-183 Aveiro, Portugal}
\affiliation{Theoretical Physics Department, CERN, 1211 Geneva 23, Switzerland}

\author{Roman~Pasechnik}
\email{Roman.Pasechnik@thep.lu.se}
\affiliation{Department of Astronomy and Theoretical Physics, Lund University, 221 00 Lund, Sweden}

\author{Werner~Porod}
\email{porod@physik.uni-wuerzburg.de}
\affiliation{Institut für Theoretische Physik und Astrophysik, Uni.~W\"urzburg, D-97074 W\"urzburg, Germany}

\begin{abstract}
A minimal extension of the Standard Model (SM) featuring two scalar leptoquarks, an SU(2) doublet with hypercharge 1/6 and a singlet with hypercharge 1/3, is proposed as an economical benchmark model for studies of an interplay between flavour physics and properties of the neutrino sector. The presence of such type of leptoquarks radiatively generates neutrino masses and offers a simultaneous explanation for the current B-physics anomalies involving $b \to c \ell \nu_\ell$ decays. The model can also accommodate both the muon magnetic moment and the recently reported $W$ mass anomalies, while complying with the most stringent lepton flavour violating observables.
\end{abstract}

\maketitle

\tableofcontents

\section{Introduction}
The Standard Model (SM) of particle physics is our current guide towards a consistent description of the subatomic phenomena, able to withstand a series of most stringent tests \cite{UA1:1983crd,CMS:2012qbp,GargamelleNeutrino:1973jyy,Hasert:1973cr,CDF:1995wbb,Parker:2018vye,Hanneke:2010au}. However, the SM does not resemble a fundamentally complete theory. It cannot explain various observations such as neutrino masses, dark matter relic density or the baryon asymmetry of the Universe. Apart from these limitations, recent anomalies have emerged in significance as of late. Specifically, the anomalous magnetic moment of the muon \cite{Muong-2:2021ojo,Muong-2:2023cdq} and hints for lepton flavour universality (LFU) violation in B meson decays, such as $R_{D^{(*)}}$ \cite{BaBar:2013mob, BaBar:2012obs, Belle:2015qfa, Belle:2016ure, Belle:2017ilt, LHCb:2015gmp}, defined as
\begin{equation}\label{eq:Rddtsar}
    R_{D^{(*)}} \equiv \frac{\textrm{Br}\left( \bar{B} \rightarrow D^{(*)} \tau \Bar{\nu}_\tau \right) }{\textrm{Br}\left( \bar{B} \rightarrow D^{(*)} l \Bar{\nu}_l \right)}, \quad \textrm{with} \quad l=\mu, e
\end{equation}
as well as tensions regarding decays of the $B_0/B_s$ mesons into a pair of muons, showcasing a 2.3$\sigma$ deviation from the SM prediction \cite{Altmannshofer:2021qrr}. Some previous results on $R_{K^{(*)}}$ \cite{BELLE:2019xld,Belle:2019oag,LHCb:2017avl,LHCb:2021trn} indicated a tension, but recently \cite{LHCb:2022december} it was shown to be consistent with the SM. There is also the recently reported CDF-II precision measurement of the $W$ mass indicating a 7.0$\sigma$ deviation from the SM prediction \cite{CDF:2022hxs}, whose new physics (NP) effects can be parameterized in a modification to the oblique $T$ parameter \cite{Strumia:2022qkt}. Attempts to address these anomalies have been extensively reported in the literature (see, e.g.~\cite{Bauer:2015knc,Altmannshofer:2017poe,Das:2016vkr,Angelescu:2018tyl,Altmannshofer:2020axr,Belanger:2021smw,Becker:2021sfd,Crivellin:2022mff,Crivellin:2020tsz,Crivellin:2018yvo,Blanke:2018sro,Calibbi:2017qbu,Crivellin:2019dwb,Carvunis:2021dss,Coy:2021hyr}) but are often treated in isolation rather than being simultaneously resolved in the same model. In a recent article \cite{Marzocca:2021azj}, the B-physics anomalies and the anomalous magnetic moment of the muon were shown to be simultaneously accommodated in an economical framework solely featuring a leptoquark (LQ) and a charged scalar singlet. An explanation for neutrino properties is also well known to be a tantalizing possibility in LQ models as was discussed in~\cite{Saad:2020ucl,Chowdhury:2022dps,Chen:2022hle,Dorsner:2017wwn,AristizabalSierra:2007nf,Zhang:2021dgl,Pas:2015hca,Cai:2017jrq,Babu:2010vp,Cata:2019wbu,Popov:2016fzr,Nomura:2021yjb,Chang:2021axw,Nomura:2021oeu,Babu:2019mfe,Faber:2018afz,Faber:2018qon,Bigaran:2019bqv,Gargalionis:2019drk,Gargalionis:2020xvt,Saad:2020ihm,Julio:2022bue,Julio:2022ton,Crivellin:2020mjs,Cai:2017wry}. Particularly relevant are \cite{Dorsner:2017wwn,AristizabalSierra:2007nf,Zhang:2021dgl,Pas:2015hca,Cai:2017jrq,Babu:2010vp,Cata:2019wbu} where a minimal two-LQ scenario featuring a weak-singlet $S\sim(\overline{\textbf{3}}, \textbf{1})_{1/3}$ and a doublet $R\sim(\textbf{3}, \textbf{2})_{1/6}$, offers the simplest known framework for radiative neutrino mass generation. However, a complete analysis of such an economical setting in the light of current flavour anomalies is lacking.

Furthermore, while minimal models often imply that fits to experimental data can become rather challenging, they also represent an opportunity for concrete and falsifiable predictions. In this paper, we then propose an inclusive study where B-physics, the muon $ a_\mu \equiv \tfrac{1}{2}(g-2)_\mu$ and the CDF-II $W$ mass anomalies are simultaneously explained alongside neutrino masses and mixing while keeping lepton flavour violating (LFV) observables under control. We further inspire our model on the flavoured grand unified framework first introduced by some of the authors in \cite{Morais:2020ypd,Morais:2020odg} in order to motivate the presence of a baryon number parity defined as $\mathbb{P}_\mathrm{B} = (-1)^{3 \mathrm{B} + 2 S}$, with $B$ being the baryon number and $S$ the spin. Such a parity forbids di-quark type interactions for the $S$ LQ  otherwise responsible for fast proton decay.

In this model, the $R_{D,D^*}$ observables are explained via the tree-level exchange of the $S$ LQ as in diagram (a). Noteworthy, the mixing between the $S$ and $R$ doublet induces radiative generation of neutrino masses at one-loop level, while a splitting between the two components of the $R$ doublet can modify the $W$ mass.

In what follows, we present the model and demonstrate how the fields contribute to each of the relevant observables and the main experimental constraints that affect the allowed parameter space. We then discuss the regions of parameter space where all anomalies and constraints are realized within experimental bounds. Finally, we summarize our results.

\section{The minimal LQ model}
The interactions of the singlet and doublet LQs with the SM fermion sector invariant both under the gauge symmetry and the $\mathbb{P}_\mathrm{B}$ parity are described by the following terms
\begin{equation}\label{eq:lagYukSMa}
\begin{aligned}
\mathcal{L}_{\mathrm{Y}} = \hphantom{.}& {\red \Theta_{ij}} \bar{Q}^{c}_j  L_i S + {\blue \Omega_{ij}} \bar{L}_i d_j R^\dagger + {\green \Upsilon_{ij}}  \bar{u}^c_j e_i S + \mathrm{h.c.}\,.
\end{aligned}
\end{equation}
As usual, $Q$ and $L$ are the left-handed quark and lepton SU(2) doublets, respectively, whereas $d$ and $e$ are the right-handed down quark and charged lepton SU(2) singlets. All Yukawa couplings, ${\red \Theta}$, ${\blue \Omega}$ and ${ \green \Upsilon}$, are complex $3\times 3$ matrices. Here, $\mathrm{SU(2)}$ contractions are also left implicit. For example, $\bar{Q^c}L \equiv \epsilon_{\alpha\beta}\bar{Q}^{c, \alpha} L^\beta$, with $\epsilon_{\alpha\beta}$ being the Levi-Civita symbol in two dimensions and $c$ indicating charge conjugation. 


The relevant part of the scalar potential reads as 
\begin{equation}\label{eq:scalar_potential_1}
\begin{aligned}
V \supset \hphantom{.}& -\mu^2 \abs{H}^2 + \mu_S^2 \abs{S}^2 + \mu_R^2 \abs{R}^2  + \lambda (H^\dagger H)^2 + g_{HR} (H^\dagger H) (R^\dagger R) + {\color{violet} g^\prime_{HR}} (H^\dagger R) (R^\dagger H) + g_{HS} (H^\dagger H)(S^\dagger S) + \\ &\left( {\amber a_{1}} R S H^\dagger + \mathrm{h.c.} \right) \,.
\end{aligned}
\end{equation}
Once the Higgs doublet gains a vacuum expectation value (VEV), which in the unitary gauge corresponds to $\expval{H} = \begin{bmatrix}0 & (v + h)/\sqrt{2} \end{bmatrix}^T$ and $v \approx 246~\mathrm{GeV}$, the mass for the Higgs field remains identical to that of the Standard model (SM), $m_h^2 = 2\lambda v^2$. One of the components of the $R$ doublet mixes with the $S$ field (corresponding to the LQs with an electrical charge of $1/3e$) via the ${\amber a_1}$ interaction term in Eq.~\eqref{eq:scalar_potential_1}, resulting in the squared mass matrix
\begin{equation}\label{eq:LQ_13}
M^2_{LQ^{1/3}} = \begin{bmatrix}
\mu_S^2 + \dfrac{g_{HS}v^2}{2} & \dfrac{v {\color{amber} a_1}}{\sqrt{2}} \\
\dfrac{v {\color{amber} a_1}}{\sqrt{2}} & \mu_R^2 + \dfrac{G v^2}{2}
\end{bmatrix}
\end{equation}
where $G = (g_{HR} + {\color{violet}g_{HR}^\prime})$ and we assume that ${\color{amber} a_1}$ is a real parameter. The eigenvalues of the mass matrix read
\begin{equation}\label{eq:eigenvalues_LQ_13}
\begin{aligned}
m^2_{S_1^{1/3}} = \frac{1}{4} \Bigg(2\mu_R^2 + 2\mu_S^2 + v^2(G + g_{HS}) - \sqrt{(2\mu_R^2 - 2\mu_S^2 + (G - g_{HS})v^2)^2 + 8{\color{amber} a_1}^2v^2}\Bigg),\\
m^2_{S_2^{1/3}} = \frac{1}{4} \Bigg(2\mu_R^2 + 2\mu_S^2 + v^2(G + g_{HS})  + \sqrt{(2\mu_R^2 - 2\mu_S^2 + (G - g_{HS})v^2)^2 + 8{\color{amber} a_1}^2v^2}\Bigg), \\
\end{aligned}
\end{equation}
where we adopt the notation for the mass eigenstates of $S^{1/3}_1$ and $S^{1/3}_2$. Do note that one can diagonalise the matrix in Eq.~\eqref{eq:LQ_13} via a bi-unitary transformation, that is,
\begin{equation}\label{eq:biunitary_LQ^(1/3)}
M^{\mathrm{diag}}_{LQ^{1/3}} = Z^H M^2_{LQ^{1/3}} Z^{H, \dagger},
\end{equation}
where $Z^H$ is an unitary matrix and $M^{\mathrm{diag}}_{LQ^{1/3}}$ is the LQ mass matrix in the diagonal form. Since this is a $2\times 2$ matrix, the mixing can be parameterized by a single angle, which in terms of the mass eigenstates it is given by
\begin{equation}\label{eq:mix_angle}
\sin(2\theta) = \frac{\sqrt{2}v {\color{amber} a_1}}{m_{S_1^{1/3}}^2 - m_{S_2^{1/3}}^2},
\end{equation}
where $\theta$ is a mixing angle. This relation necessarily implies the condition $-1 \leq (\sqrt{2}v{\color{amber} a_1}/(m_{S_1}^2 - m_{S_2}^2)) \leq 1$. The remainder LQ does not mix with the others and its tree level mass reads as
\begin{equation}\label{eq:eigenvalues_LQ_23}
m^2_{S^{2/3}} = \mu_R^2 + \frac{g_{HR} v^2}{2},
\end{equation}
where we adopt the nomenclature for the $2/3 e$ one as $S^{2/3}$. The relations in \eqref{eq:eigenvalues_LQ_13} can be inverted such that the physical masses of the LQ can be given as input in the numerical scan. Solving with the system of equations with respect to $\mu_R^2$ and $\mu_S^2$, one obtains
\begin{equation}\label{eq:inverted_equations_LQs}
\begin{aligned}
&\mu_S^2 = \frac{1}{2} (m^2_{S_1^{1/3}} + m^2_{S_2^{1/3}} - g_{HS} v^2 + \sqrt{(m^2_{S_1^{1/3}} -  m^2_{S_2^{1/3}})^2 - 2 {\color{amber} a_1}^2 v^2}), \\
&\mu_R^2 = \frac{1}{2} (m^2_{S_1^{1/3}} + m^2_{S_2^{1/3}} - (g_{HR} + {\color{violet} g_{HR}^\prime}) v^2 - \sqrt{(m^2_{S_1^{1/3}} - m^2_{S_2^{1/3}})^2 - 2 {\color{amber} a_1}^2 v^2}).
\end{aligned}
\end{equation}
Note that the mass of the $(2/3)e$ LQ is not given as input and is determined from ${\color{violet} g_{HR}}$ and the calculated value of $\mu_R^2$. As one can note from both equations \eqref{eq:LQ_13} and \eqref{eq:eigenvalues_LQ_13}, in the limit of small mixing (${\color{amber} a_1} \rightarrow 0$) the $2/3e$ LQ is approximately degenerate with the heaviest $1/3e$ LQ, \textit{i.e.}~$S_2^{1/3}$, with $m_{S_2^{1/3}}$ and $m_{S^{2/3}}$ differing only by a factor of ${\color{violet}g_{HR}^\prime}$. This means that the majority of cases feature a $2/3e$ state with mass close to one of the two LQ masses used as input. On the other hand, if the mixing is large, then we should obtain a sizeable mass splitting, but not significant enough to deviate from  $m_{S_1^{1/3}}$. Therefore, ${\color{amber} a_1}$ and ${\color{violet}g_{HR}^\prime}$ are responsible for generating a mass splitting between the two components of the $R$ doublet, providing a contribution to the CDF-II $W$ mass discrepancy.

A similar analysis can be conducted in both the quark and lepton sectors. For simplicity of the numerical analysis, one assumes a flavour diagonal basis for the up-type quarks such that the Cabibbo–Kobayashi–Maskawa (CKM) mixing resides entirely within the down-quark sector. Additionally, we assume that the charged lepton mass matrix to be diagonal. With this in kind, we can express the fermion Yukawa matrices as 
\begin{equation}
\begin{aligned}
\bm{Y_d} = \frac{\sqrt{2}}{v} \bm{V}^\dagger \bm{M_d}^\mathrm{diag}, \quad\quad \bm{Y_u} = \frac{\sqrt{2}}{v} \bm{M_u}^\mathrm{diag}, \quad\quad \bm{Y_e} = \frac{\sqrt{2}}{v} \bm{M_e}^\mathrm{diag}.
\end{aligned}
\end{equation}
where $\bm{V}$ is the Cabibbo–Kobayashi–Maskawa (CKM)  mixing matrix and $\bm{M_{f}}^\mathrm{diag}$ are the diagonal mass matrices for $f = u,d,l$ fermions.

The mixing parameter ${\amber a_1}$ is also responsible for enabling radiative generation of neutrino masses at one-loop level via the diagram 
\begin{equation}
	\qty(M_{\nu})_{ij} = \adjincludegraphics[valign=c, height=1.8cm, raise=2.89\baselineskip]{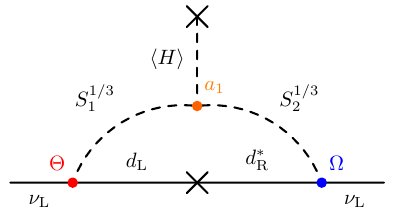}
\end{equation}
which for simplicity one assumes a flavour diagonal basis for the up-type quarks such that the CKM mixing resides entirely within the down-quark sector. Therefore, one can express the components of the neutrino masses as
\begin{equation}\label{eq:loop_integral}
\begin{aligned}
(M_\nu)_{ij} = &\frac{3}{16\pi^2(m_{S^{1/3}_2}^2 - m_{S^{1/3}_1}^2)}\frac{v {\amber a_1}}{\sqrt{2}}\ln(\frac{m_{S^{1/3}_2}^2}{m_{S^{1/3}_1}^2}) \sum_{m,a} \qty(m_d)_a V_{am} \qty({\red \Theta_{im}} {\blue \Omega_{ja}} + {\red \Theta_{jm}} {\blue \Omega_{ia}}) \,,
\end{aligned}
\end{equation}
where $V_{ab}$ denote the CKM matrix elements and $(m_d)_a$ are the down-type quark masses. In the limit of vanishing LQ mixing, $i.e.$ ${\amber a_1} \rightarrow 0$, the loop contribution goes to zero. Indeed, mixing between the doublet and singlet LQs is a necessary aspect for a viable phenomenology. As in the previous cases, it can be inverted such that the neutrino mass differences as well as the mixing angles can be given as input. In this case, however, we do not obtain a closed-form formula for the inversion in terms of the physical input parameters and instead we numerically invert equation \eqref{eq:loop_integral}.

2006.04822

\section{Setting up the problem: Anomalies}\label{sec:anomalies}

In this study, besides considering the properties of the neutrino sector, we focus our attention on the three main observables: (i) the anomalous magnetic moment of the muon, (ii) the flavour universality ratio $R_{D,D^*}$ as well as (iii) the $W$-mass anomaly. We do note that, for the later, no independent experimental verification of this anomaly has been made, hence, a healthy dose of scepticism is advised. On the same note, the muon anomaly is also not consensual if lattice results from the BMW collaboration \cite{Borsanyi:2020mff} are taken at face value, which have now been independently verified by different lattice groups \cite{ExtendedTwistedMass:2022jpw,Ce:2022kxy}.

\subsubsection{Anomalous magnetic moment of the muon}

The anomalous magnetic moment of leptons represents a deviation from the classical $g=2$ prediction of Dirac's theory, sourced by loop corrections to the electromagnetic vertex. Within the SM, these corrections can be reliably computed in QED and in weak processes involving massive vector and Higgs bosons. 
However, QCD corrections are typically the largest source of uncertainties, coming from the hadronic vacuum polarisation (HVP) and hadronic light-by-light loop-induced diagrams, since a first principles' calculation is arduous and requires sophisticated computational techniques. Combining the latter contributions leads to the SM prediction \cite{aoyama:2012wk,Aoyama:2019ryr,czarnecki:2002nt,gnendiger:2013pva,davier:2017zfy,keshavarzi:2018mgv,colangelo:2018mtw,hoferichter:2019gzf,davier:2019can,keshavarzi:2019abf,kurz:2014wya,melnikov:2003xd,masjuan:2017tvw,Colangelo:2017fiz,hoferichter:2018kwz,gerardin:2019vio,bijnens:2019ghy,colangelo:2019uex,Blum:2019ugy,colangelo:2014qya,Aoyama:2020ynm}. The precision measurement of $a_\ell\equiv (g-2)_\ell/2$ is the goal of several experimental efforts, and not only for the electron ($\ell=e$) but also for other particles such as the muon ($\ell=\mu$). The latter has gained a particular interest due to a combined result from the Brookhaven National Laboratory (BNL) \cite{Muong-2:2006rrc} and the Fermi National Laboratory (FNAL) \cite{Muong-2:2021ojo,Muong-2:2023cdq}, showing a $5.0\sigma$ deviation from the SM prediction as
 \begin{equation}
    \begin{cases}
        a_\mu^{{\scriptsize \mathrm{FNAL}}} = \left( 116 \ 592 \ 055 \pm 24 \right) \times 10^{-11} \\
        a_\mu^{{\scriptsize \mathrm{BNL}}} = \left( 116 \ 592 \ 089 \pm 63 \right) \times 10^{-11} \\
        a_\mu^{2023} = \left( 116 \ 592 \ 059 \pm 22 \right) \times 10^{-11}
    \end{cases} \,,
    \quad a_\mu^{{\scriptsize \mathrm{SM}}} = \left( 116 \ 591 \ 810 \pm 43 \right) \times 10^{-11} \,,
\end{equation}
with $a_\mu^{2023}$ representing the world average as of 2023. Here we note that the SM theoretical result $a_\mu^{\mathrm{SM}}$ is primarily driven by the R-ratio approach, which relies on data-driven methods \cite{Aoyama:2020ynm}. The results obtained in this approach are not in agreement with those obtained by the lattice QCD community \cite{Borsanyi:2020mff,ExtendedTwistedMass:2022jpw,Ce:2022kxy}. Given that the most recent FNAL result reaches a discrepancy between the SM prediction in \cite{Aoyama:2020ynm} and the experimental value at the $5\sigma$ level, the importance of clarifying the correct SM theoretical calculation becomes rather significant for the community.

\begin{figure*}[htb!]
	\centering
	\captionsetup{justification=raggedright}
	\begin{subfigure}{0.32\textwidth}
		\centering
		\includegraphics[width=\textwidth]{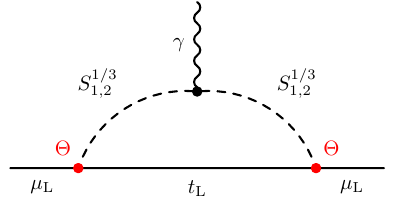}
		\caption{}
	\end{subfigure}
    \hspace*{3em}
	\begin{subfigure}{0.32\textwidth}
		\centering
		\includegraphics[width=\textwidth]{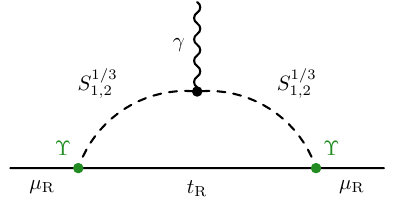}
		\caption{}
	\end{subfigure}\\
	\begin{subfigure}{0.32\textwidth}
		\centering
		\includegraphics[width=\textwidth]{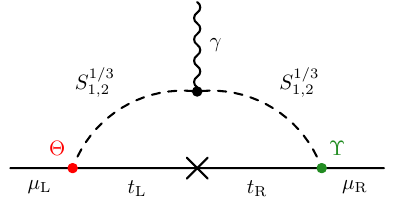}
		\caption{}
	\end{subfigure}
    \hspace*{3em}
 	\begin{subfigure}{0.32\textwidth}
		\centering
		\includegraphics[width=\textwidth]{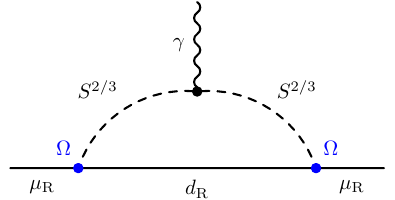}
		\caption{}
	\end{subfigure}
	\caption{One-loop Feynman diagrams that contribute to the anomalous magnetic moment of the muon, involving the new LQs. In the left and top-right diagrams, we show contributions from $1/3e$ LQs, while the bottom-right diagram represents the contribution from the $2/3e$ LQ. Four additional diagrams with the photon attached to the quark propagators are also considered.}
	\label{fig:muon_g2}
\end{figure*}

The usage of scalar LQs to address the anomalous magnetic moment of the muon is not a novel idea (for earlier studies see, for example, \cite{Dorsner:2019itg,Arcadi:2021cwg,Perez:2021ddi,Zhang:2021dgl,Nomura:2021oeu}). As was discussed in previous works, the dominant contributions to $a_\mu$ arise from chirality flipping of the internal fermionic propagators. The latter in turn leads to a correction that scales as $a_\mu \propto m_{q_i}/m_\mu$, where $m_{q_i}$ and $m_\mu$ denote the SM quarks and muon masses respectively. This makes the top contribution the most important. At one-loop level, the relevant contributions in our model are shown in Fig.~\ref{fig:muon_g2}, where it must be noted that the photon can also be attached to the quark propagators. 

With this in mind, we write each individual contribution to the anomalous magnetic moment as follows \cite{Zhang:2021dgl}:
\begin{equation}\label{eq:muon_g2_LQs}
    \begin{aligned}
        &\Delta a_\mu^{S^{1/3}_1} = -\frac{3m_\mu \cos^2\theta}{36\pi^2 m^2_{S_1^{1/3}}} \qty[2m_t \mathrm{Re}\qty{ {\color{red}\Theta_{\mu t}}{\color{ForestGreen}\Upsilon_{\mu t}}} \mathcal{A}\Bigg(\frac{m_t^2}{m^2_{S^{1/3}_1}}\Bigg) - m_\mu\qty( \abs{{\color{red}\Theta_{\mu t}}}^2 + \abs{{\color{ForestGreen}\Upsilon_{\mu t}}}^2 ) \mathcal{B}\Bigg(\frac{m_t^2}{m^2_{S^{1/3}_1}}\Bigg) ]\,, \\
        &\Delta a_\mu^{S^{1/3}_2} = -\frac{3m_\mu \sin^2\theta}{36\pi^2 m^2_{S_2^{1/3}}} \qty[2m_t \mathrm{Re}\qty{ {\color{red}\Theta_{\mu t}}{\color{ForestGreen}\Upsilon_{\mu t}}} \mathcal{A}\Bigg(\frac{m_t^2}{m^2_{S^{1/3}_2}}\Bigg) - m_\mu\qty( \abs{{\color{red}\Theta_{\mu t}}}^2 + \abs{{\color{ForestGreen}\Upsilon_{\mu t}}}^2 ) \mathcal{B}\Bigg(\frac{m_t^2}{m^2_{S^{1/3}_2}}\Bigg) ]\,, \\
        &\Delta a_\mu^{S^{2/3}} = \frac{3m_\mu^2 \abs{{\color{blue}\Omega_{d\mu}}}^2}{36\pi^2 m^2_{S^{2/3}}} \mathcal{C}\Bigg(\frac{m_d^2}{m^2_{S^{2/3}}}\Bigg)\,,
    \end{aligned}
\end{equation}
where $d = d, s, b$, and $m_t$ is the top quark mass. Here, the loop functions are defined as
\begin{equation}
\label{eq:loop_functions}
\begin{aligned}
&\mathcal{A}(x) = \frac{7-8x+x^2+2(2+x)\ln x}{(1-x)^3}\,,\\
&\mathcal{B}(x) = \frac{1 + 4x - 5x^2 + 2x(2+x)\ln x}{(1-x)^4}\,, \\
&\mathcal{C}(x) = \frac{x(5 - 4x - x^2 + (2+4x)\ln x)}{(1-x)^4}\,.
\end{aligned}
\end{equation}
Note that the contributions from the $1/3e$ LQs play the dominant role, as they contain contributions enhanced by $m_t/m_\mu$ as can be seen from \eqref{eq:muon_g2_LQs}.. Additionally, in the scenarios where the mixing is small (${\color{amber} a_1} \rightarrow 0$), then only the first eigenstate contributes, since the contribution of the second one scales with $\sin^2\theta$. Note that the presence of diagrams such as the ones in Fig.~\ref{fig:muon_g2} implies that LFV graphs also exist (and amount to replacing the external muons with any other combination of charged leptons), leading to transitions such as e.g.~$\mu\rightarrow e\gamma$ or $\tau\rightarrow \mu\gamma$. Therefore, sizeable chirality flipping contributions proportional to $e.g.$~${\color{red} \Theta_{et}} {\green \Upsilon_{\mu t}}$ or ${\color{red} \Theta_{\tau t}} {\color{ForestGreen} \Upsilon_{e t}}$ can efficiently generate large corrections to tightly constrained LFV observables and must be taken into account when finding viable parameter space domains.

\subsubsection{$R_{D,D^*}$ flavour anomaly}

In recent years, an intriguing set of anomalies has emerged, showing deviations from LFU predicted by the SM. The experiments conducted at BaBar \cite{BaBar:2012obs,BaBar:2013mob}, Belle \cite{Belle:2015qfa,Belle:2016ure,Belle:2017ilt} and LHCb \cite{LHCb:2015gmp} concerned tree-level decays of $B$ mesons to final states with a $\tau$ lepton, specifically,
\begin{equation}\label{eq:Rddtsar_1}
    R_{D^{(*)}} \equiv \frac{\textrm{BR}\left( B \rightarrow D^{(*)} \tau \Bar{\nu}_\tau \right) }{\textrm{BR}\left( B \rightarrow D^{(*)} l \Bar{\nu}_l \right)}, \quad \textrm{with} \quad l=\mu, e \,,
\end{equation}
with $D^{(*)}$ being an (excited state of) $D$ meson and BR -- the branching ratio. This ratio exceeds the SM predictions consistently across different experiments. The way the SM deems these processes to happen is via a $W^-$ boson exchange. The following are the averages of these results as well as the SM prediction \cite{HeavyFlavorAveragingGroup:2022wzx}, 
\begin{align}
    &R_D = 0.339 \pm 0.026 \ \textrm{(stat)}  \pm 0.014 \ \textrm{(syst)}, \quad R_{D, {\scriptsize \mathrm{SM}}} = 0.298 \pm 0.004 \\
    &R_{D^*} = 0.295 \pm 0.010 \textrm{(stat)}  \pm 0.010 \ \textrm{(syst)}, \quad R_{D^*, {\scriptsize \mathrm{SM}}} = 0.254 \pm 0.005
\end{align}
measured with a dilepton invariant mass squared between $0 < q^2 < 10~\mathrm{GeV^2}$, showing a discrepancy of 1.4$\sigma$ and 2.8$\sigma$, respectively, when compared to the values predicted by the SM. Here, the SM/BSM prediction is taken from \texttt{flavio} package \cite{Straub:2018kue}, which is based on \cite{MILC:2015uhg,Bigi:2016mdz}. Taking into account the correlated nature of these observables, the difference between the experiment and the SM amounts to 3.3$\sigma$.
\begin{figure*}[htb!]
	\centering
	\captionsetup{justification=raggedright}
	\begin{subfigure}{0.35\textwidth}
		\centering
		\includegraphics[width=\textwidth]{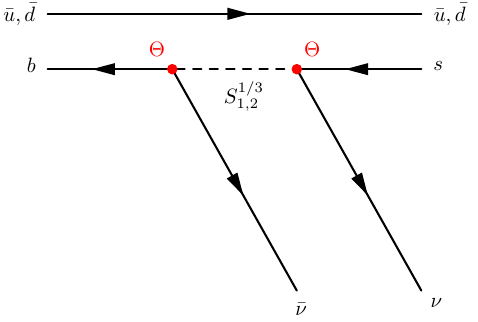}
		\caption{}
	\end{subfigure}
    \hspace*{2em}
	\begin{subfigure}{0.35\textwidth}
		\centering
		\includegraphics[width=\textwidth]{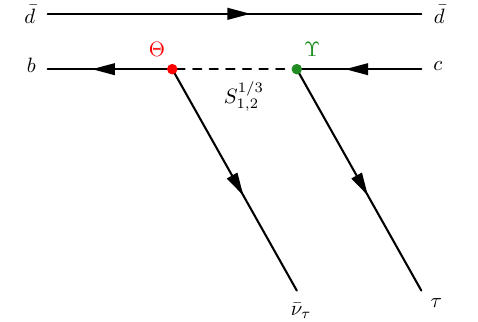}
		\caption{}
	\end{subfigure}
    \caption{On the left, we show an additional contribution from the LQ to the transition $B \rightarrow K^+ \bar{\nu}\nu$, while on the right we present the model's impact on the $B\rightarrow D^{(*)}\tau\bar{\nu}_\tau$ decay.}
    \label{Image:RD}
\end{figure*}

Within the context of the LQ model, this tension can be alleviated through a tree-level exchange of the two $1/3e$ LQs, as shown in Fig.~\ref{Image:RD}. Similarly to $a_\mu$, how much each of them contributes to this observable depends on the size of ${\amber a_1}$. Here, we note that the $S^{2/3}$ does not contribute to this observable. The same Yukawa matrices ${\color{red} \Theta}$ and ${\color{ForestGreen} \Upsilon}$ that played a role in $a_\mu$ are also present here, albeit through distinct matrix elements. As noted in \cite{Bardhan:2016uhr}, $R_D$ and $R_{D^*}$ are impacted by different operators, namely, the $R_D$ transition is dominated by scalar operator $(\bar{c}b_\mathrm{L})(\bar{\tau}_\mathrm{R}\nu_\tau)$, which in turns implies that this observable is enhanced by real couplings, while the $R_{D^*}$ transition is primarily driven by the pseudo-scalar operator $(\bar{c}\gamma_5b_\mathrm{L})(\bar{\tau}_\mathrm{R}\nu_\tau)$ which prefers imaginary couplings. Hence, a complex parametrisation of both ${\color{red} \Theta}$ and ${\color{ForestGreen} \Upsilon}$ allows for an easier fit of both observables. Simultaneously, the $R_{K,K^*}^{\nu\nu}$ observable, which is defined as the ratio between the model prediction for $\mathrm{BR}(B^{+(0)}\rightarrow K^{*+(0)}\nu\nu)$ and the corresponding SM prediction, is also induced at tree-level via the virtual exchange of the same LQs, through the ${\color{red}\Theta}$ Yukawa couplings. In turn, maximizing $R_{D^{*}}$ can also result in larger contributions to $R_{K,K^*}^{\nu\nu}$, in particular, if ${\color{red} \Theta}$ contains additional sizeable entries. Notice that a recent measurement by the Belle II Collaboration \cite{BelleII_Knunu} points towards a deviation of the $B^{+(0)}\rightarrow K^{*+(0)}\nu\nu$ branching ratio, whose value is measured to be higher than that of the SM prediction. Indeed, the preference for a larger $R_D$ favours an enhancement of $\mathrm{BR}(B^{+(0)}\rightarrow K^{*+(0)}\nu\nu)$ in our model due to the presence of a shared coupling as can be seen in Fig.~\ref{Image:RD}. This suggests good prospects for accommodating the new result. However, our numerical analysis was performed before the recent announcement and therefore one has considered a $R_{K,K^*}^{\nu\nu}$ to be SM-like, leaving a dedicated analysis for future work.

\subsubsection{CDF $W$-mass anomaly}

A recent measurement by the CDF collaboration seemed to indicate that a substantial tension between the experimental value of the $W$ mass, and the corresponding SM prediction \cite{CDF:2022hxs}, amounting to a $7\sigma$ deviation, well above the threshold for discovery. However, no independent measurement with such level of precision has so far been made, while the other existing measurements \cite{ATLAS:2017rzl,LHCb:2021bjt,ATLAS:2023fsi} point towards a consistent description of the SM. Either way, combining the CDF result with the earlier measurements leads to a tension of $3.7\sigma$ \cite{deBlas:2022hdk}, which is still below the discovery threshold.

Corrections to the $W$-mass can be parametrised through deviations of the EW precision observables $S$, $T$ and $U$ \cite{Strumia:2022qkt}, particularly, the $T$ parameter. While only the $T$ parameter is needed to analyse the $W$-mass deviations, the other parameters are also impacted and are taken into account in the numerical analysis, especially since they are strongly correlated with each other. Alterations to the $T$ parameter can be expressed in terms of corrections to the self-energy of the $W$ boson as 
\begin{equation}\label{eq:self_energy}
    \Pi_{WW}(0) = \adjincludegraphics[valign=c, height=1.4cm, raise=2.89\baselineskip]{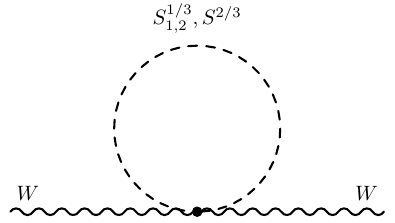} + \adjincludegraphics[valign=c, height=1.4cm, raise=-0.3\baselineskip]{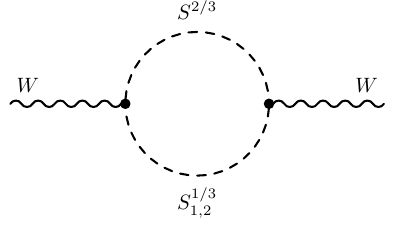} \,,
\end{equation}
such that the $T$ parameter scales as \cite{Crivellin:2020ukd}
\begin{equation}\label{eq:T_par}
    T \sim \frac{1}{\alpha M_W^2} \ln(\frac{m_{S^{2/3}}}{m_{S_a^{1/3}}})\qty(\frac{m_{S^{2/3}}}{m_{S_a^{1/3}}}-1)^{-1} \,, 
\end{equation}
where $\alpha\sim1/137$ is the fine-structure constant. Hence, for non-zero $T$ required by the CDF experiment, we must have that $m_{S^{2/3}} \neq m_{S_a^{1/3}}$. Following the discussion above, the degeneracy between the doublet components can be lifted either by having a non-zero mixing ${\color{amber} a_1}$ (which is always true, since a non-zero value is needed for a viable neutrino description), or a non-zero value for the quartic coupling ${\color{violet} g_{HR}^\prime}$.

\subsubsection{Constraints on the parameter space}

Besides the observables so far discussed, there is a plethora of other constraints that must be taken into account. Here, we shall discuss only the stringiest ones, while the full list used in our numerical calculations is shown in Tab.~\ref{tab:flav_obs_exp}. Notice that we have not assumed any flavour ansatz (see e.g.~\cite{Marzocca:2021azj}), which implies that ${\color{red}\Theta}$, ${\color{blue} \Omega}$ and $\color{ForestGreen} \Upsilon$ are taken to be generic $3\times 3$ complex matrices. While assuming texture zeros would simplify our analysis, these would be artificial as no symmetry in the Lagrangian is present to protect them from being radiatively generated. Indeed, ${\color{red}\Theta}$ and ${\color{blue} \Omega}$ need to have a generic structure if one wishes to explain neutrino physics\footnote{While most elements need to be non-zero, it is possible to have some zero entries, as long as at least two neutrinos remain massive. In this work, we have not explored what are the minimal textures that can still lead to viable neutrino phenomenology.}.

\begin{figure*}[htb!]
	\centering
	\captionsetup{justification=raggedright}
	\begin{subfigure}{0.25\textwidth}
		\centering
		\includegraphics[width=\textwidth]{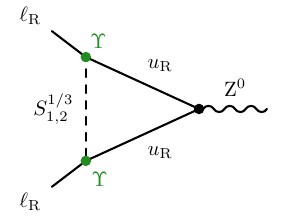}
		\caption{}
	\end{subfigure}
    \hspace*{1em}
	\begin{subfigure}{0.25\textwidth}
		\centering
		\includegraphics[width=\textwidth]{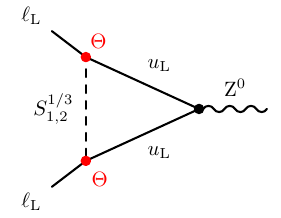}
		\caption{}
	\end{subfigure}
    \hspace*{1em}
	\begin{subfigure}{0.25\textwidth}
		\centering
		\includegraphics[width=\textwidth]{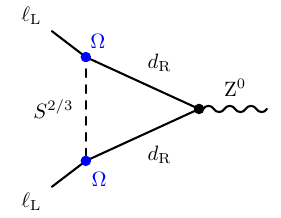}
		\caption{}
	\end{subfigure}
    \caption{Some of the one-loop contributions mediated by the model's LQs to the flavour conserving and non-conserving decays $\mathrm{Z^0}\rightarrow \ell\ell^\prime$. There exist additional wave contributions to the one-loop amplitude (see e.g.~\cite{Arnan:2019olv}) as well as similar diagrams to those shown Fig.~\ref{fig:muon_g2}, which are not shown here but are taken into account in the numerical calculations.}
    \label{fig:Zboson_decays}
\end{figure*}

Besides constraints from LFV such as $\mu\rightarrow e\gamma$, which are generated through topologies identical to those of Fig.~\ref{fig:muon_g2}, there are also constraints coming from LFV decays of the $\mathrm{Z^0}$ boson such as e.g.~$\mathrm{Z^0}\rightarrow \mu \tau$, where our model's main contributions are displayed in Fig.~\ref{fig:Zboson_decays}. Not only that, we also need to worry about the flavour conserving cases as those are very well measured at LEP \cite{ALEPH:2005ab} and tightly constrain LQ couplings. For this, we have considered the full one-loop expressions as determined by P.~Arnan \textit{et.~al} \cite{Arnan:2019olv}. Higgs LFV decays are also relevant and are considered in the analysis. The diagrams are identical to those shown in Fig.~\ref{fig:Zboson_decays} by replacing $\mathrm{Z^0}$ with the Higgs boson.

\begin{figure*}[htb!]
	\centering
	\captionsetup{justification=raggedright}
	\begin{subfigure}{0.32\textwidth}
		\centering
		\includegraphics[width=\textwidth]{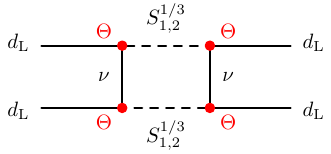}
		\caption{}
	\end{subfigure}
	\begin{subfigure}{0.32\textwidth}
		\centering
		\includegraphics[width=\textwidth]{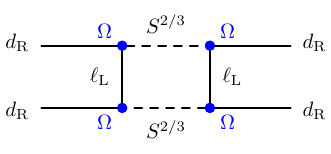}
		\caption{}
	\end{subfigure}    
	\begin{subfigure}{0.23\textwidth}
		\centering
		\includegraphics[width=\textwidth]{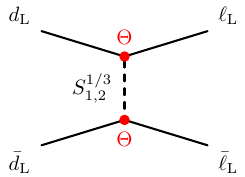}
		\caption{}
	\end{subfigure}
    \caption{Dominant one-loop box contributions to CP-sensitive meson mixing constraints (first two diagrams), and the dominant tree-level graph that contributes to the Kaon decays (third graph).}
    \label{fig:one_box_Meson}
\end{figure*}

Due to the complex parametrisation of the Yukawa couplings, strong constraints also come from CP-sensitive observables as well as from quark flavour violating (QFV) decays. In the former, the electric dipole moments (EDMs) of the charged leptons represent a strong constraint on the allowed sizes of the imaginary parts of the Yukawas couplings. Contributions to these observables come at one-loop level via identical diagrams to the ones shown in Fig.~\ref{fig:muon_g2}, with the only difference being that the EDMs are proportional to the imaginary part of Yukawa couplings, and not to the real part as in the anomalous magnetic moment. On the other hand, QFV decays strongly constrain the allowed couplings, in particular, for the ${\color{blue} \Omega}$ and ${\color{red} \Theta}$ matrices. The main constraints come from the meson mixing observables ($\Delta M_d$, $\Delta M_s$, $\epsilon_k$, $\epsilon^\prime/\epsilon$ and $\phi_s$), which are sensitive to the additional sources of CP violation coming from the Yukawa couplings. These observables are impacted through one-loop box diagrams involving the exchange of virtual LQ states, with some examples seen in Fig.~\ref{fig:one_box_Meson}. Besides this, fully leptonic rare Kaon decays such as e.g.~$K_L^0 \rightarrow \mu^+\mu^-$ or semi-leptonic ones such as $K^+ \rightarrow \pi^0 \mu^+ \nu$ are particularly important. These decays can be written as functions of the Wilson coefficients for the semi-leptonic operators $(\bar{L}\gamma_\mu L)(\bar{Q}_\mathrm{L}\gamma^\mu Q_L)$ (for the full list of relevant operators, see Tab.1 of \cite{Bobeth:2017ecx}), which are generated already at tree-level in our LQ model, via the last diagram shown in Fig.~\ref{fig:one_box_Meson}. Atomic parity constraints \cite{Dorsner:2016wpm,Crivellin:2021egp} are also included in the numerical analysis. 

\begin{figure*}[b!]
	\centering
	\captionsetup{justification=raggedright,singlelinecheck=false}
	\includegraphics[width=\textwidth]{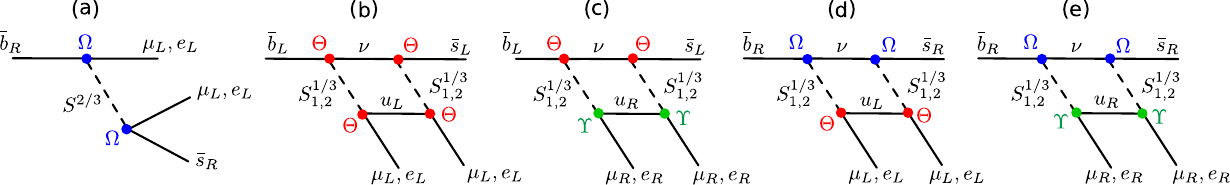} 
	\caption{Box and tree-level diagrams responsible for generating the LQ contributions to $R_{K,K^*}$ and $B_s/B_0 \rightarrow \mu^+\mu^-$ processes. }
	\label{fig:Bphysics_diagrams}
\end{figure*}

There are additional constraints coming from $B$-physics. Namely, we consider the current limits on $\mathrm{BR}(B_s/B_0 \rightarrow \mu^+\mu^-)$ as well as the LFU observable $R_{K,K^*}$. The $b\rightarrow s\ell\ell$ observables are impacted via both tree-level and box diagrams involving the virtual exchange of the $S$ LQ as shown in Fig.~\ref{fig:Bphysics_diagrams}. These can be parameterised in terms of the Wilson operators \mbox{$O_9^{\ell} \propto C_9^{bs\ell \ell} (\bar{s} \gamma^\mu P_\mathrm{L} b)(\bar{\ell} \gamma_\mu \ell)$} and $O_{10}^{\ell} \propto C_{10}^{bs\ell \ell} (\bar{s} \gamma^\mu P_\mathrm{L} b)(\bar{\ell} \gamma_\mu \gamma^5 \ell)$ for diagrams (b) and (c) and \mbox{$O_9^{\prime \ell} \propto C_9^{\prime bs\ell \ell} (\bar{s} \gamma^\mu P_\mathrm{R} b)(\bar{\ell} \gamma_\mu \ell)$} and $O_{10}^{\prime \ell} \propto C_{10}^{\prime bs\ell \ell} (\bar{s} \gamma^\mu P_\mathrm{R} b)(\bar{\ell} \gamma_\mu \gamma^5 \ell)$ for diagrams (a), (d) and (e). As usual, the $C$-factors are the Wilson coefficients and $\ell = e,\mu$. 

To finalise, since no positive results have been reported at colliders, direct searches for LQs also pose limits on their allowed masses. Constraints coming from pair production channels at the ATLAS and CMS experiments \cite{ATLAS:2021jyv,ATLAS:2021oiz,CMS:2018iye,CMS:2018oaj} provide a lower bound, approximately, between 1 and 1.5 TeV, considered in this work.

\section{Numerical methodology}\label{sec:methods}
We perform a parameter space scan considering a plethora of different observables as listed in Appendix~\ref{appendix}. The experimental limits were taken from the latest PDG review \cite{ParticleDataGroup:2022pth}. For an extensive analysis featuring a large number of observables we have implemented the model in \texttt{SARAH} \cite{Staub:2013tta}, where interaction vertices and one-loop contributions relevant for such observables were determined. Outputs were then generated for numerical evaluation in \texttt{SPheno} \cite{Porod:2011nf}, where the particle spectrum and the necessary Wilson coefficients to be used in \texttt{flavio} \cite{Straub:2018kue} were calculated. \texttt{SPheno} calculates the Wilson coefficients in the WET basis, where the LFV coefficients are evaluated at the $\mathrm{Z^0}$ mass scale ($\mu = 91~\mathrm{GeV}$) and the QFV coefficients are evaluated at the top mass scale ($\mu = 160~\mathrm{GeV}$). Renormalisation-group running between these scales and those of the low-energy observables is done \texttt{flavio} through a interface with the \texttt{wilson} \cite{Aebischer:2018bkb} package. With this in consideration, we have constructed a $\chi^2$ function, defined as \cite{Altmannshofer:2021qrr}
\begin{equation}\label{eq:likelihood}
\chi^2 = (\mathcal{O}_{\mathrm{exp}} - \mathcal{O}_{\mathrm{th}})^{\mathrm{T}} (\bm{\Sigma}_{\mathrm{th}} + \bm{\Sigma}_{\mathrm{exp}})^{-1} (\mathcal{O}_{\mathrm{exp}} - \mathcal{O}_{\mathrm{th}})    
\end{equation}
using the observables indicated in Appendix~\ref{appendix}. Notice that the method used to calculate each of the observales considered in this work is indicated in the first column of Tabs.~\ref{tab:flav_obs} and \ref{tab:flav_obs_1}. In \eqref{eq:likelihood} $\mathcal{O}_{\mathrm{exp}}$ and $\mathcal{O}_{\mathrm{th}}$ represent vectors of experimental values and the model prediction, respectively, while $\bm{\Sigma}_{\mathrm{exp}}$ is the experimental covariance and $\bm{\Sigma}_{\mathrm{th}}$ is the theoretical one. Both covariance matrices can be computed using well-known formulas
\begin{equation}
\begin{aligned}
\bm{\Sigma}_{\mathrm{th}} = \sigma_\mathrm{th} \rho_\mathrm{th} \sigma_\mathrm{th}, \quad\quad\mathrm{and}\quad\quad \bm{\Sigma}_{\mathrm{exp}} = \sigma_\mathrm{exp} \rho_\mathrm{exp} \sigma_\mathrm{exp},
\end{aligned}
\end{equation}
where $\sigma_{\mathrm{th}}$ ($\sigma_{\mathrm{exp}}$) are diagonal matrices whose entries are the 1$\sigma$ theoretical (experimental) errors and $\rho_{\mathrm{th}}$ ($\rho_{\mathrm{exp}}$) are the theoretical (experimental) correlation matrices. For the experimental inputs, the experimental uncertainties can be easily extracted from literature, while for the experimental correlations, we extract those that are available and neglect if those do not exist. The various uncertainties and correlations were taken from the references inside Tab.~\ref{tab:flav_obs_exp}.

For the theoretical inputs, the errors can be computed inside \texttt{flavio} \cite{Straub:2018kue}, with the function \texttt{flavio.np\_uncertainty} for each of the observables of interested. This also takes into account potential hadronic uncertainties that exist for observables sensitive to these. As for the theoretical correlations, those can be computed from our entire dataset using standard methods available in statistics libraries. In our case, we have use Pearson's algorithm through the \texttt{pandas} package \cite{mckinney-proc-scipy-2010}. Since the LQ mass scale is well above the scale of observables that we analyse, one needs to run the various couplings to the appropriate scales, which is done with the  \texttt{wilson} \cite{Aebischer:2018bkb} package.

With this in mind, a numerical scan over all relevant parameters of the model is then conducted. In particular, we perform an inclusive logarithmic scan over the various parameters within the ranges shown in Tab.~\ref{tab:sample}.

\begin{table}[htb!]
	\centering
    \captionsetup{justification=raggedright}
	\begin{tabular}{c|c|c|c}
		\toprule
		$m_{S^{1/3}_1}$, $m_{S^{1/3}_2}$  (TeV) & $g_{HS}, g_{HR}, {\color{violet}g_{HR}^\prime}$ & ${\color{ForestGreen} \abs{\Upsilon}}, {\color{red} \abs{\Theta}}, {\color{blue} \abs{\Omega}}$ & ${\color{amber}a_1}$ (GeV)  \\
		\midrule
		$  \left[ 1.5 , 10 \right]$ & $ \left[10^{-8}, 4 \pi\right]$ &  $\left[10^{-8}, \sqrt{4 \pi} \right]$ & $\left[10^{-8}, 100\right]$
		\\
		\bottomrule
	\end{tabular}
	\caption{Ranges used for the free parameters during the numerical scan. The values for the masses of the SM fields and corresponding mixings were varied within the allowed experimental ranges.}
	\label{tab:sample}
\end{table}

Once valid solutions are found within the first initial random scan, we then use these points as seeds for finding new solutions in subsequent runs, by perturbing around the valid couplings/masses in order to find new consistent points. Do note that not all ${\color{red}\Theta}$ and ${\color{blue}\Omega}$ Yukawas are free parameters, with some being calculated through the inversion procedure of the neutrino mass matrix. In this regard, within the GitHub page (\url{https://github.com/Mrazi09/LQ-flavour-project}) one can find auxiliary \texttt{jupyter} notebooks, which demonstrate how to numerically implement the inversion procedure for the neutrinos/quark/charged leptons and LQs (named \texttt{Neutrino\_inversion.ipynb}) as well as how to utilise the data to extract the relevant neutrino observables (named \texttt{Read\_neutrino.ipynb}). 

We have performed parameter space scans considering three cases: a) $a_\mu$ and $m_W$ both consistent with the SM, b) only $m_W$ consistent with the SM and c) neither of them consistent with the SM prediction. In all three scenarios we do take into account the LFU deviation in $R_{D,D^*}$ as well as keeping the remaining constraints (for a complete list, see Appendix~\ref{appendix}) under control. The generic parameterization of our couplings also implies that kaon decays \cite{Mandal:2019gff} and atomic parity violating constraints \cite{Dorsner:2016wpm,Crivellin:2021egp} are relevant for our parameter scan. We use as input parameters the quark and charged lepton masses as well as the CKM and PMNS mixing matrices, which we allow to vary within their two sigma uncertainty. Regarding neutrino masses, we focus on a normal ordering scenario with three massive states.

\section{Numerical results}\label{sec:results}

\begin{figure*}[htb!]
	\centering
	\captionsetup{justification=raggedright}
	\begin{subfigure}{0.32\textwidth}
		\centering
		\includegraphics[width=\textwidth]{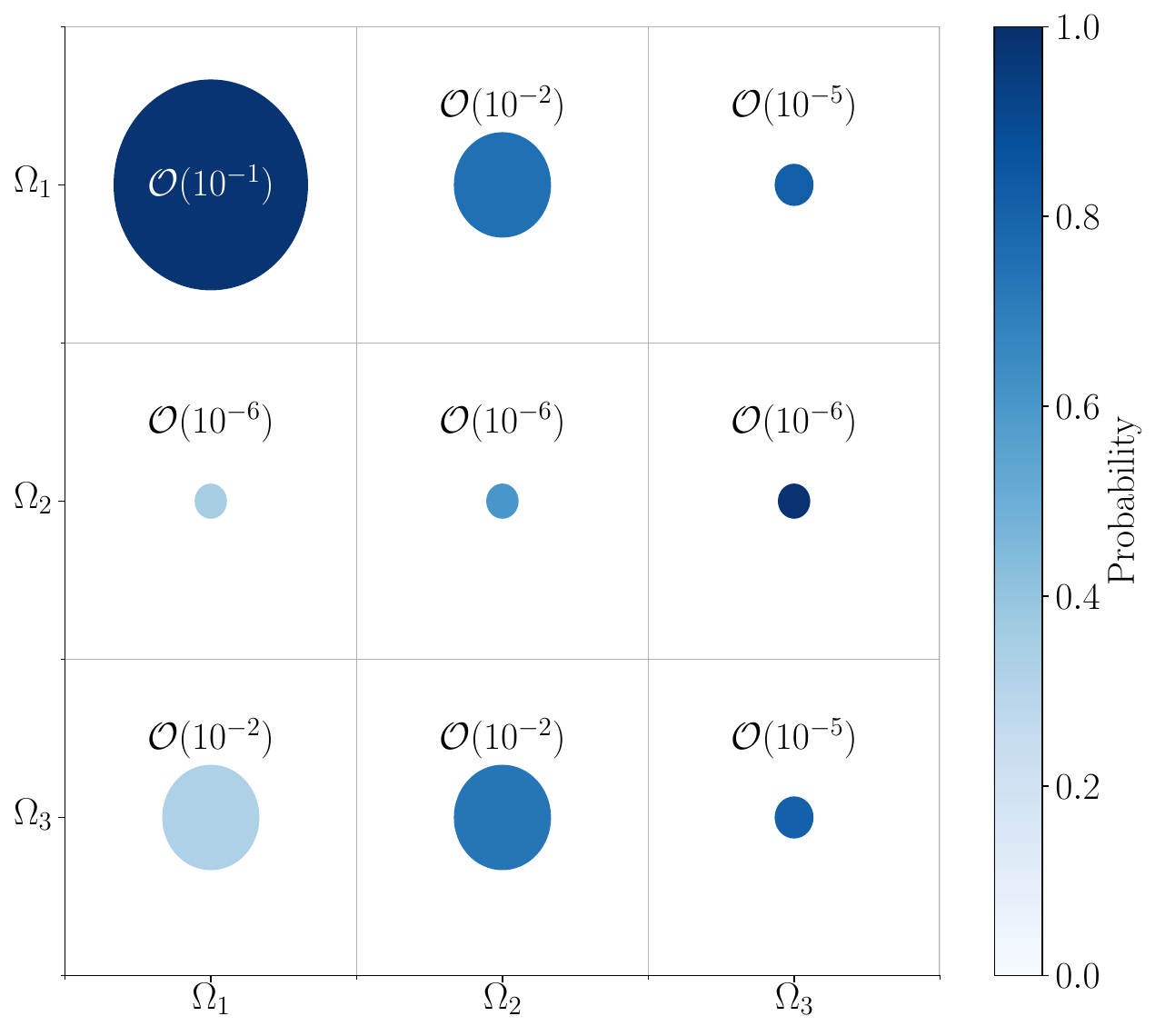}
		\caption{}
	\end{subfigure}
	\begin{subfigure}{0.32\textwidth}
		\centering
		\includegraphics[width=\textwidth]{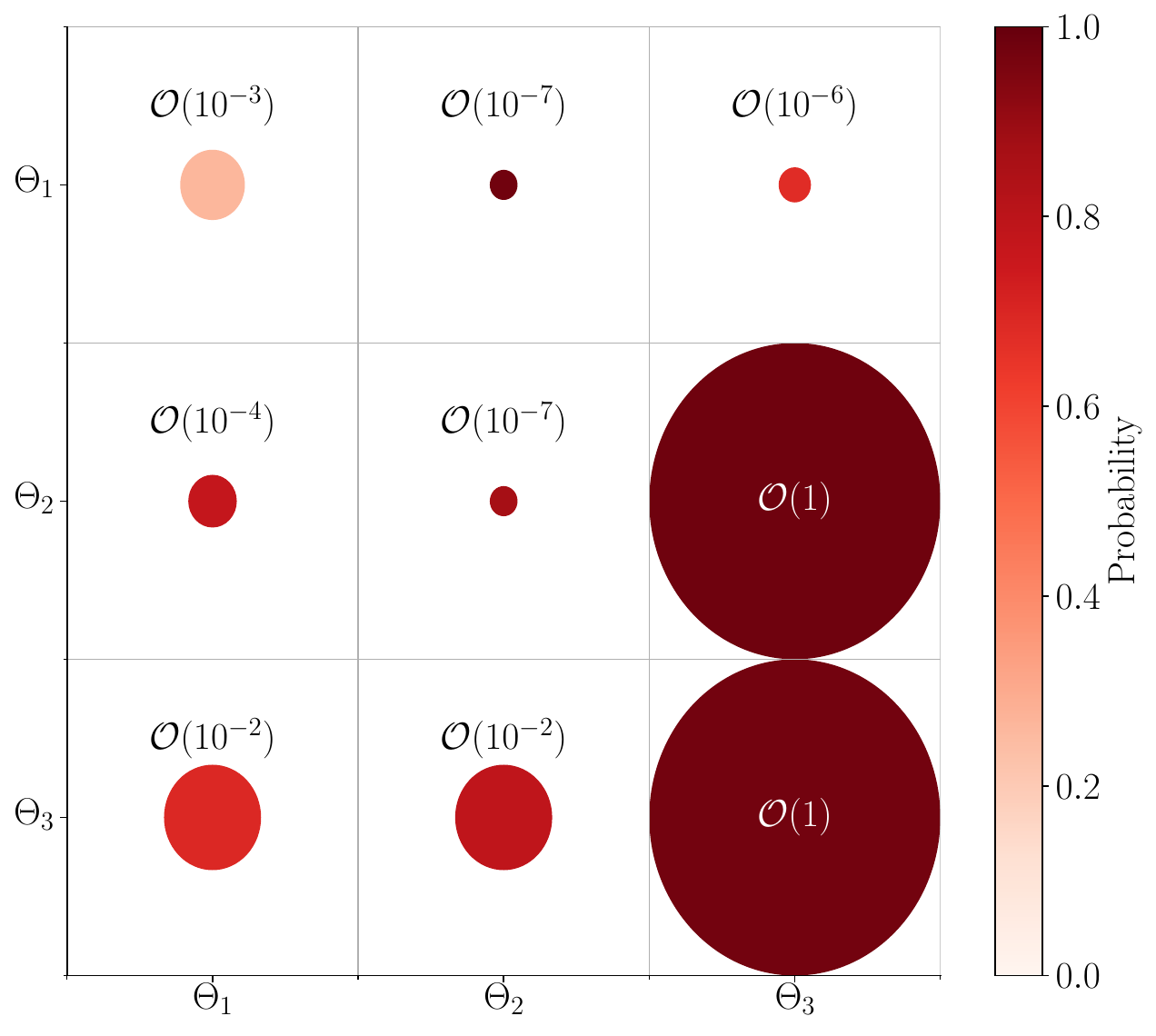}
		\caption{}
	\end{subfigure}    
	\begin{subfigure}{0.32\textwidth}
		\centering
		\includegraphics[width=\textwidth]{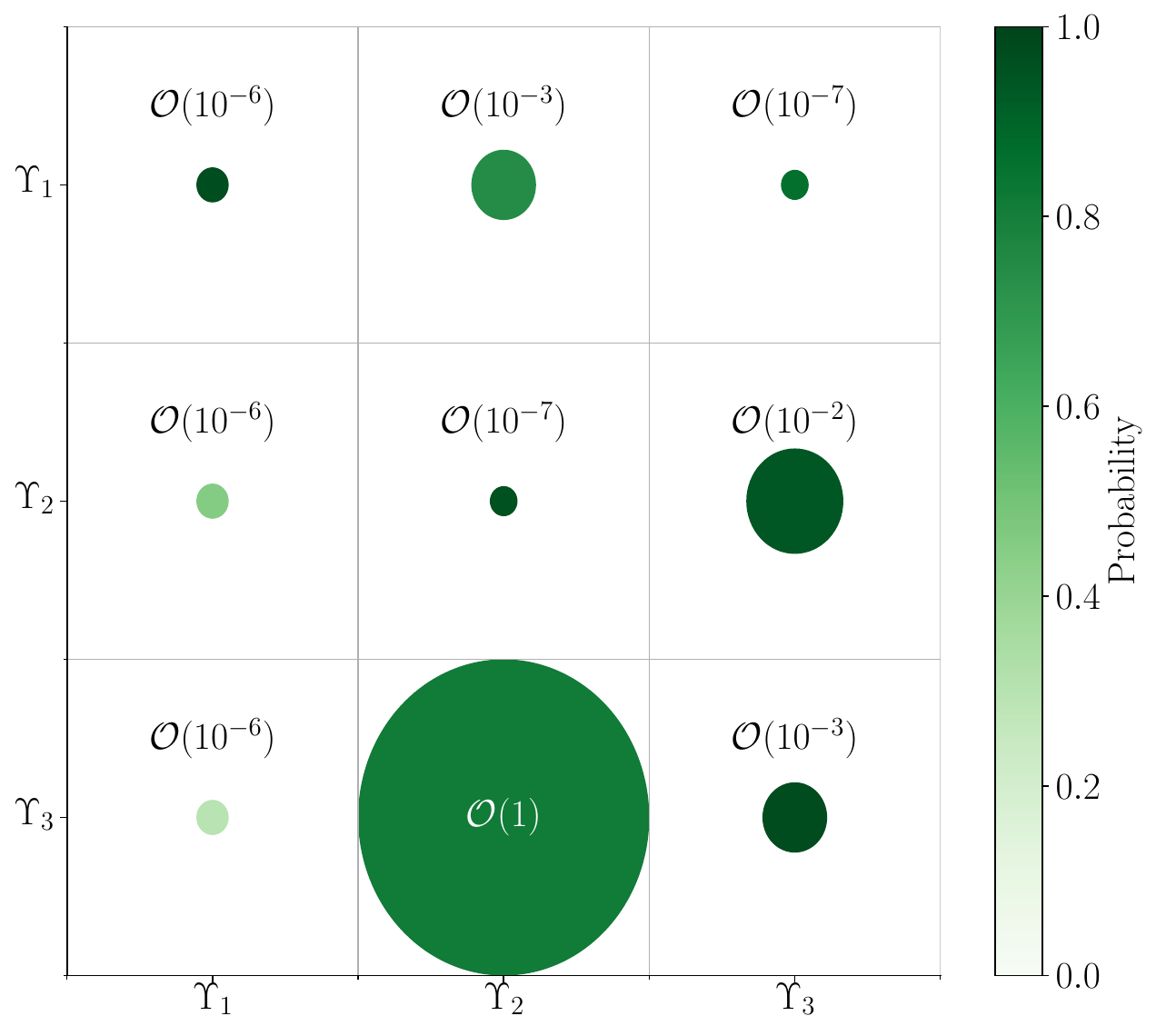}
		\caption{}
	\end{subfigure}
	\caption{Preferred sizes for each of the LQ Yukawas couplings: (a) {\blue $\Omega$}, (b) {\red $\Theta$} and (c) {\green $\Upsilon$}. The radius of the circumference represents the size of the absolute value of the coupling while the color gradation describes how frequent such a magnitude appears in the scan, $i.e.$~darker shades indicate more preferred sizes, according the probability $\mathcal{P}(X) = \tfrac{N(X)}{N_\mathrm{tot}}$ with $N(X)$ the number of points with order $X$ in our data.}
	\label{fig:textures}
\end{figure*}

\begin{figure*}[htb!]
	\centering
	\captionsetup{justification=raggedright}
	\begin{subfigure}{0.40\textwidth}
		\centering
		\includegraphics[width=1.10\textwidth]{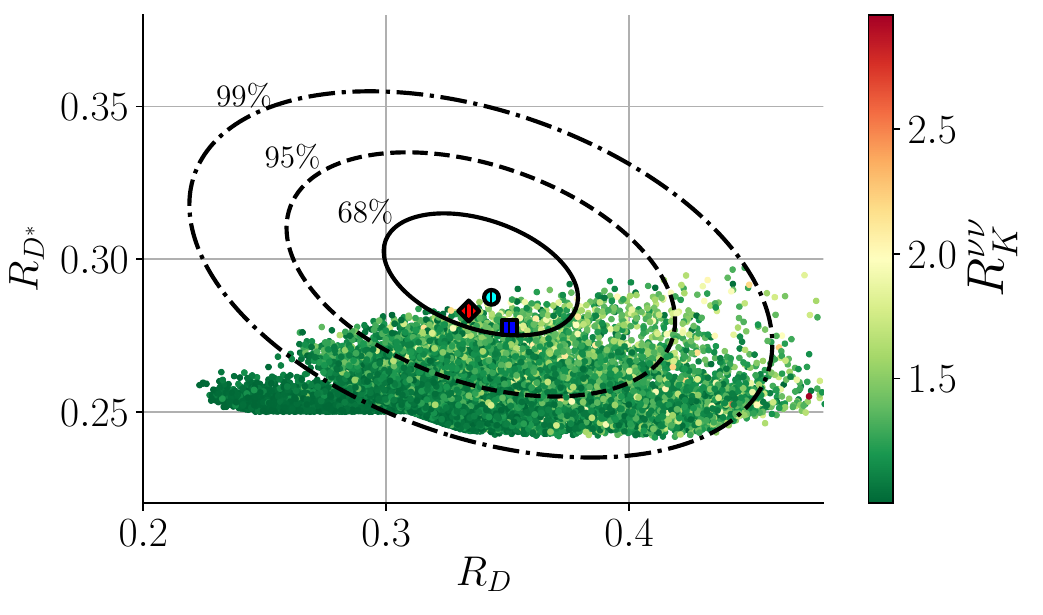}
		\caption{}
	\end{subfigure}
	\hspace*{5em}
	\begin{subfigure}{0.40\textwidth}
		\centering
		\includegraphics[width=1.10\textwidth]{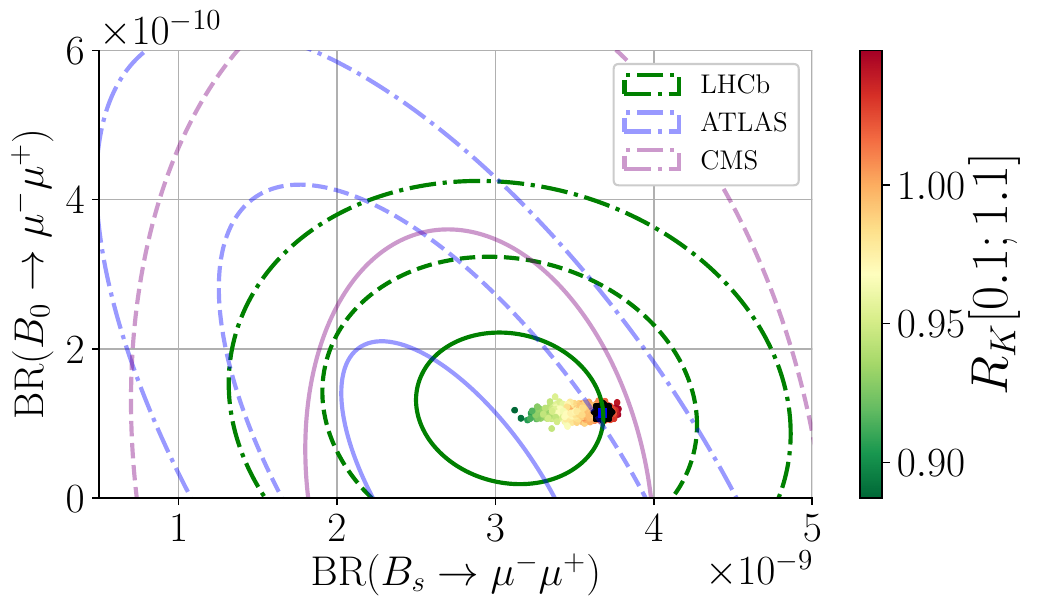}
		\caption{}
	\end{subfigure}
	\hspace*{0.3em}
	\begin{subfigure}{0.40\textwidth}
		\centering
		\includegraphics[width=1.05\textwidth]{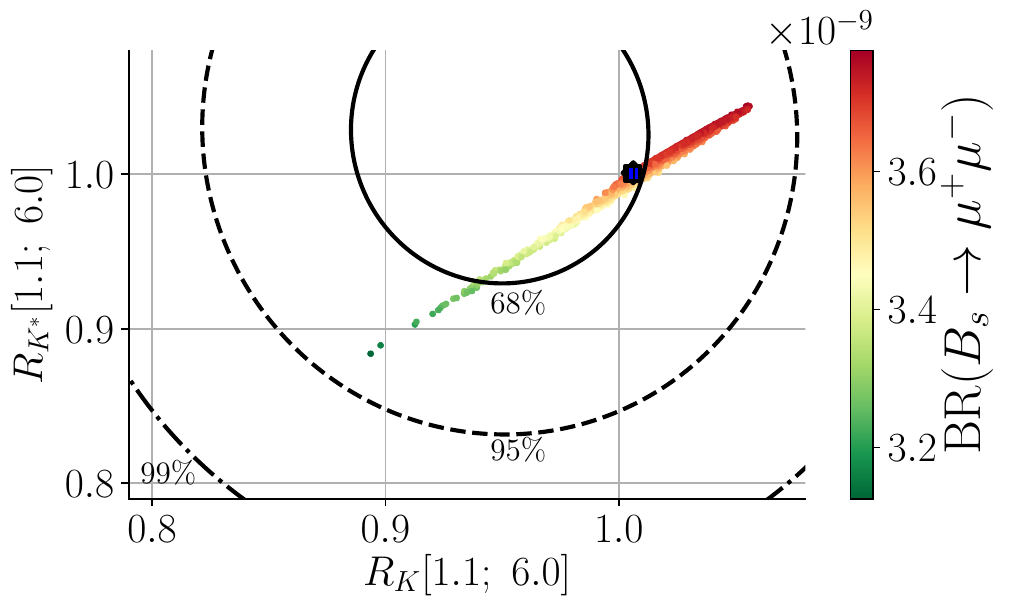}
		\caption{}
	\end{subfigure}
	\hspace*{5em}
	\begin{subfigure}{0.40\textwidth}
		\centering
		\hspace*{-1.2em}
		\includegraphics[width=1.05\textwidth]{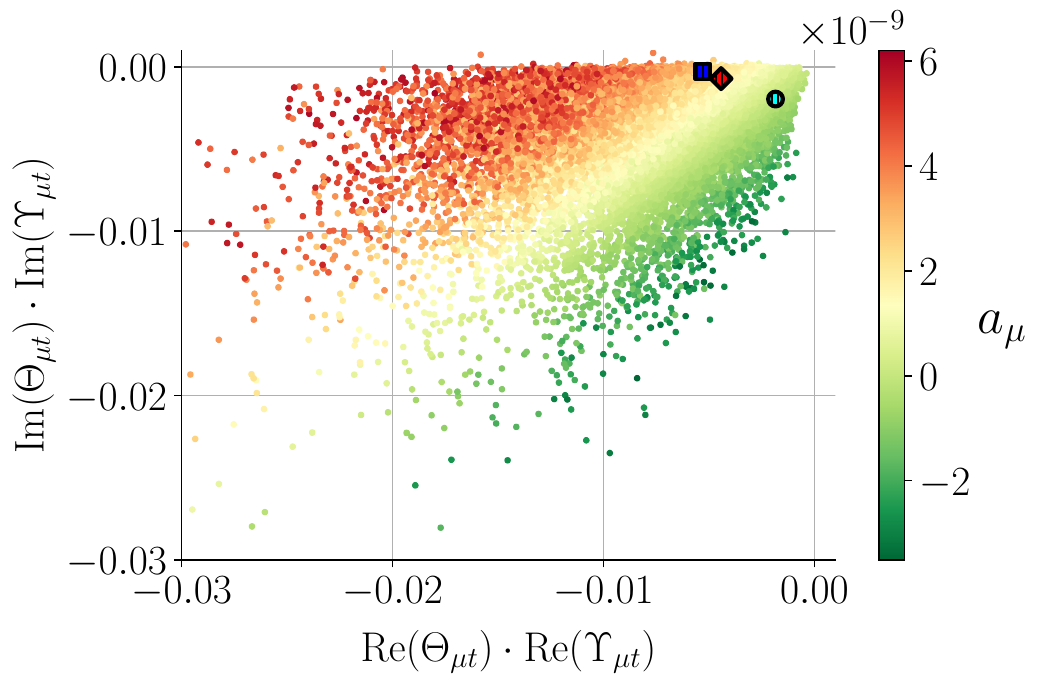}
		\caption{}
	\end{subfigure}
	\hspace*{-0.9em}
	\begin{subfigure}{0.40\textwidth}
		\centering
		\includegraphics[width=1.12\textwidth]{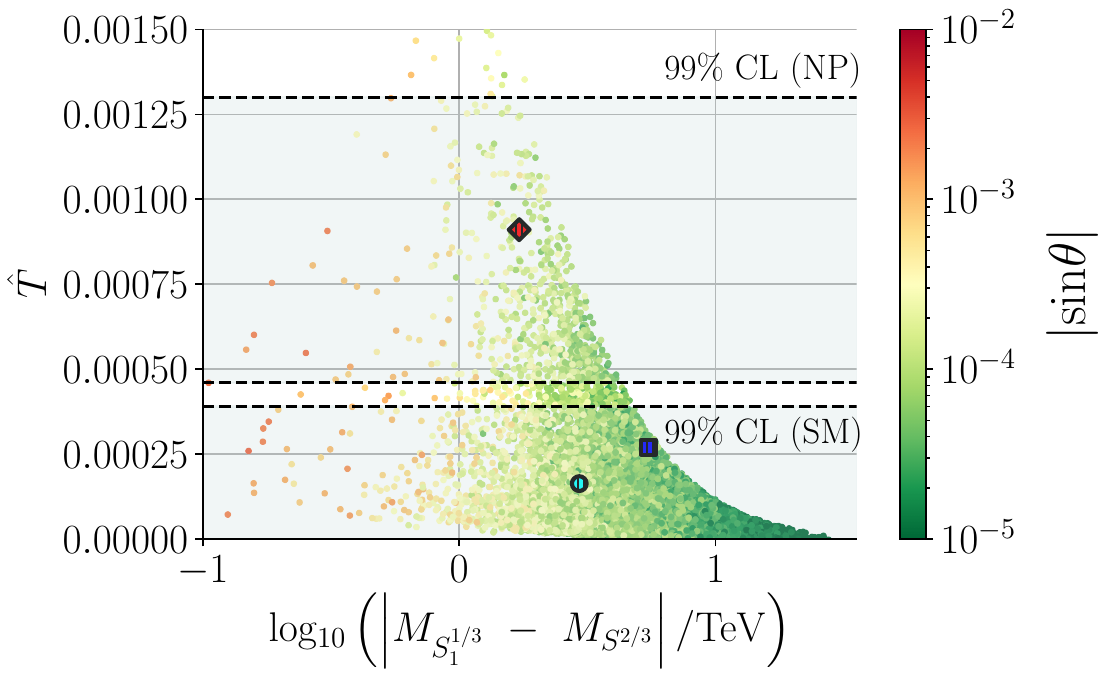}
		\caption{}
	\end{subfigure}
    \hspace*{5em}
 	\begin{subfigure}{0.40\textwidth}
		\centering
		\includegraphics[width=1.00\textwidth]{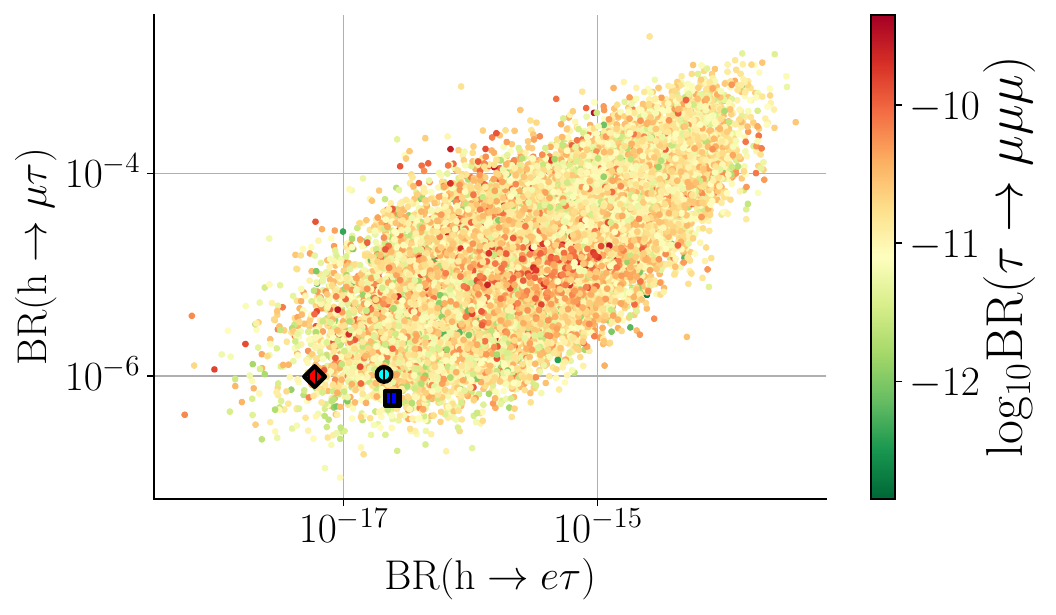}
		\caption{}
	\end{subfigure}
	\caption{Scatter plots of selected observables analysed in this work. In (a) we plot the $R_{D^*}$ as a function of $R_D$ with $R_K^{\nu\nu}$ in the color scale. In (b), the branching ratios of $B_0\rightarrow \mu^-\mu^+$ and $B_s\rightarrow \mu^-\mu^+$, with $R_K[1.1;~6.0]$ in the color scale. In (c) we plot $R_{K^*}$ vs. $R_{K}$ with $\mathrm{BR}(B_s \to \mu \mu)$ in the color scale. In (d) the product of the real and imaginary parts of ${\color{ForestGreen} \Upsilon_{\mu t}}{\color{red} \Theta_{\mu t}}$ are shown with $a_\mu$ in the color gradation while in (e) the $T$ parameter as function of the logarithm of the mass difference between the masses of $S_1^{1/3}$ and $S^{2/3}$ is presented, where the color scale represents the LQ mixing angle and in (f) we show $\mathrm{BR}(\mathrm{Z^0}\rightarrow \mu\tau)$ versus $R_K^{\nu\nu}$ with $\mathrm{log}_{10}\mathrm{BR}(\tau\rightarrow \mu \mu \mu)$ in the colour axis. Areas of phenomenological interest lie inside the contours. For the $T$ parameter, we show the areas of interest for both NP and SM-like cases. The relevant range  for $a_\mu$ lies within $(251 \pm 59)\times 10^{-11}$, while $\mathrm{BR}(\mathrm{Z^0}\rightarrow \mu\tau) < 1.2 \times 10^{-5}$, $R_K^{\nu\nu} < 4.35$ and $\mathrm{BR}(\tau\rightarrow \mu\mu\mu) < 2.1\times 10^{-8}$. The best fit points are marked with a blue square (scenario a), a cyan circle (scenario b) and a red diamond (scenario c).}
	\label{fig:Flav_plots}
\end{figure*}

\begin{table}[htb!]
	\centering
    \captionsetup{justification=raggedright,singlelinecheck=false}
    \begin{tabular}{c|c|c|c|}
		  & $\chi^2/\mathrm{d.o.f}$ & $\chi^{2,\mathrm{SM}}/\mathrm{d.o.f}$ & $(m_{S_1^{1/3}}, m_{S_2^{1/3}}, m_{S^{2/3}})~\mathrm{TeV}$  \\[0.1em] \hline
        Scenario a)  & $1.16$   & $1.26$ & $(1.53\,, 7.02\,, 7.00)$  \\
        Scenario b)  & $1.17$   & $1.66$ & $(1.58\,, 4.50\,, 4.52)$  \\
        Scenario c)  & $1.37$   & $2.46$ & $(1.63\,, 3.30\,, 3.35)$  
	\end{tabular}
	\caption{\label{tab:best_fits} $\chi^2/\mathrm{d.o.f}$ for the obtained best fit points (first column) and for the SM limit (second column) with $\mathrm{d.o.f} = 45$. The LQ masses (in TeV) are indicated in the third column.}
\end{table}

Employing our $\chi^2$ analysis, the results we obtain are summarised in Tab.~\ref{tab:best_fits} where we show the $\chi^2$ and the LQ masses for each scenario. We note that in all three cases the model predictions offer a better fit than that of the SM limit. A numerical scan in the couplings and masses of the LQs is conducted and the main results are highlighted in Figs.~\ref{fig:textures} and \ref{fig:Flav_plots}. In Fig.~\ref{fig:textures} we show the preferred sizes that were found to simultaneously address the studied anomalies and are consistent with neutrino physics and flavor constraints. While the darker shades offer a conclusive estimate the lighter ones allow for some dispersion. This information in combination with the best-fit points can be relevant in proposing searches for LQs at colliders. In particular, taking ${\red \Theta_{\mu t}} \sim \mathcal{O}(1)$, the $\mu^+ \mu^- \to t\bar{t}$ $t$-channel $S^{1/3}$ LQ exchange can be seen as a smoking-gun benchmark scenario of the considered model and a physics case for the future muon collider. For the case of the $S^{2/3}$ LQ, its couplings to $d$-quark can be as large as ${\blue \Omega_{e d}}  \sim {\blue \Omega_{\tau d}} \sim \mathcal{O}(10^{-1})$, which might be sufficiently large to be tested at the LHC in the $t$-channel LQ exchange for $e e$, $\tau \tau$ and $e \tau$ pair production \cite{Faroughy:2016osc,Greljo:2017vvb}. Furthermore, such LQ can be searched for at future hadronic machines such as the HE-LHC or the FCC. In particular, for the best-fit point \eqref{eq:yukawa_numbers_NP}, the future $50~\mathrm{TeV}$ FCC-eh collider offers an opportunity for the $s$-channel process $e d \rightarrow S^{2/3} \rightarrow t\mu$.

In Fig.~\ref{fig:Flav_plots} we demonstrate that for all displayed observables the data can be well accommodated. In particular, we show the three best fit points marked as colored polygons, with the blue, cyan and red denoting scenarios a), b) and c) respectively. In panel (a) one sees that both $R_{D,D^*}$ and $R_K^{\nu \nu}$ can be reconciled simultaneously, which is also displayed in panel (f). In (c), a linear correlation between $R_{K,K^*}$ observables is found, in consistency with previous literature \cite{Altmannshofer:2021qrr,DAmico:2017mtc}. Furthermore, $B_{s,0} \to \mu \mu$ is well-fitted with a strong correlation with  $R_{K,K^*}$, as expected. In panel (d), we note that the combination of ${\color{ForestGreen} \Upsilon_{\mu t}}$ and ${\color{red} \Theta_{\mu t}}$ is the dominant source for the contribution for $a_\mu$ since these couplings induce chirality flipping of the top quark in the internal propagator. In panel (e) we show how the $\hat{T}$ parameter depends on the mass difference between the LQs that originate from the doublet. In colour we show the LQ mixing. We note that for most of the generated points, $S^{1/3}_1$ is essentially the $S$ singlet, as indicated by the yellow and green regions.

\section{Conclusions}
In this paper, we have studied the most economical extension of the SM with two scalar LQs, representing the minimal scenario capable of addressing all measured flavour anomalies as well as explaining neutrino masses and their mixing structure. Additionally, the model can accommodate the measured value of the muon anomalous magnetic moment as well as opening the door for alleviating the CDF-II W mass anomaly, if both observables are confirmed to be inconsistent with
the SM predictions. For the best-fit points the lightest LQ can have a mass around 1.6 TeV, which should be accessible at the high-luminosity phase of the LHC. In this regard, our numerical results have highlighted the preferred sizes for the LQ Yukawa couplings which will be relevant in pinpointing the direction for future searches.

\begin{acknowledgments}
J.G., F.F.F., and A.P.M. are supported by the Center for Research and Development in Mathematics and Applications (CIDMA) through the Portuguese Foundation for Science and Technology (FCT - Funda\c{c}\~{a}o para a Ci\^{e}ncia e a Tecnologia), references UIDB/04106/2020 and UIDP/04106/2020. A.P.M., F.F.F. and J.G. are supported by the projects PTDC/FIS-PAR/31000/2017 amd CERN/FIS-PAR/0021/2021. A.P.M. and J.G. are also supported by the projects CERN/FIS-PAR/0019/2021 and CERN/FIS-PAR/0025/2021.
A.P.M.~is also supported by national funds (OE), through FCT, I.P., in the scope of the framework contract foreseen in the numbers 4, 5 and 6 of the article 23, of the Decree-Law 57/2016, of August 29, changed by Law 57/2017, of July 19.
J.G is also directly funded by FCT through a doctoral program grant with the reference 2021.04527.BD.
R.P.~is supported in part by the Swedish Research Council grant, contract number 2016-05996, as well as by the European Research Council (ERC) under the European Union's Horizon 2020 research and innovation programme (grant agreement No 668679).
\end{acknowledgments}

\appendix

\section{Numerical benchmarks for best fit points}
\label{appendix}

If we take that both the $W$ mass and the anomalous magnetic moment of the muon are SM-like, scenario a), then the best fit point found in the scan is

\begin{equation}\label{eq:yukawa_numbers_SMl}
\begin{aligned}
&{\color{ForestGreen} \Upsilon} = \begin{pmatrix}
-7.273 \times 10^{-7}+4.372 \times 10^{-7}i & 0.001174-0.000872i & -2.575 \times 10^{-8}-7.646 \times 10^{-8}i \\
1.862 \times 10^{-6}+1.4 \times 10^{-7}i & -8.78 \times 10^{-8}+6.04 \times 10^{-8}i & -0.00299-0.01222i \\
-7.74 \times 10^{-8}-2.695 \times 10^{-7}i & -0.467+2.298i & 0.0007761+0.0002683i
\end{pmatrix}, \\[0.1em]
&{\color{blue} \Omega} = \begin{pmatrix}
0.00483+0.09936i & -0.01122+0.01725i & 7.48 \times 10^{-6}-1.338 \times 10^{-5}i \\
2.238 \times 10^{-6}+5.57 \times 10^{-7}i & 6.502 \times 10^{-7}-6.579 \times 10^{-7}i & 6.548 \times 10^{-7}-1.702 \times 10^{-7}i \\
-0.2284+0.1277i & -0.000803-0.001714i & -1.698 \times 10^{-6}-7.62 \times 10^{-7}i
\end{pmatrix}, \\[0.1em]
&{\color{red} \Theta} = \begin{pmatrix}
0.004857-0.000927i & 5.827 \times 10^{-8}+1.049 \times 10^{-8}i & -1.82 \times 10^{-8}+3.979 \times 10^{-7}i \\
-0.0001236+0.0002241i & 1.601 \times 10^{-8}-4.832 \times 10^{-8}i & 0.6154+0.1603i \\
0.01319-0.01262i & 0.001651-0.004844i & 0.174+1.843i
\end{pmatrix}\,,
\end{aligned}
\end{equation}
with the mixing parameter ${\amber a_1} = 36.72~\mathrm{GeV}$. This point correspond to the blue diamond in the scatter plots plots of the main text. On the other hand, if we assume the $W$ boson mass to take the SM value, but the muon $a_\mu$ anomaly requires a NP explanation, scenario b), the following best fit point is obtained

\begin{equation}\label{eq:yukawa_numbers_SMl_g2}
\begin{aligned}
&{\color{ForestGreen} \Upsilon} = \begin{pmatrix}
-2.319 \times 10^{-7}+6.392 \times 10^{-7}i & 6.77 \times 10^{-7}+5.516 \times 10^{-6}i & -9.08 \times 10^{-9}-6.355 \times 10^{-8}i \\
0.000526-0.002488i & -6.06 \times 10^{-8}+1.78 \times 10^{-7}i & -0.01325-0.00789i \\
-6.79 \times 10^{-8}-2.811 \times 10^{-7}i & -0.315+2.623i & 0.001025+0.000288i
\end{pmatrix}, \\[0.1em]
&{\color{blue} \Omega} = \begin{pmatrix}
0.0447+0.1871i & -0.001551-0.002839i & -2.442 \times 10^{-6}+2.446 \times 10^{-6}i \\
5.394 \times 10^{-6}+3.39 \times 10^{-7}i & 4.237 \times 10^{-6}-5.81 \times 10^{-7}i & 1.882 \times 10^{-6}-1.63 \times 10^{-7}i \\
0.006595+0.006782i & -0.00344-0.01522i & -3.41 \times 10^{-6}+3.156 \times 10^{-6}i
\end{pmatrix}, \\[0.1em]
&{\color{red} \Theta} = \begin{pmatrix}
0.000683-0.002639i & 1.075 \times 10^{-7}-1.834 \times 10^{-7}i & 1.048 \times 10^{-6}+8.8 \times 10^{-8}i \\
-6.09 \times 10^{-5}+0.0003344i & -2.281 \times 10^{-8}+2.002 \times 10^{-8}i & 0.3998+0.0348i \\
0.02393-0.00698i & -0.0295-0.0612i & 0.239+2.042i
\end{pmatrix}
\end{aligned}
\end{equation}
and ${\amber a_1}  = 9.85~\mathrm{GeV}$. This point correspond to the cyan diamond in the scatter plots of the main text. Last but not least, if we assume that both the $W$ mass and $a_{\mu}$ require a NP explanation, scenario c), then the best fit point is
\begin{equation}\label{eq:yukawa_numbers_NP}
\begin{aligned}
&{\color{ForestGreen} \Upsilon} = \begin{pmatrix}
-5.9 \times 10^{-7}+8.26 \times 10^{-7}i & 4.842 \times 10^{-6}+4.7 \times 10^{-8}i & -1.352 \times 10^{-8}-3.261 \times 10^{-8}i \\
0.000319-0.002549i & -3.34 \times 10^{-8}+1.449 \times 10^{-7}i & -0.00759-0.01239i \\
-1.524 \times 10^{-7}-9.06 \times 10^{-8}i & -0.15+2.51i & 0.0002995+0.0004214i
\end{pmatrix}, \\[0.1em]
&{\color{blue} \Omega} = \begin{pmatrix}
0.0567+0.1529i & -0.0032+0.000344i & -5.832 \times 10^{-7}-6.36 \times 10^{-8}i \\
3.488 \times 10^{-6}+2.24 \times 10^{-7}i & 2.34 \times 10^{-6}-3.97 \times 10^{-7}i & 1.354 \times 10^{-6}-1.34 \times 10^{-7}i \\
-0.02495-0.00425i & -0.00879-0.01723i & 1.72 \times 10^{-6}+7.294 \times 10^{-6}i
\end{pmatrix}, \\[0.1em]
&{\color{red} \Theta} = \begin{pmatrix}
0.00135-0.002976i & 1.444 \times 10^{-7}-9.32 \times 10^{-8}i & 7.894 \times 10^{-7}+2.147 \times 10^{-7}i \\
-5.44 \times 10^{-5}+0.0002732i & -1.592 \times 10^{-8}+1.964 \times 10^{-8}i & 0.5828+0.0578i \\
0.02236+0.00576i & -0.01707-0.02831i & 0.408+2.085i
\end{pmatrix}\,,
\end{aligned}
\end{equation}
with ${\amber a_1} = 6.69~\mathrm{GeV}$. This point correspond to the red diamond in the scatter plots of the main text. For each of these cases, the LQ masses are indicated in the main text. These benchmarks were determined by minimizing the $\chi^2$ function in Eq.~\eqref{eq:likelihood}, whose input observables are showcased in Tab.~\ref{tab:flav_obs_exp}. In Tabs.~\ref{tab:flav_obs} and \ref{tab:flav_obs_1} we indicate the predictions for the observables for each of the benchmark scenarios.

\begin{table*}[htb!] 
	\centering
    \captionsetup{justification=raggedright}
	\resizebox{0.95\columnwidth}{!}{
		\begin{tabular}{|c|c|c|c|c|}
			\hline
			\textbf{Observable} & \textbf{Experimental measurement}   \\[0.2em] \hline
			$(g-2)_\mu$          & $(251 \pm 59)\times 10^{-11}$ \cite{Muong-2:2021ojo,Muong-2:2023cdq}  \\[0.2em]
			$\hat{T}$            & $(0.88 \pm 0.14)\times 10^{-3}$ \cite{Strumia:2022qkt}  \\[0.2em]
			$R_K[1.1, 6.0]$         & $0.949^{+0.042+0.022}_{-0.041-0.022}$ \cite{LHCb:2022december} \\[0.2em]
			$R_{K^*}[1.1, 6.0]$         & $1.027^{+0.072 + 0.027}_{-0.068 - 0.026}$  \cite{LHCb:2022december} \\[0.2em] 
			$R_{K}[0.1, 1.1]$          & $0.994^{+0.090 + 0.029}_{-0.082 - 0.027}$  \cite{LHCb:2022december}  \\[0.2em]
            $R_{K^*}[0.1, 1.1]$          & $0.927^{+0.093 + 0.036}_{-0.087 - 0.035}$  \cite{LHCb:2022december} \\[0.2em]
			$R_D$          &   $0.340\pm 0.027 \pm 0.013$ \cite{HFLAV:2019otj} \\[0.2em]
			$R_{D^*}$          & $0.295 \pm 0.011 \pm 0.008$ \cite{HFLAV:2019otj}\\[0.2em]
			$\mathrm{BR}(h\rightarrow e \mu)$          & $< 6.1 \times 10^{-5}~[95\%~\mathrm{CL}]$ \cite{ParticleDataGroup:2022pth}  \\[0.2em]
			$\mathrm{BR}(h\rightarrow e \tau)$          & $< 4.7 \times 10^{-3}~[95\%~\mathrm{CL}]$ \cite{ParticleDataGroup:2022pth} \\[0.2em]    
			$\mathrm{BR}(h\rightarrow \mu \tau)$          & $< 2.5 \times 10^{-3}~[95\%~\mathrm{CL}]$ \cite{ParticleDataGroup:2022pth} \\[0.2em] 
			$\mathrm{BR}(\mu\rightarrow e \gamma)$          & $< 4.2 \times 10^{-13}~[90\%~\mathrm{CL}]$ \cite{ParticleDataGroup:2022pth}  \\[0.2em]
			$\mathrm{BR}(\mu\rightarrow e e e)$          & $< 1.0 \times 10^{-12}~[90\%~\mathrm{CL}]$ \cite{ParticleDataGroup:2022pth}  \\[0.2em]
			$\mathrm{BR}(\tau\rightarrow e \gamma)$            & $< 3.3 \times 10^{-8}~[90\%~\mathrm{CL}]$ \cite{ParticleDataGroup:2022pth}   \\[0.2em]
			$\mathrm{BR}(\tau\rightarrow \mu \gamma)$          & $< 4.4 \times 10^{-8}~[90\%~\mathrm{CL}]$ \cite{ParticleDataGroup:2022pth}  \\[0.2em]
			$\mathrm{BR}(\tau\rightarrow e e e)$          & $< 2.7 \times 10^{-8}~[90\%~\mathrm{CL}]$ \cite{ParticleDataGroup:2022pth} \\[0.2em]
			$\mathrm{BR}(\tau\rightarrow  e \mu \mu)$          & $< 2.7 \times 10^{-8}~[90\%~\mathrm{CL}]$ \cite{ParticleDataGroup:2022pth}  \\[0.2em]    
			$\mathrm{BR}(\tau\rightarrow  \mu e e)$          & $< 1.5 \times 10^{-8}~[90\%~\mathrm{CL}]$ \cite{ParticleDataGroup:2022pth} \\[0.2em]
   			$\mathrm{BR}(\tau\rightarrow  \mu \mu \mu)$          & $< 2.1 \times 10^{-8}~[90\%~\mathrm{CL}]$ \cite{ParticleDataGroup:2022pth} \\[0.2em]
			$\mathrm{BR}(Z\rightarrow  \mu e)$            & $< 7.5  \times 10^{-7}~[95\%~\mathrm{CL}]$ \cite{ParticleDataGroup:2022pth}\\[0.2em]
			$\mathrm{BR}(Z\rightarrow  \tau e)$          & $< 9.8 \times 10^{-6}~[95\%~\mathrm{CL}]$ \cite{ParticleDataGroup:2022pth} \\[0.2em]
			$\mathrm{BR}(Z\rightarrow  \mu \tau)$          & $< 1.2 \times 10^{-5}~[95\%~\mathrm{CL}]$ \cite{ParticleDataGroup:2022pth}\\[0.2em]
			$\mathrm{BR}(\tau \rightarrow  \pi e)$          & $ < 8.0 \times 10^{-8}~[90\%~\mathrm{CL}]$ \cite{ParticleDataGroup:2022pth}\\[0.2em]			
			$\mathrm{BR}(\tau \rightarrow  \pi \mu)$          & $ < 1.1 \times 10^{-7}~[90\%~\mathrm{CL}]$ \cite{ParticleDataGroup:2022pth}\\[0.2em]
			$\mathrm{BR}(\tau \rightarrow  \phi e)$          & $ < 3.1 \times 10^{-8}~[90\%~\mathrm{CL}]$ \cite{ParticleDataGroup:2022pth}\\[0.2em]
			$\mathrm{BR}(\tau \rightarrow  \phi \mu)$          & $ < 8.4 \times 10^{-8}~[90\%~\mathrm{CL}]$ \cite{ParticleDataGroup:2022pth}\\[0.2em]			
			$\mathrm{BR}(\tau \rightarrow  \rho e)$          & $ < 1.8 \times 10^{-8}~[90\%~\mathrm{CL}]$ \cite{ParticleDataGroup:2022pth}\\[0.2em]			
			$\mathrm{BR}(\tau \rightarrow  \rho \mu)$          & $ < 1.2 \times 10^{-8}~[90\%~\mathrm{CL}]$ \cite{ParticleDataGroup:2022pth}\\[0.2em]
			$d_e$          & $ < 1.1\times 10^{-29}~\mathrm{e.cm}~[90\%~\mathrm{CL}]$ \cite{ParticleDataGroup:2022pth}  \\[0.2em]
			$d_\mu$          & $ < 1.8\times 10^{-19}~\mathrm{e.cm}~[95\%~\mathrm{CL}]$ \cite{ParticleDataGroup:2022pth} \\[0.2em]    
			$d_\tau$          & $< (1.15 \pm 1.70)\times 10^{-17}~\mathrm{e.cm}~[95\%~\mathrm{CL}]$ \cite{Belle:2002nla} \\[0.2em] 
			$\mathrm{BR}(B^0\rightarrow  \mu \mu)$          & $(0.56\pm 0.70)\times 10^{-10}$ \cite{Altmannshofer:2021qrr} \\[0.2em] 
			$\mathrm{BR}(B_s\rightarrow  \mu \mu)$          & $(2.93\pm 0.35)\times 10^{-9}$ \cite{Altmannshofer:2021qrr}  \\[0.2em] 
			$\mathrm{R}(B\rightarrow  \chi_s \gamma)$          & $1.009\pm 0.075$  \\[0.2em] 
			$R_K^{\nu\nu}$         & $< 4.65$ [$95\%$ CL] \cite{Belle:2017oht} \\[0.2em] 
			$R_{K^*}^{\nu\nu}$          & $< 3.22$ [$95\%$ CL] \cite{Belle:2017oht}  \\[0.2em]	
			$|\mathrm{Re}~\delta g_R^e|$         & $\leq 2.9\times 10^{-4}$ \cite{Saad:2020ucl,ALEPH:2005ab} \\[0.2em] 
			$|\mathrm{Re}~\delta g_L^e|$          & $\leq 3.0\times 10^{-4}$ \cite{Saad:2020ucl,ALEPH:2005ab}  \\[0.2em]	
			$|\mathrm{Re}~\delta g_R^\mu|$        & $\leq 1.3\times 10^{-3}$ \cite{Saad:2020ucl,ALEPH:2005ab} \\[0.2em] 
			$|\mathrm{Re}~\delta g_L^\mu|$          & $\leq 1.1\times 10^{-3}$ \cite{Saad:2020ucl,ALEPH:2005ab}  \\[0.2em]	
			$|\mathrm{Re}~\delta g_R^\tau|$        & $\leq 6.2\times 10^{-4}$ \cite{Saad:2020ucl,ALEPH:2005ab} \\[0.2em] 
			$|\mathrm{Re}~\delta g_L^\tau|$          & $\leq 5.8\times 10^{-4}$ \cite{Saad:2020ucl,ALEPH:2005ab}  \\[0.2em]
			$\mathrm{R}(\epsilon_k)$            & $1.234\pm 0.144$ \\[0.2em]
		    $\mathrm{R}(\Delta M_d)$            & $0.838\pm 0.109$ \\[0.2em]
	    	$\mathrm{R}(\Delta M_s)$           & $0.935\pm 0.054$ \\[0.2em]
	    	$\mathrm{R} (\mathrm{Re}(\epsilon^\prime/\epsilon))$            & $0.868\pm 0.496$ \\[0.2em]
	    	$Q_W(p^+)$           & $0.0719\pm 0.045$ \\[0.2em]
	    	$Q_W(\mathrm{Cs^{133}})$            & $-72.82\pm 0.42$ \\[0.2em]
			\hline
		\end{tabular}
		\hspace{2em}
		\begin{tabular}{|c|c|c|c|c|}
		\hline
		\textbf{Observable} & \textbf{Experimental measurement}   \\[0.2em] \hline
        $\phi_{s}$          & $-0.008 \pm 0.019$ \cite{HeavyFlavorAveragingGroup:2022wzx}  \\[0.2em]
        $A_{CP} (B^0\rightarrow K^{*0}\mu\mu)$          & $-0.035 \pm 0.024 \pm 0.003$ \cite{LHCb:2014mit}  \\[0.2em]
        $A_{CP} (B^+\rightarrow K^{+}\mu\mu)$          & $0.012 \pm 0.017 \pm 0.001$ \cite{LHCb:2014mit}  \\[0.2em]
        $A_{CP} (B\rightarrow \chi_{s+d} \gamma)$          & $0.032\pm 0.034$ \cite{HeavyFlavorAveragingGroup:2022wzx}  \\[0.2em]
		$F_L (B^+\rightarrow K\mu\mu)$          & $0.34\pm 0.10\pm 0.06$ \cite{LHCb:2020gog}  \\[0.2em]
		$S_3 (B^+\rightarrow K\mu\mu)$            & $0.14^{+0.15+0.02}_{-0.14-0.02}$ \cite{LHCb:2020gog}  \\[0.2em]
		$S_4 (B^+\rightarrow K\mu\mu)$         & $-0.04^{+0.17+0.04}_{-0.16-0.04}$ \cite{LHCb:2020gog} \\[0.2em]
		$S_5 (B^+\rightarrow K\mu\mu)$         & $0.24^{+0.12+0.04}_{-0.15-0.04}$ \cite{LHCb:2020gog} \\[0.2em] 
		$A_{FB} (B^+\rightarrow K\mu\mu)$          & $-0.05\pm 0.12\pm 0.03$  \cite{LHCb:2020gog} \\[0.2em]
		$S_7 (B^+\rightarrow K\mu\mu)$         &   $-0.01^{+0.19+0.01}_{-0.17-0.01}$ \cite{LHCb:2020gog} \\[0.2em]
		$S_8 (B^+\rightarrow K\mu\mu)$          & $0.21^{+0.22+0.05}_{-0.20-0.05}$ \cite{LHCb:2020gog}\\[0.2em]
		$S_9 (B^+\rightarrow K\mu\mu)$          & $0.28^{+0.25+0.06}_{-0.12-0.06}$ \cite{LHCb:2020gog}  \\[0.2em]
		$P_1 (B^+\rightarrow K\mu\mu)$          & $0.44^{+0.38+0.11}_{-0.40-0.11}$ \cite{LHCb:2020gog} \\[0.2em]    
		$P_2 (B^+\rightarrow K\mu\mu)$          & $-0.05\pm 0.12\pm 0.03$ \cite{LHCb:2020gog} \\[0.2em] 
		$P_3 (B^+\rightarrow K\mu\mu)$          & $-0.42^{+0.20+0.05}_{-0.21-0.05}$ \cite{LHCb:2020gog}  \\[0.2em]
		$P^\prime_4 (B^+\rightarrow K\mu\mu)$          & $-0.092^{+0.36+0.12}_{-0.35-0.12}$ \cite{LHCb:2020gog}  \\[0.2em]
		$P^\prime_5 (B^+\rightarrow K\mu\mu)$            & $0.51^{+0.30+0.12}_{-0.28-0.12}$ \cite{LHCb:2020gog}   \\[0.2em]
		$P^\prime_6 (B^+\rightarrow K\mu\mu)$          & $-0.02^{+0.40+0.06}_{-0.34-0.06}$ \cite{LHCb:2020gog}  \\[0.2em]
		$P^\prime_8 (B^+\rightarrow K\mu\mu)$          & $-0.45^{+0.50+0.09}_{-0.39-0.09}$ \cite{LHCb:2020gog} \\[0.2em]
		$F_L (B^0\rightarrow K\mu\mu)$          & $0.255 \pm 0.032 \pm 0.007$ \cite{LHCb:2020lmf}  \\[0.2em]
		$S_3 (B^0\rightarrow K\mu\mu)$            & $0.034\pm 0.044\pm 0.003$ \cite{LHCb:2020lmf}  \\[0.2em]
		$S_4 (B^0\rightarrow K\mu\mu)$         & $0.059\pm 0.050\pm 0.004$ \cite{LHCb:2020lmf} \\[0.2em]
		$S_5 (B^0\rightarrow K\mu\mu)$         & $0.227\pm 0.041\pm 0.008$  \cite{LHCb:2020lmf} \\[0.2em] 
		$A_{FB} (B^0\rightarrow K\mu\mu)$          & $-0.004\pm 0.040\pm 0.004$  \cite{LHCb:2020lmf} \\[0.2em]
		$S_7 (B^0\rightarrow K\mu\mu)$         &   $0.006\pm 0.042\pm 0.002$ \cite{LHCb:2020lmf} \\[0.2em]
		$S_8 (B^0\rightarrow K\mu\mu)$          & $-0.003\pm 0.051\pm 0.001$ \cite{LHCb:2020lmf}\\[0.2em]
		$S_9 (B^0\rightarrow K\mu\mu)$          & $-0.055\pm 0.041\pm 0.002$ \cite{LHCb:2020lmf}  \\[0.2em]
		$P_1 (B^0\rightarrow K\mu\mu)$          & $0.090\pm 0.119\pm 0.009$ \cite{LHCb:2020lmf} \\[0.2em]    
		$P_2 (B^0\rightarrow K\mu\mu)$          & $-0.003\pm 0.038\pm 0.003$ \cite{LHCb:2020lmf} \\[0.2em] 
		$P_3 (B^0\rightarrow K\mu\mu)$          & $-0.073\pm 0.057\pm 0.003$ \cite{LHCb:2020lmf}  \\[0.2em]
		$P^\prime_4 (B^0\rightarrow K\mu\mu)$          & $-0.135\pm 0.118\pm 0.003$\cite{LHCb:2020lmf}  \\[0.2em]
		$P^\prime_5 (B^0\rightarrow K\mu\mu)$            & $-0.521\pm 0.095\pm 0.024$ \cite{LHCb:2020lmf}   \\[0.2em]
		$P^\prime_6 (B^0\rightarrow K\mu\mu)$          & $-0.015\pm 0.094\pm 0.007$ \cite{LHCb:2020lmf}  \\[0.2em]
		$P^\prime_8 (B^0\rightarrow K\mu\mu)$          & $-0.007\pm 0.122\pm 0.002$ \cite{LHCb:2020lmf} \\[0.2em]
		$\mathrm{R}(K^+\rightarrow \pi^0 \mu^+ \nu)$            & $0.989\pm 0.016$ \\[0.2em]
		$\mathrm{R}(K^+\rightarrow \pi^0 e^+ \nu)$           & $0.988\pm 0.014$ \\[0.2em]
		$\mathrm{R}(K_L^0 \rightarrow \pi^+ \mu^- \nu)$            & $0.997\pm 0.011$ \\[0.2em]
		$\mathrm{R}(K_L^0 \rightarrow \pi^+ e^- \nu)$            & $0.991\pm 0.029$ \\[0.2em]
		$\mathrm{R}(K_S^0 \rightarrow \pi^\pm e^\mp \nu)$            & $0.982\pm 0.015$ \\[0.2em]
		$\mathrm{R}(K_L^0 \rightarrow \mu^+ \mu^-)$            & $0.918\pm 0.158$ \\[0.2em]
		$\mathrm{R}(K_L^0 \rightarrow e^+ e^-)$            & $1.000\pm 0.598$ \\[0.2em]
		$\mathrm{BR}(K_L^0 \rightarrow e^\mp \mu^\pm)$           & $< 4.7 \times 10^{-12}~[90\%~\mathrm{CL}]$ \cite{ParticleDataGroup:2022pth} \\[0.2em]
		$\mathrm{BR}(K_S^0 \rightarrow \mu^+ \mu^-)$           & $< 8.0 \times 10^{-10}~[90\%~\mathrm{CL}]$ \cite{ParticleDataGroup:2022pth} \\[0.2em]
		$\mathrm{BR}(K_S^0 \rightarrow e^+ e^-)$           & $< 9.0 \times 10^{-12}~[90\%~\mathrm{CL}]$ \cite{ParticleDataGroup:2022pth} \\[0.2em]
		$\mathrm{R}(K^+ \rightarrow \mu^+ \nu)$           & $1.008 \pm 0.015$ \\[0.2em]
		$\mathrm{R}(K^+ \rightarrow e^+ \nu)$           & $1.014 \pm 0.016$ \\[0.2em]
		$\mathrm{R}(K^+ \rightarrow \pi^+ \nu \nu)$           & $1.840\pm 1.202$ \\[0.2em]
		$\mathrm{BR}(K_L^0 \rightarrow \pi^0 \nu \nu)$           & $< 3.0 \times 10^{-9}~[90\%~\mathrm{CL}]$ \cite{ParticleDataGroup:2022pth} \\[0.2em]
		\hline
	\end{tabular}
	}
	\caption{\label{tab:flav_obs_exp} Set of observables used as input for the $\chi^2$ function, as well as the experimental measured value. The observables $F_L$, $A_{FB}$, $S_i$, $P_i$ and $P_i^\prime$ are relative to the $[0.10, 0.98]~\mathrm{GeV^2}$ $q^2$ bins. Observables starting with ``$\mathrm{R}$'' are defined as the ratio between the experimental value for the observable, taken from \cite{ParticleDataGroup:2022pth} and the SM prediction, determined in \texttt{flavio}. The total uncertainty is taken by error propagation, taking into account both the experimental and theoretical errors, with the latter determined also in \texttt{flavio}.}
\end{table*}

\begin{table*}[htb!] 
	\centering
	\resizebox{0.925\columnwidth}{!}{
	\begin{tabular}{|c|c|c|c|c|}
		\hline
		\textbf{Observable} & \textbf{Theoretical prediction: \eqref{eq:yukawa_numbers_SMl}} & \textbf{Theoretical prediction: \eqref{eq:yukawa_numbers_SMl_g2}}  & \textbf{Theoretical prediction: \eqref{eq:yukawa_numbers_NP}}   \\[0.2em] \hline
		$a_\mu$ (\texttt{sph})         &    $-5.891 \times 10^{-11}$            & $2.649 \times 10^{-9}$               & $1.879 \times 10^{-9}$\\[0.2em]
		$\hat{T}$ (\texttt{sph})           &    $0.0002702$             &      $0.0001631$            & $0.0009105$\\[0.2em]
		$R_K[1.1, 6.0]$ (\texttt{fla})          &    $1.006$             &      $1.006$            & $1.006$ \\[0.2em]
		$R_{K*}[1.1, 6.0]$ (\texttt{fla})          &    $1.000$             &      $1.001$            & $1.001$\\[0.2em]
		$R_K[0.1, 1.1]$ (\texttt{fla})          &    $0.9987$             &      $0.9984$            & $0.999$ \\[0.2em]
		$R_{K*}[0.1, 1.1]$ (\texttt{fla})          &    $1.006$             &      $1.006$            & $1.006$\\[0.2em]
		$R_D$ (\texttt{fla})         &    $0.3508$             &      $0.3434$            & $0.334$\\[0.2em]
		$R_{D^*}$ (\texttt{fla})         &    $0.2776$             &      $0.2875$            & $0.283$ \\[0.2em]
		$\mathrm{BR}(h\rightarrow e \mu)$ (\texttt{sph})         &    $3.008 \times 10^{-18}$             &      $9.692 \times 10^{-19}$            & $5.98 \times 10^{-19}$ \\[0.2em]
		$\mathrm{BR}(h\rightarrow e \tau)$ (\texttt{sph})         &    $2.435 \times 10^{-17}$             &      $2.086 \times 10^{-17}$            & $5.931 \times 10^{-18}$\\[0.2em]    
		$\mathrm{BR}(h\rightarrow \mu \tau)$ (\texttt{sph})         &    $6.072 \times 10^{-7}$             &      $1.039 \times 10^{-6}$            & $9.904 \times 10^{-7}$           \\[0.2em] 
		$\mathrm{BR}(\mu\rightarrow e \gamma)$ (\texttt{fla})         &    $3.074 \times 10^{-16}$             &      $2.916 \times 10^{-15}$            & $1.44 \times 10^{-15}$ \\[0.2em]
		$\mathrm{BR}(\mu\rightarrow e e e)$ (\texttt{fla})         &    $2.456 \times 10^{-18}$             &      $1.953 \times 10^{-17}$            & $9.509 \times 10^{-18}$\\[0.2em]
		$\mathrm{BR}(\tau\rightarrow e \gamma)$ (\texttt{fla})           &    $1.925 \times 10^{-17}$             &      $1.17 \times 10^{-17}$            & $1.479 \times 10^{-17}$\\[0.2em]
		$\mathrm{BR}(\tau\rightarrow \mu \gamma)$ (\texttt{fla})         &    $6.713 \times 10^{-9}$             &      $3.294 \times 10^{-9}$            & $4.864 \times 10^{-9}$ \\[0.2em]
		$\mathrm{BR}(\tau\rightarrow e e e)$ (\texttt{sph})         &    $2.611 \times 10^{-14}$             &      $5.161 \times 10^{-15}$            & $3.055 \times 10^{-15}$ \\[0.2em]
		$\mathrm{BR}(\tau\rightarrow  e \mu \mu)$ (\texttt{sph})         &    $1.779 \times 10^{-14}$             &      $3.516 \times 10^{-15}$            & $2.089 \times 10^{-15}$\\[0.2em]    
		$\mathrm{BR}(\tau\rightarrow  \mu e e)$ (\texttt{sph})         &    $5.824 \times 10^{-27}$             &      $3.924 \times 10^{-28}$            & $3.579 \times 10^{-28}$               \\[0.2em] 
		$\mathrm{BR}(Z\rightarrow  \mu e)$ (\texttt{sph})            &    $7.003 \times 10^{-21}$             &      $9.718 \times 10^{-21}$            & $1.074 \times 10^{-20}$ \\[0.2em]
		$\mathrm{BR}(Z\rightarrow  \tau e)$ (\texttt{sph})         &    $1.265 \times 10^{-15}$             &      $2.605 \times 10^{-16}$            & $1.007 \times 10^{-16}$\\[0.2em]
		$\mathrm{BR}(Z\rightarrow  \mu \tau)$ (\texttt{sph})         &    $4.873 \times 10^{-8}$             &      $2.244 \times 10^{-8}$            & $4.64 \times 10^{-8}$\\[0.2em]
		$\mathrm{BR}(\tau \rightarrow  \pi e)$ (\texttt{fla})          &    $3.94 \times 10^{-12}$             &      $1.173 \times 10^{-13}$            & $1.631 \times 10^{-12}$\\[0.2em]
		$\mathrm{BR}(\tau \rightarrow  \pi \mu)$ (\texttt{fla})         &    $1.401 \times 10^{-22}$             &      $7.932 \times 10^{-22}$            & $3.776 \times 10^{-22}$ \\[0.2em]
		$\mathrm{BR}(\tau \rightarrow  \phi e)$ (\texttt{fla})         &    $1.349 \times 10^{-16}$             &      $3.556 \times 10^{-16}$            & $1.534 \times 10^{-15}$ \\[0.2em]		
		$\mathrm{BR}(\tau \rightarrow  \phi \mu)$ (\texttt{fla})        &    $1.869 \times 10^{-12}$             &      $9.174 \times 10^{-13}$            & $1.354 \times 10^{-12}$\\[0.2em]
		$\mathrm{BR}(\tau \rightarrow  \rho e)$ (\texttt{fla})         &    $8.828 \times 10^{-12}$             &      $2.743 \times 10^{-13}$            & $4.809 \times 10^{-12}$\\[0.2em]
		$\mathrm{BR}(\tau \rightarrow  \rho \mu)$ (\texttt{fla})         &    $1.688 \times 10^{-11}$             &      $8.282 \times 10^{-12}$            & $1.223 \times 10^{-11}$ \\[0.2em]	
		$d_e$ (\texttt{sph})         &    $3.283 \times 10^{-33}$             &      $2.75 \times 10^{-32}$            & $7.79 \times 10^{-33}$ \\[0.2em]
		$d_\mu$ (\texttt{sph})         &    $1.248 \times 10^{-22}$             &      $5.388 \times 10^{-23}$            & $1.094 \times 10^{-22}$\\[0.2em]    
		$d_\tau$ (\texttt{sph})         &    $2.446 \times 10^{-23}$             &      $1.638 \times 10^{-23}$            & $1.999 \times 10^{-24}$               \\[0.2em] 
		$\mathrm{BR}(B^0\rightarrow  \mu \mu)$ (\texttt{fla})         &    $1.139 \times 10^{-10}$             &      $1.14 \times 10^{-10}$            & $1.148 \times 10^{-10}$                \\[0.2em] 
		$\mathrm{BR}(B_s\rightarrow  \mu \mu)$ (\texttt{fla})         &    $3.673 \times 10^{-9}$             &      $3.685 \times 10^{-9}$            & $3.679 \times 10^{-9}$              \\[0.2em] 
		$\mathrm{R}(B\rightarrow  \chi_s \gamma)$ (\texttt{fla})         &    $1.000$             &      $1.000$            & $1.000$              \\[0.2em] 
		$R_K^{\nu\nu}$ (\texttt{fla})        &    $1.793$             &      $1.227$            & $1.476$               \\[0.2em] 
		$R_{K^*}^{\nu\nu}$ (\texttt{fla})         &    $1.793$             &      $1.227$            & $1.476$   \\[0.2em] 
		$|\mathrm{Re}~\delta g_R^e|$ (\texttt{ind})        &    $4.902 \times 10^{-8}$             &      $-8.815 \times 10^{-8}$            & $7.09 \times 10^{-8}$\\[0.2em] 
		$|\mathrm{Re}~\delta g_L^e|$ (\texttt{ind})         &    $6.788 \times 10^{-9}$             &      $5.207 \times 10^{-8}$            & $6.277 \times 10^{-8}$\\[0.2em]	
		$|\mathrm{Re}~\delta g_R^\mu|$ (\texttt{ind})       &    $3.891 \times 10^{-9}$             &      $6.56 \times 10^{-7}$            & $1.456 \times 10^{-7}$\\[0.2em] 
		$|\mathrm{Re}~\delta g_L^\mu|$ (\texttt{ind})         &    $0.002152$             &      $0.002661$            & $0.002322$\\[0.2em]	
		$|\mathrm{Re}~\delta g_R^\tau|$ (\texttt{ind})       &    $0.0005011$             &      $0.0005977$            & $0.0005954$\\[0.2em] 
		$|\mathrm{Re}~\delta g_L^\tau|$ (\texttt{ind})       &    $3.732 \times 10^{-8}$             &      $1.049 \times 10^{-8}$            & $6.595 \times 10^{-9}$\\[0.2em]
		$\mathrm{R}(\epsilon_k)$ (\texttt{fla})       &    $1.135$             &      $1.326$            & $1.107$\\[0.2em]
		$\mathrm{R}(\Delta M_d)$ (\texttt{fla})       &    $0.9361$             &      $0.7878$            & $1.08$\\[0.2em]
		$\mathrm{R}(\Delta M_s)$ (\texttt{fla})       &    $0.8263$             &      $1.01$            & $0.9117$\\[0.2em]
		$\mathrm{R}(\mathrm{Re}(\epsilon^\prime/\epsilon))$ (\texttt{fla})       &    $1.177$             &      $1.013$            & $1.425$\\[0.2em]
		$Q_W(p^+)$ (\texttt{ind})       &    $0.071$             &      $0.071$            & $0.071$\\[0.2em]
		$Q_W(\mathrm{Cs^{133}})$ (\texttt{ind})       &    $-73.33$             &      $-73.33$            & $-73.33$\\[0.2em]
		\hline
	\end{tabular}
}
	\caption{\label{tab:flav_obs} Theoretical predictions for the each of the benchmarks. In the first column we indicate how each observable is computed, with \texttt{fla} being \texttt{flavio}, \texttt{sph} being \texttt{SPheno} and \texttt{ind} indicates that is based in our own implementation.}
\end{table*}

\begin{table*}[htb!] 
	\centering
	\resizebox{0.90\columnwidth}{!}{
	\begin{tabular}{|c|c|c|c|c|}
		\hline
		\textbf{Observable} & \textbf{Theoretical prediction: \eqref{eq:yukawa_numbers_SMl}} & \textbf{Theoretical prediction: \eqref{eq:yukawa_numbers_SMl_g2}} & \textbf{Theoretical prediction: \eqref{eq:yukawa_numbers_NP}}   \\[0.2em] \hline
         $\phi_{s}$ (\texttt{fla})         &    $-0.0092$             &      $-0.00296$            & $-0.0585$ \\[0.2em]
         $A_{CP} (B^0\rightarrow K^{*0}\mu\mu)$ (\texttt{fla})         &    $0.000167$             &      $0.000172$            & $0.000187$ \\[0.2em]
         $A_{CP} (B^+\rightarrow K^{+}\mu\mu)$ (\texttt{fla})         &    $0.001771$             &      $0.001774$            & $0.001791$ \\[0.2em]
         $A_{CP} (B\rightarrow \chi_{s+d} \gamma)$ (\texttt{fla})         &    $3.603\times 10^{-7}$             &      $7.202\times 10^{-7}$            & $6.588\times 10^{-8}$ \\[0.2em]
		$F_L (B^+\rightarrow K\mu\mu)$ (\texttt{fla})         &    $0.3041$             &      $0.304$            & $0.3042$ \\[0.2em]
		$S_3 (B^+\rightarrow K\mu\mu)$ (\texttt{fla})           &    $0.01081$             &      $0.0108$            & $0.01081$\\[0.2em]
		$S_4 (B^+\rightarrow K\mu\mu)$ (\texttt{fla})         &    $0.08939$             &      $0.08907$            & $0.08927$\\[0.2em]
		$S_5 (B^+\rightarrow K\mu\mu)$ (\texttt{fla})        &    $0.2591$             &      $0.2597$            & $0.2593$\\[0.2em] 
		$A_{FB} (B^+\rightarrow K\mu\mu)$ (\texttt{fla})         &    $-0.097$             &      $-0.09717$            & $-0.09707$\\[0.2em]
		$S_7 (B^+\rightarrow K\mu\mu)$ (\texttt{fla})        &    $-0.0179$             &      $-0.01793$            & $-0.01792$ \\[0.2em]
		$S_8 (B^+\rightarrow K\mu\mu)$ (\texttt{fla})        &    $-0.01217$             &      $-0.01215$            & $-0.01216$ \\[0.2em]
		$S_9 (B^+\rightarrow K\mu\mu)$ (\texttt{fla})         &    $-0.0007138$             &      $-0.0007131$            & $-0.0007135$ \\[0.2em]
		$P_1 (B^+\rightarrow K\mu\mu)$ (\texttt{fla})         &    $0.04543$             &      $0.04538$            & $0.04541$ \\[0.2em]    
		$P_2 (B^+\rightarrow K\mu\mu)$ (\texttt{fla})         &    $-0.1359$             &      $-0.1361$            & $-0.136$ \\[0.2em] 
		$P_3 (B^+\rightarrow K\mu\mu)$ (\texttt{fla})         &    $0.0015$             &      $0.001498$            & $0.001499$ \\[0.2em]
		$P^\prime_4 (B^+\rightarrow K\mu\mu)$ (\texttt{fla})         &    $0.235$             &      $0.2341$            & $0.2346$ \\[0.2em]
		$P^\prime_5 (B^+\rightarrow K\mu\mu)$ (\texttt{fla})           &    $0.6811$             &      $0.6826$            & $0.6816$ \\[0.2em]
		$P^\prime_6 (B^+\rightarrow K\mu\mu)$ (\texttt{fla})         &    $-0.04706$             &      $-0.04713$            & $-0.04709$ \\[0.2em]
		$P^\prime_8 (B^+\rightarrow K\mu\mu)$ (\texttt{fla})         &    $-0.03198$             &      $-0.03193$            & $-0.03196$ \\[0.2em]
		$F_L (B^0\rightarrow K\mu\mu)$ (\texttt{fla})         &    $0.2972$             &      $0.2971$            & $0.2973$ \\[0.2em]
		$S_3 (B^0\rightarrow K\mu\mu)$ (\texttt{fla})           &    $0.01083$             &      $0.01082$            & $0.01082$ \\[0.2em]
		$S_4 (B^0\rightarrow K\mu\mu)$ (\texttt{fla})        &    $0.09582$             &      $0.09549$            & $0.0957$ \\[0.2em]
		$S_5 (B^0\rightarrow K\mu\mu)$ (\texttt{fla})        &    $0.2605$             &      $0.2611$            & $0.2608$ \\[0.2em] 
		$A_{FB} (B^0\rightarrow K\mu\mu)$ (\texttt{fla})         &    $-0.09668$             &      $-0.09686$            & $-0.09675$ \\[0.2em]
		$S_7 (B^0\rightarrow K\mu\mu)$ (\texttt{fla})        &    $-0.02056$             &      $-0.02059$            & $-0.02057$ \\[0.2em] 
		$S_8 (B^0\rightarrow K\mu\mu)$ (\texttt{fla})         &    $-0.002203$             &      $-0.002182$            & $-0.002197$ \\[0.2em]
		$S_9 (B^0\rightarrow K\mu\mu)$ (\texttt{fla})         &    $-0.0006991$             &      $-0.0006985$            & $-0.0006988$ \\[0.2em]
		$P_1 (B^0\rightarrow K\mu\mu)$ (\texttt{fla})         &    $0.04449$             &      $0.04444$            & $0.04447$ \\[0.2em]    
		$P_2 (B^0\rightarrow K\mu\mu)$ (\texttt{fla})         &    $-0.1324$             &      $-0.1326$            & $-0.1325$ \\[0.2em] 
		$P_3 (B^0\rightarrow K\mu\mu)$ (\texttt{fla})         &    $0.001436$             &      $0.001434$            & $0.001436$ \\[0.2em]
		$P^\prime_4 (B^0\rightarrow K\mu\mu)$ (\texttt{fla})         &    $0.2519$             &      $0.2511$            & $0.2516$ \\[0.2em]
		$P^\prime_5 (B^0\rightarrow K\mu\mu)$ (\texttt{fla})           &    $0.685$             &      $0.6864$            & $0.6855$ \\[0.2em]
		$P^\prime_6 (B^0\rightarrow K\mu\mu)$ (\texttt{fla})         &    $-0.05405$             &      $-0.05413$            & $-0.05408$ \\[0.2em]
		$P^\prime_8 (B^0\rightarrow K\mu\mu)$ (\texttt{fla})         &    $-0.005791$             &      $-0.005738$            & $-0.005776$ \\[0.2em]
		$C_{9}^{bs\mu\mu}$ (\texttt{fla})          &    $0.02261$             &      $0.0142$            & $0.02012$ \\[0.2em]
		$C_{10}^{bs\mu\mu}$ (\texttt{fla})         &    $0.0009718$             &      $-0.00568$            & $-0.002634$ \\[0.2em]
		$C_{9}^{\prime bs\mu\mu}$ (\texttt{fla})         &    $1.007 \times 10^{-7}$             &      $3.501 \times 10^{-8}$            & $6.651 \times 10^{-8}$ \\[0.2em]
		$C_{10}^{\prime bs\mu\mu}$ (\texttt{fla})         &    $-1.007 \times 10^{-7}$             &      $-3.496 \times 10^{-8}$            & $-6.645 \times 10^{-8}$\\[0.2em]    
		$C_9^{bsee}$ (\texttt{fla})         &    $-1.714 \times 10^{-6}$             &      $-1.652 \times 10^{-7}$            & $-5.143 \times 10^{-7}$ \\[0.2em] 
		$C_{10}^{bsee}$ (\texttt{fla})           &    $1.561 \times 10^{-6}$             &      $1.379 \times 10^{-7}$            & $4.929 \times 10^{-7}$\\[0.2em]
		$\mathrm{R}(K^+\rightarrow \pi^0 \mu^+ \nu)$ (\texttt{fla})           &    $1.000$             &      $1.000$            & $1.000$\\[0.2em]
		$\mathrm{R}(K^+\rightarrow \pi^0 e^+ \nu)$ (\texttt{fla})           &    $1.000$             &      $1.000$            & $1.000$\\[0.2em]
		$\mathrm{R}(K_L^0 \rightarrow \pi^+ \mu^- \nu)$ (\texttt{fla})           &    $1.000$             &      $1.000$            & $1.000$\\[0.2em]
		$\mathrm{R}(K_L^0 \rightarrow \pi^+ e^- \nu)$ (\texttt{fla})           &    $1.000$             &      $1.000$            & $1.000$\\[0.2em]
		$\mathrm{R}(K_S^0 \rightarrow \pi^\pm e^\mp \nu)$ (\texttt{fla})           &    $1.000$             &      $1.000$            & $1.000$\\[0.2em]
		$\mathrm{R}(K_L^0 \rightarrow \mu^+ \mu^-)$ (\texttt{fla})           &    $1.001$             &      $1.000$            & $1.000$\\[0.2em]
		$\mathrm{R}(K_L^0 \rightarrow e^+ e^-)$ (\texttt{fla})           &    $0.9472$             &      $1.058$            & $1.021$\\[0.2em]
		$\mathrm{BR}(K_L^0 \rightarrow e^\mp \mu^\pm)$ (\texttt{fla})           &    $4.822 \times 10^{-14}$             &      $6.372 \times 10^{-16}$            & $3.13 \times 10^{-15}$\\[0.2em]
		$\mathrm{BR}(K_S^0 \rightarrow \mu^+ \mu^-)$ (\texttt{fla})           &    $5.168 \times 10^{-12}$             &      $5.172 \times 10^{-12}$            & $5.169 \times 10^{-12}$\\[0.2em]
		$\mathrm{BR}(K_S^0 \rightarrow e^+ e^-)$ (\texttt{fla})           &    $1.684 \times 10^{-16}$             &      $1.637 \times 10^{-16}$            & $1.762 \times 10^{-16}$\\[0.2em]
		$\mathrm{R}(K^+ \rightarrow \mu^+ \nu)$ (\texttt{fla})           &    $1.000$             &      $1.000$            & $1.000$\\[0.2em]
		$\mathrm{R}(K^+ \rightarrow e^+ \nu)$ (\texttt{fla})           &    $1.000$             &      $1.000$            & $1.000$\\[0.2em]
		$\mathrm{R}(K^+ \rightarrow \pi^+ \nu \nu)$ (\texttt{fla})           &    $1.261$             &      $0.8588$            & $1.939$\\[0.2em]
		$\mathrm{BR}(K_L^0 \rightarrow \pi^0 \nu \nu)$ (\texttt{fla})           &    $2.573 \times 10^{-10}$             &      $4.059 \times 10^{-11}$            & $2.184 \times 10^{-10}$\\[0.2em]
		\hline
	\end{tabular}
}
	\caption{\label{tab:flav_obs_1} Theoretical predictions for the each of the benchmarks. In the first column we indicate how each observable is computed, with \texttt{fla} being \texttt{flavio}. The computation of $\phi_s$ observable is not available in the current version of \texttt{flavio} and needs to be added. The necessary functions for the implementation can be found in the GitHub page.}
\end{table*}

\cleardoublepage

\bibliography{main}

\begin{thebibliography}{129}
\expandafter\ifx\csname natexlab\endcsname\relax\def\natexlab#1{#1}\fi
\expandafter\ifx\csname bibnamefont\endcsname\relax
  \def\bibnamefont#1{#1}\fi
\expandafter\ifx\csname bibfnamefont\endcsname\relax
  \def\bibfnamefont#1{#1}\fi
\expandafter\ifx\csname citenamefont\endcsname\relax
  \def\citenamefont#1{#1}\fi
\expandafter\ifx\csname url\endcsname\relax
  \def\url#1{\texttt{#1}}\fi
\expandafter\ifx\csname urlprefix\endcsname\relax\def\urlprefix{URL }\fi
\providecommand{\bibinfo}[2]{#2}
\providecommand{\eprint}[2][]{\url{#2}}

\bibitem[{\citenamefont{Arnison et~al.}(1983)}]{UA1:1983crd}
\bibinfo{author}{\bibfnamefont{G.}~\bibnamefont{Arnison}} \bibnamefont{et~al.}
  (\bibinfo{collaboration}{UA1}), \bibinfo{journal}{Phys. Lett. B}
  \textbf{\bibinfo{volume}{122}}, \bibinfo{pages}{103} (\bibinfo{year}{1983}).

\bibitem[{\citenamefont{Chatrchyan et~al.}(2012)}]{CMS:2012qbp}
\bibinfo{author}{\bibfnamefont{S.}~\bibnamefont{Chatrchyan}}
  \bibnamefont{et~al.} (\bibinfo{collaboration}{CMS}), \bibinfo{journal}{Phys.
  Lett. B} \textbf{\bibinfo{volume}{716}}, \bibinfo{pages}{30}
  (\bibinfo{year}{2012}), \eprint{1207.7235}.

\bibitem[{\citenamefont{Hasert
  et~al.}(1973{\natexlab{a}})}]{GargamelleNeutrino:1973jyy}
\bibinfo{author}{\bibfnamefont{F.~J.} \bibnamefont{Hasert}}
  \bibnamefont{et~al.} (\bibinfo{collaboration}{Gargamelle Neutrino}),
  \bibinfo{journal}{Phys. Lett. B} \textbf{\bibinfo{volume}{46}},
  \bibinfo{pages}{138} (\bibinfo{year}{1973}{\natexlab{a}}).

\bibitem[{\citenamefont{Hasert et~al.}(1973{\natexlab{b}})}]{Hasert:1973cr}
\bibinfo{author}{\bibfnamefont{F.~J.} \bibnamefont{Hasert}}
  \bibnamefont{et~al.}, \bibinfo{journal}{Phys. Lett. B}
  \textbf{\bibinfo{volume}{46}}, \bibinfo{pages}{121}
  (\bibinfo{year}{1973}{\natexlab{b}}).

\bibitem[{\citenamefont{Abe et~al.}(1995)}]{CDF:1995wbb}
\bibinfo{author}{\bibfnamefont{F.}~\bibnamefont{Abe}} \bibnamefont{et~al.}
  (\bibinfo{collaboration}{CDF}), \bibinfo{journal}{Phys. Rev. Lett.}
  \textbf{\bibinfo{volume}{74}}, \bibinfo{pages}{2626} (\bibinfo{year}{1995}),
  \eprint{hep-ex/9503002}.

\bibitem[{\citenamefont{Parker et~al.}(2018)\citenamefont{Parker, Yu, Zhong,
  Estey, and M\"uller}}]{Parker:2018vye}
\bibinfo{author}{\bibfnamefont{R.~H.} \bibnamefont{Parker}},
  \bibinfo{author}{\bibfnamefont{C.}~\bibnamefont{Yu}},
  \bibinfo{author}{\bibfnamefont{W.}~\bibnamefont{Zhong}},
  \bibinfo{author}{\bibfnamefont{B.}~\bibnamefont{Estey}}, \bibnamefont{and}
  \bibinfo{author}{\bibfnamefont{H.}~\bibnamefont{M\"uller}},
  \bibinfo{journal}{Science} \textbf{\bibinfo{volume}{360}},
  \bibinfo{pages}{191} (\bibinfo{year}{2018}), \eprint{1812.04130}.

\bibitem[{\citenamefont{Hanneke et~al.}(2011)\citenamefont{Hanneke,
  Hoogerheide, and Gabrielse}}]{Hanneke:2010au}
\bibinfo{author}{\bibfnamefont{D.}~\bibnamefont{Hanneke}},
  \bibinfo{author}{\bibfnamefont{S.~F.} \bibnamefont{Hoogerheide}},
  \bibnamefont{and}
  \bibinfo{author}{\bibfnamefont{G.}~\bibnamefont{Gabrielse}},
  \bibinfo{journal}{Phys. Rev. A} \textbf{\bibinfo{volume}{83}},
  \bibinfo{pages}{052122} (\bibinfo{year}{2011}), \eprint{1009.4831}.

\bibitem[{\citenamefont{Abi et~al.}(2021)}]{Muong-2:2021ojo}
\bibinfo{author}{\bibfnamefont{B.}~\bibnamefont{Abi}} \bibnamefont{et~al.}
  (\bibinfo{collaboration}{Muon g-2}), \bibinfo{journal}{Phys. Rev. Lett.}
  \textbf{\bibinfo{volume}{126}}, \bibinfo{pages}{141801}
  (\bibinfo{year}{2021}), \eprint{2104.03281}.

\bibitem[{\citenamefont{Aguillard et~al.}(2023)}]{Muong-2:2023cdq}
\bibinfo{author}{\bibfnamefont{D.~P.} \bibnamefont{Aguillard}}
  \bibnamefont{et~al.} (\bibinfo{collaboration}{Muon g-2})
  (\bibinfo{year}{2023}), \eprint{2308.06230}.

\bibitem[{\citenamefont{Lees et~al.}(2013)}]{BaBar:2013mob}
\bibinfo{author}{\bibfnamefont{J.~P.} \bibnamefont{Lees}} \bibnamefont{et~al.}
  (\bibinfo{collaboration}{BaBar}), \bibinfo{journal}{Phys. Rev. D}
  \textbf{\bibinfo{volume}{88}}, \bibinfo{pages}{072012}
  (\bibinfo{year}{2013}), \eprint{1303.0571}.

\bibitem[{\citenamefont{Lees et~al.}(2012)}]{BaBar:2012obs}
\bibinfo{author}{\bibfnamefont{J.~P.} \bibnamefont{Lees}} \bibnamefont{et~al.}
  (\bibinfo{collaboration}{BaBar}), \bibinfo{journal}{Phys. Rev. Lett.}
  \textbf{\bibinfo{volume}{109}}, \bibinfo{pages}{101802}
  (\bibinfo{year}{2012}), \eprint{1205.5442}.

\bibitem[{\citenamefont{Huschle et~al.}(2015)}]{Belle:2015qfa}
\bibinfo{author}{\bibfnamefont{M.}~\bibnamefont{Huschle}} \bibnamefont{et~al.}
  (\bibinfo{collaboration}{Belle}), \bibinfo{journal}{Phys. Rev. D}
  \textbf{\bibinfo{volume}{92}}, \bibinfo{pages}{072014}
  (\bibinfo{year}{2015}), \eprint{1507.03233}.

\bibitem[{\citenamefont{Sato et~al.}(2016)}]{Belle:2016ure}
\bibinfo{author}{\bibfnamefont{Y.}~\bibnamefont{Sato}} \bibnamefont{et~al.}
  (\bibinfo{collaboration}{Belle}), \bibinfo{journal}{Phys. Rev. D}
  \textbf{\bibinfo{volume}{94}}, \bibinfo{pages}{072007}
  (\bibinfo{year}{2016}), \eprint{1607.07923}.

\bibitem[{\citenamefont{Hirose et~al.}(2018)}]{Belle:2017ilt}
\bibinfo{author}{\bibfnamefont{S.}~\bibnamefont{Hirose}} \bibnamefont{et~al.}
  (\bibinfo{collaboration}{Belle}), \bibinfo{journal}{Phys. Rev. D}
  \textbf{\bibinfo{volume}{97}}, \bibinfo{pages}{012004}
  (\bibinfo{year}{2018}), \eprint{1709.00129}.

\bibitem[{\citenamefont{Aaij et~al.}(2015)}]{LHCb:2015gmp}
\bibinfo{author}{\bibfnamefont{R.}~\bibnamefont{Aaij}} \bibnamefont{et~al.}
  (\bibinfo{collaboration}{LHCb}), \bibinfo{journal}{Phys. Rev. Lett.}
  \textbf{\bibinfo{volume}{115}}, \bibinfo{pages}{111803}
  (\bibinfo{year}{2015}), \bibinfo{note}{[Erratum: Phys.Rev.Lett. 115, 159901
  (2015)]}, \eprint{1506.08614}.

\bibitem[{\citenamefont{Altmannshofer and
  Stangl}(2021)}]{Altmannshofer:2021qrr}
\bibinfo{author}{\bibfnamefont{W.}~\bibnamefont{Altmannshofer}}
  \bibnamefont{and} \bibinfo{author}{\bibfnamefont{P.}~\bibnamefont{Stangl}},
  \bibinfo{journal}{Eur. Phys. J. C} \textbf{\bibinfo{volume}{81}},
  \bibinfo{pages}{952} (\bibinfo{year}{2021}), \eprint{2103.13370}.

\bibitem[{\citenamefont{Choudhury et~al.}(2021)}]{BELLE:2019xld}
\bibinfo{author}{\bibfnamefont{S.}~\bibnamefont{Choudhury}}
  \bibnamefont{et~al.} (\bibinfo{collaboration}{BELLE}),
  \bibinfo{journal}{JHEP} \textbf{\bibinfo{volume}{03}}, \bibinfo{pages}{105}
  (\bibinfo{year}{2021}), \eprint{1908.01848}.

\bibitem[{\citenamefont{Abdesselam et~al.}(2021)}]{Belle:2019oag}
\bibinfo{author}{\bibfnamefont{A.}~\bibnamefont{Abdesselam}}
  \bibnamefont{et~al.} (\bibinfo{collaboration}{Belle}),
  \bibinfo{journal}{Phys. Rev. Lett.} \textbf{\bibinfo{volume}{126}},
  \bibinfo{pages}{161801} (\bibinfo{year}{2021}), \eprint{1904.02440}.

\bibitem[{\citenamefont{Aaij et~al.}(2017)}]{LHCb:2017avl}
\bibinfo{author}{\bibfnamefont{R.}~\bibnamefont{Aaij}} \bibnamefont{et~al.}
  (\bibinfo{collaboration}{LHCb}), \bibinfo{journal}{JHEP}
  \textbf{\bibinfo{volume}{08}}, \bibinfo{pages}{055} (\bibinfo{year}{2017}),
  \eprint{1705.05802}.

\bibitem[{\citenamefont{Aaij et~al.}(2022{\natexlab{a}})}]{LHCb:2021trn}
\bibinfo{author}{\bibfnamefont{R.}~\bibnamefont{Aaij}} \bibnamefont{et~al.}
  (\bibinfo{collaboration}{LHCb}), \bibinfo{journal}{Nature Phys.}
  \textbf{\bibinfo{volume}{18}}, \bibinfo{pages}{277}
  (\bibinfo{year}{2022}{\natexlab{a}}), \eprint{2103.11769}.

\bibitem[{\citenamefont{{LHCb Collaboration}}(2022)}]{LHCb:2022december}
\bibinfo{author}{\bibnamefont{{LHCb Collaboration}}} (\bibinfo{year}{2022}),
  \bibinfo{note}{pre-print}, \eprint{2212.09153}.

\bibitem[{\citenamefont{Aaltonen et~al.}(2022)}]{CDF:2022hxs}
\bibinfo{author}{\bibfnamefont{T.}~\bibnamefont{Aaltonen}} \bibnamefont{et~al.}
  (\bibinfo{collaboration}{CDF}), \bibinfo{journal}{Science}
  \textbf{\bibinfo{volume}{376}}, \bibinfo{pages}{170} (\bibinfo{year}{2022}).

\bibitem[{\citenamefont{Strumia}(2022)}]{Strumia:2022qkt}
\bibinfo{author}{\bibfnamefont{A.}~\bibnamefont{Strumia}}
  (\bibinfo{year}{2022}), \bibinfo{note}{pre-print},
  \eprint{hep-ph/2204.04191}.

\bibitem[{\citenamefont{Bauer and Neubert}(2016)}]{Bauer:2015knc}
\bibinfo{author}{\bibfnamefont{M.}~\bibnamefont{Bauer}} \bibnamefont{and}
  \bibinfo{author}{\bibfnamefont{M.}~\bibnamefont{Neubert}},
  \bibinfo{journal}{Phys. Rev. Lett.} \textbf{\bibinfo{volume}{116}},
  \bibinfo{pages}{141802} (\bibinfo{year}{2016}), \eprint{1511.01900}.

\bibitem[{\citenamefont{Altmannshofer et~al.}(2017)\citenamefont{Altmannshofer,
  Bhupal~Dev, and Soni}}]{Altmannshofer:2017poe}
\bibinfo{author}{\bibfnamefont{W.}~\bibnamefont{Altmannshofer}},
  \bibinfo{author}{\bibfnamefont{P.~S.} \bibnamefont{Bhupal~Dev}},
  \bibnamefont{and} \bibinfo{author}{\bibfnamefont{A.}~\bibnamefont{Soni}},
  \bibinfo{journal}{Phys. Rev. D} \textbf{\bibinfo{volume}{96}},
  \bibinfo{pages}{095010} (\bibinfo{year}{2017}), \eprint{1704.06659}.

\bibitem[{\citenamefont{Das et~al.}(2016)\citenamefont{Das, Hati, Kumar, and
  Mahajan}}]{Das:2016vkr}
\bibinfo{author}{\bibfnamefont{D.}~\bibnamefont{Das}},
  \bibinfo{author}{\bibfnamefont{C.}~\bibnamefont{Hati}},
  \bibinfo{author}{\bibfnamefont{G.}~\bibnamefont{Kumar}}, \bibnamefont{and}
  \bibinfo{author}{\bibfnamefont{N.}~\bibnamefont{Mahajan}},
  \bibinfo{journal}{Phys. Rev. D} \textbf{\bibinfo{volume}{94}},
  \bibinfo{pages}{055034} (\bibinfo{year}{2016}), \eprint{1605.06313}.

\bibitem[{\citenamefont{Angelescu et~al.}(2018)\citenamefont{Angelescu,
  Be\v{c}irevi\'c, Faroughy, and Sumensari}}]{Angelescu:2018tyl}
\bibinfo{author}{\bibfnamefont{A.}~\bibnamefont{Angelescu}},
  \bibinfo{author}{\bibfnamefont{D.}~\bibnamefont{Be\v{c}irevi\'c}},
  \bibinfo{author}{\bibfnamefont{D.~A.} \bibnamefont{Faroughy}},
  \bibnamefont{and}
  \bibinfo{author}{\bibfnamefont{O.}~\bibnamefont{Sumensari}},
  \bibinfo{journal}{JHEP} \textbf{\bibinfo{volume}{10}}, \bibinfo{pages}{183}
  (\bibinfo{year}{2018}), \eprint{1808.08179}.

\bibitem[{\citenamefont{Altmannshofer et~al.}(2020)\citenamefont{Altmannshofer,
  Dev, Soni, and Sui}}]{Altmannshofer:2020axr}
\bibinfo{author}{\bibfnamefont{W.}~\bibnamefont{Altmannshofer}},
  \bibinfo{author}{\bibfnamefont{P.~S.~B.} \bibnamefont{Dev}},
  \bibinfo{author}{\bibfnamefont{A.}~\bibnamefont{Soni}}, \bibnamefont{and}
  \bibinfo{author}{\bibfnamefont{Y.}~\bibnamefont{Sui}},
  \bibinfo{journal}{Phys. Rev. D} \textbf{\bibinfo{volume}{102}},
  \bibinfo{pages}{015031} (\bibinfo{year}{2020}), \eprint{2002.12910}.

\bibitem[{\citenamefont{Belanger et~al.}(2022)}]{Belanger:2021smw}
\bibinfo{author}{\bibfnamefont{G.}~\bibnamefont{Belanger}}
  \bibnamefont{et~al.}, \bibinfo{journal}{JHEP} \textbf{\bibinfo{volume}{02}},
  \bibinfo{pages}{042} (\bibinfo{year}{2022}), \eprint{2111.08027}.

\bibitem[{\citenamefont{Becker et~al.}(2021)\citenamefont{Becker, D\"oring,
  Karmakar, and P\"as}}]{Becker:2021sfd}
\bibinfo{author}{\bibfnamefont{M.}~\bibnamefont{Becker}},
  \bibinfo{author}{\bibfnamefont{D.}~\bibnamefont{D\"oring}},
  \bibinfo{author}{\bibfnamefont{S.}~\bibnamefont{Karmakar}}, \bibnamefont{and}
  \bibinfo{author}{\bibfnamefont{H.}~\bibnamefont{P\"as}},
  \bibinfo{journal}{Eur. Phys. J. C} \textbf{\bibinfo{volume}{81}},
  \bibinfo{pages}{1053} (\bibinfo{year}{2021}), \eprint{2103.12043}.

\bibitem[{\citenamefont{Crivellin et~al.}(2022)\citenamefont{Crivellin, Fuks,
  and Schnell}}]{Crivellin:2022mff}
\bibinfo{author}{\bibfnamefont{A.}~\bibnamefont{Crivellin}},
  \bibinfo{author}{\bibfnamefont{B.}~\bibnamefont{Fuks}}, \bibnamefont{and}
  \bibinfo{author}{\bibfnamefont{L.}~\bibnamefont{Schnell}}
  (\bibinfo{year}{2022}), \eprint{2203.10111}.

\bibitem[{\citenamefont{Crivellin
  et~al.}(2021{\natexlab{a}})\citenamefont{Crivellin, Mueller, and
  Saturnino}}]{Crivellin:2020tsz}
\bibinfo{author}{\bibfnamefont{A.}~\bibnamefont{Crivellin}},
  \bibinfo{author}{\bibfnamefont{D.}~\bibnamefont{Mueller}}, \bibnamefont{and}
  \bibinfo{author}{\bibfnamefont{F.}~\bibnamefont{Saturnino}},
  \bibinfo{journal}{Phys. Rev. Lett.} \textbf{\bibinfo{volume}{127}},
  \bibinfo{pages}{021801} (\bibinfo{year}{2021}{\natexlab{a}}),
  \eprint{2008.02643}.

\bibitem[{\citenamefont{Crivellin et~al.}(2019)\citenamefont{Crivellin, Greub,
  M\"uller, and Saturnino}}]{Crivellin:2018yvo}
\bibinfo{author}{\bibfnamefont{A.}~\bibnamefont{Crivellin}},
  \bibinfo{author}{\bibfnamefont{C.}~\bibnamefont{Greub}},
  \bibinfo{author}{\bibfnamefont{D.}~\bibnamefont{M\"uller}}, \bibnamefont{and}
  \bibinfo{author}{\bibfnamefont{F.}~\bibnamefont{Saturnino}},
  \bibinfo{journal}{Phys. Rev. Lett.} \textbf{\bibinfo{volume}{122}},
  \bibinfo{pages}{011805} (\bibinfo{year}{2019}), \eprint{1807.02068}.

\bibitem[{\citenamefont{Blanke and Crivellin}(2018)}]{Blanke:2018sro}
\bibinfo{author}{\bibfnamefont{M.}~\bibnamefont{Blanke}} \bibnamefont{and}
  \bibinfo{author}{\bibfnamefont{A.}~\bibnamefont{Crivellin}},
  \bibinfo{journal}{Phys. Rev. Lett.} \textbf{\bibinfo{volume}{121}},
  \bibinfo{pages}{011801} (\bibinfo{year}{2018}), \eprint{1801.07256}.

\bibitem[{\citenamefont{Calibbi et~al.}(2018)\citenamefont{Calibbi, Crivellin,
  and Li}}]{Calibbi:2017qbu}
\bibinfo{author}{\bibfnamefont{L.}~\bibnamefont{Calibbi}},
  \bibinfo{author}{\bibfnamefont{A.}~\bibnamefont{Crivellin}},
  \bibnamefont{and} \bibinfo{author}{\bibfnamefont{T.}~\bibnamefont{Li}},
  \bibinfo{journal}{Phys. Rev. D} \textbf{\bibinfo{volume}{98}},
  \bibinfo{pages}{115002} (\bibinfo{year}{2018}), \eprint{1709.00692}.

\bibitem[{\citenamefont{Crivellin
  et~al.}(2020{\natexlab{a}})\citenamefont{Crivellin, M\"uller, and
  Saturnino}}]{Crivellin:2019dwb}
\bibinfo{author}{\bibfnamefont{A.}~\bibnamefont{Crivellin}},
  \bibinfo{author}{\bibfnamefont{D.}~\bibnamefont{M\"uller}}, \bibnamefont{and}
  \bibinfo{author}{\bibfnamefont{F.}~\bibnamefont{Saturnino}},
  \bibinfo{journal}{JHEP} \textbf{\bibinfo{volume}{06}}, \bibinfo{pages}{020}
  (\bibinfo{year}{2020}{\natexlab{a}}), \eprint{1912.04224}.

\bibitem[{\citenamefont{Carvunis et~al.}(2022)\citenamefont{Carvunis,
  Crivellin, Guadagnoli, and Gangal}}]{Carvunis:2021dss}
\bibinfo{author}{\bibfnamefont{A.}~\bibnamefont{Carvunis}},
  \bibinfo{author}{\bibfnamefont{A.}~\bibnamefont{Crivellin}},
  \bibinfo{author}{\bibfnamefont{D.}~\bibnamefont{Guadagnoli}},
  \bibnamefont{and} \bibinfo{author}{\bibfnamefont{S.}~\bibnamefont{Gangal}},
  \bibinfo{journal}{Phys. Rev. D} \textbf{\bibinfo{volume}{105}},
  \bibinfo{pages}{L031701} (\bibinfo{year}{2022}), \eprint{2106.09610}.

\bibitem[{\citenamefont{Coy and Frigerio}(2022)}]{Coy:2021hyr}
\bibinfo{author}{\bibfnamefont{R.}~\bibnamefont{Coy}} \bibnamefont{and}
  \bibinfo{author}{\bibfnamefont{M.}~\bibnamefont{Frigerio}},
  \bibinfo{journal}{Phys. Rev. D} \textbf{\bibinfo{volume}{105}},
  \bibinfo{pages}{115041} (\bibinfo{year}{2022}), \eprint{2110.09126}.

\bibitem[{\citenamefont{Marzocca and Trifinopoulos}(2021)}]{Marzocca:2021azj}
\bibinfo{author}{\bibfnamefont{D.}~\bibnamefont{Marzocca}} \bibnamefont{and}
  \bibinfo{author}{\bibfnamefont{S.}~\bibnamefont{Trifinopoulos}},
  \bibinfo{journal}{Phys. Rev. Lett.} \textbf{\bibinfo{volume}{127}},
  \bibinfo{pages}{061803} (\bibinfo{year}{2021}), \eprint{2104.05730}.

\bibitem[{\citenamefont{Saad and Thapa}(2020)}]{Saad:2020ucl}
\bibinfo{author}{\bibfnamefont{S.}~\bibnamefont{Saad}} \bibnamefont{and}
  \bibinfo{author}{\bibfnamefont{A.}~\bibnamefont{Thapa}},
  \bibinfo{journal}{Phys. Rev. D} \textbf{\bibinfo{volume}{102}},
  \bibinfo{pages}{015014} (\bibinfo{year}{2020}), \eprint{2004.07880}.

\bibitem[{\citenamefont{Chowdhury and Saad}(2022)}]{Chowdhury:2022dps}
\bibinfo{author}{\bibfnamefont{T.~A.} \bibnamefont{Chowdhury}}
  \bibnamefont{and} \bibinfo{author}{\bibfnamefont{S.}~\bibnamefont{Saad}}
  (\bibinfo{year}{2022}), \bibinfo{note}{pre-print},
  \eprint{hep-ph/2205.03917}.

\bibitem[{\citenamefont{SChen et~al.}(2022)\citenamefont{SChen, Jiang, and
  Liu}}]{Chen:2022hle}
\bibinfo{author}{\bibfnamefont{S.-L.} \bibnamefont{SChen}},
  \bibinfo{author}{\bibfnamefont{W.-w.} \bibnamefont{Jiang}}, \bibnamefont{and}
  \bibinfo{author}{\bibfnamefont{Z.-K.} \bibnamefont{Liu}}
  (\bibinfo{year}{2022}), \bibinfo{note}{pre-print},
  \eprint{hep-ph/2205.15794}.

\bibitem[{\citenamefont{Dor\v{s}ner et~al.}(2017)\citenamefont{Dor\v{s}ner,
  Fajfer, and Ko\v{s}nik}}]{Dorsner:2017wwn}
\bibinfo{author}{\bibfnamefont{I.}~\bibnamefont{Dor\v{s}ner}},
  \bibinfo{author}{\bibfnamefont{S.}~\bibnamefont{Fajfer}}, \bibnamefont{and}
  \bibinfo{author}{\bibfnamefont{N.}~\bibnamefont{Ko\v{s}nik}},
  \bibinfo{journal}{Eur. Phys. J. C} \textbf{\bibinfo{volume}{77}},
  \bibinfo{pages}{417} (\bibinfo{year}{2017}), \eprint{1701.08322}.

\bibitem[{\citenamefont{Aristizabal~Sierra
  et~al.}(2008)\citenamefont{Aristizabal~Sierra, Hirsch, and
  Kovalenko}}]{AristizabalSierra:2007nf}
\bibinfo{author}{\bibfnamefont{D.}~\bibnamefont{Aristizabal~Sierra}},
  \bibinfo{author}{\bibfnamefont{M.}~\bibnamefont{Hirsch}}, \bibnamefont{and}
  \bibinfo{author}{\bibfnamefont{S.~G.} \bibnamefont{Kovalenko}},
  \bibinfo{journal}{Phys. Rev. D} \textbf{\bibinfo{volume}{77}},
  \bibinfo{pages}{055011} (\bibinfo{year}{2008}), \eprint{0710.5699}.

\bibitem[{\citenamefont{Zhang}(2021)}]{Zhang:2021dgl}
\bibinfo{author}{\bibfnamefont{D.}~\bibnamefont{Zhang}},
  \bibinfo{journal}{JHEP} \textbf{\bibinfo{volume}{07}}, \bibinfo{pages}{069}
  (\bibinfo{year}{2021}), \eprint{2105.08670}.

\bibitem[{\citenamefont{P\"as and Schumacher}(2015)}]{Pas:2015hca}
\bibinfo{author}{\bibfnamefont{H.}~\bibnamefont{P\"as}} \bibnamefont{and}
  \bibinfo{author}{\bibfnamefont{E.}~\bibnamefont{Schumacher}},
  \bibinfo{journal}{Phys. Rev. D} \textbf{\bibinfo{volume}{92}},
  \bibinfo{pages}{114025} (\bibinfo{year}{2015}), \eprint{1510.08757}.

\bibitem[{\citenamefont{Cai et~al.}(2017{\natexlab{a}})\citenamefont{Cai,
  Herrero-Garc\'\i{}a, Schmidt, Vicente, and Volkas}}]{Cai:2017jrq}
\bibinfo{author}{\bibfnamefont{Y.}~\bibnamefont{Cai}},
  \bibinfo{author}{\bibfnamefont{J.}~\bibnamefont{Herrero-Garc\'\i{}a}},
  \bibinfo{author}{\bibfnamefont{M.~A.} \bibnamefont{Schmidt}},
  \bibinfo{author}{\bibfnamefont{A.}~\bibnamefont{Vicente}}, \bibnamefont{and}
  \bibinfo{author}{\bibfnamefont{R.~R.} \bibnamefont{Volkas}},
  \bibinfo{journal}{Front. in Phys.} \textbf{\bibinfo{volume}{5}},
  \bibinfo{pages}{63} (\bibinfo{year}{2017}{\natexlab{a}}),
  \eprint{1706.08524}.

\bibitem[{\citenamefont{Babu and Julio}(2010)}]{Babu:2010vp}
\bibinfo{author}{\bibfnamefont{K.~S.} \bibnamefont{Babu}} \bibnamefont{and}
  \bibinfo{author}{\bibfnamefont{J.}~\bibnamefont{Julio}},
  \bibinfo{journal}{Nucl. Phys. B} \textbf{\bibinfo{volume}{841}},
  \bibinfo{pages}{130} (\bibinfo{year}{2010}), \eprint{1006.1092}.

\bibitem[{\citenamefont{Cat\`a and Mannel}(2019)}]{Cata:2019wbu}
\bibinfo{author}{\bibfnamefont{O.}~\bibnamefont{Cat\`a}} \bibnamefont{and}
  \bibinfo{author}{\bibfnamefont{T.}~\bibnamefont{Mannel}}
  (\bibinfo{year}{2019}), \eprint{1903.01799}.

\bibitem[{\citenamefont{Popov and White}(2017)}]{Popov:2016fzr}
\bibinfo{author}{\bibfnamefont{O.}~\bibnamefont{Popov}} \bibnamefont{and}
  \bibinfo{author}{\bibfnamefont{G.~A.} \bibnamefont{White}},
  \bibinfo{journal}{Nucl. Phys. B} \textbf{\bibinfo{volume}{923}},
  \bibinfo{pages}{324} (\bibinfo{year}{2017}), \eprint{1611.04566}.

\bibitem[{\citenamefont{Nomura et~al.}(2021)\citenamefont{Nomura, Okada, and
  Orikasa}}]{Nomura:2021yjb}
\bibinfo{author}{\bibfnamefont{T.}~\bibnamefont{Nomura}},
  \bibinfo{author}{\bibfnamefont{H.}~\bibnamefont{Okada}}, \bibnamefont{and}
  \bibinfo{author}{\bibfnamefont{Y.}~\bibnamefont{Orikasa}},
  \bibinfo{journal}{Eur. Phys. J. C} \textbf{\bibinfo{volume}{81}},
  \bibinfo{pages}{947} (\bibinfo{year}{2021}), \eprint{2106.12375}.

\bibitem[{\citenamefont{Chang}(2021)}]{Chang:2021axw}
\bibinfo{author}{\bibfnamefont{W.-F.} \bibnamefont{Chang}},
  \bibinfo{journal}{JHEP} \textbf{\bibinfo{volume}{09}}, \bibinfo{pages}{043}
  (\bibinfo{year}{2021}), \eprint{2105.06917}.

\bibitem[{\citenamefont{Nomura and Okada}(2021)}]{Nomura:2021oeu}
\bibinfo{author}{\bibfnamefont{T.}~\bibnamefont{Nomura}} \bibnamefont{and}
  \bibinfo{author}{\bibfnamefont{H.}~\bibnamefont{Okada}},
  \bibinfo{journal}{Phys. Rev. D} \textbf{\bibinfo{volume}{104}},
  \bibinfo{pages}{035042} (\bibinfo{year}{2021}), \eprint{2104.03248}.

\bibitem[{\citenamefont{Babu et~al.}(2020)\citenamefont{Babu, Dev, Jana, and
  Thapa}}]{Babu:2019mfe}
\bibinfo{author}{\bibfnamefont{K.~S.} \bibnamefont{Babu}},
  \bibinfo{author}{\bibfnamefont{P.~S.~B.} \bibnamefont{Dev}},
  \bibinfo{author}{\bibfnamefont{S.}~\bibnamefont{Jana}}, \bibnamefont{and}
  \bibinfo{author}{\bibfnamefont{A.}~\bibnamefont{Thapa}},
  \bibinfo{journal}{JHEP} \textbf{\bibinfo{volume}{03}}, \bibinfo{pages}{006}
  (\bibinfo{year}{2020}), \eprint{1907.09498}.

\bibitem[{\citenamefont{Faber et~al.}(2020)\citenamefont{Faber, Hudec,
  Kole\v{s}ov\'a, Liu, Malinsk\textasciiacute{}y, Porod, and
  Staub}}]{Faber:2018afz}
\bibinfo{author}{\bibfnamefont{T.}~\bibnamefont{Faber}},
  \bibinfo{author}{\bibfnamefont{M.}~\bibnamefont{Hudec}},
  \bibinfo{author}{\bibfnamefont{H.}~\bibnamefont{Kole\v{s}ov\'a}},
  \bibinfo{author}{\bibfnamefont{Y.}~\bibnamefont{Liu}},
  \bibinfo{author}{\bibfnamefont{M.}~\bibnamefont{Malinsk\textasciiacute{}y}},
  \bibinfo{author}{\bibfnamefont{W.}~\bibnamefont{Porod}}, \bibnamefont{and}
  \bibinfo{author}{\bibfnamefont{F.}~\bibnamefont{Staub}},
  \bibinfo{journal}{Phys. Rev. D} \textbf{\bibinfo{volume}{101}},
  \bibinfo{pages}{095024} (\bibinfo{year}{2020}), \eprint{1812.07592}.

\bibitem[{\citenamefont{Faber et~al.}(2018)\citenamefont{Faber, Hudec,
  Malinsk\'y, Meinzinger, Porod, and Staub}}]{Faber:2018qon}
\bibinfo{author}{\bibfnamefont{T.}~\bibnamefont{Faber}},
  \bibinfo{author}{\bibfnamefont{M.}~\bibnamefont{Hudec}},
  \bibinfo{author}{\bibfnamefont{M.}~\bibnamefont{Malinsk\'y}},
  \bibinfo{author}{\bibfnamefont{P.}~\bibnamefont{Meinzinger}},
  \bibinfo{author}{\bibfnamefont{W.}~\bibnamefont{Porod}}, \bibnamefont{and}
  \bibinfo{author}{\bibfnamefont{F.}~\bibnamefont{Staub}},
  \bibinfo{journal}{Phys. Lett. B} \textbf{\bibinfo{volume}{787}},
  \bibinfo{pages}{159} (\bibinfo{year}{2018}), \eprint{1808.05511}.

\bibitem[{\citenamefont{Bigaran et~al.}(2019)\citenamefont{Bigaran,
  Gargalionis, and Volkas}}]{Bigaran:2019bqv}
\bibinfo{author}{\bibfnamefont{I.}~\bibnamefont{Bigaran}},
  \bibinfo{author}{\bibfnamefont{J.}~\bibnamefont{Gargalionis}},
  \bibnamefont{and} \bibinfo{author}{\bibfnamefont{R.~R.}
  \bibnamefont{Volkas}}, \bibinfo{journal}{JHEP} \textbf{\bibinfo{volume}{10}},
  \bibinfo{pages}{106} (\bibinfo{year}{2019}), \eprint{1906.01870}.

\bibitem[{\citenamefont{Gargalionis et~al.}(2020)\citenamefont{Gargalionis,
  Popa-Mateiu, and Volkas}}]{Gargalionis:2019drk}
\bibinfo{author}{\bibfnamefont{J.}~\bibnamefont{Gargalionis}},
  \bibinfo{author}{\bibfnamefont{I.}~\bibnamefont{Popa-Mateiu}},
  \bibnamefont{and} \bibinfo{author}{\bibfnamefont{R.~R.}
  \bibnamefont{Volkas}}, \bibinfo{journal}{JHEP} \textbf{\bibinfo{volume}{03}},
  \bibinfo{pages}{150} (\bibinfo{year}{2020}), \eprint{1912.12386}.

\bibitem[{\citenamefont{Gargalionis and Volkas}(2021)}]{Gargalionis:2020xvt}
\bibinfo{author}{\bibfnamefont{J.}~\bibnamefont{Gargalionis}} \bibnamefont{and}
  \bibinfo{author}{\bibfnamefont{R.~R.} \bibnamefont{Volkas}},
  \bibinfo{journal}{JHEP} \textbf{\bibinfo{volume}{01}}, \bibinfo{pages}{074}
  (\bibinfo{year}{2021}), \eprint{2009.13537}.

\bibitem[{\citenamefont{Saad}(2020)}]{Saad:2020ihm}
\bibinfo{author}{\bibfnamefont{S.}~\bibnamefont{Saad}}, \bibinfo{journal}{Phys.
  Rev. D} \textbf{\bibinfo{volume}{102}}, \bibinfo{pages}{015019}
  (\bibinfo{year}{2020}), \eprint{2005.04352}.

\bibitem[{\citenamefont{Julio et~al.}(2022{\natexlab{a}})\citenamefont{Julio,
  Saad, and Thapa}}]{Julio:2022bue}
\bibinfo{author}{\bibfnamefont{J.}~\bibnamefont{Julio}},
  \bibinfo{author}{\bibfnamefont{S.}~\bibnamefont{Saad}}, \bibnamefont{and}
  \bibinfo{author}{\bibfnamefont{A.}~\bibnamefont{Thapa}}
  (\bibinfo{year}{2022}{\natexlab{a}}), \bibinfo{note}{pre-print},
  \eprint{2203.15499}.

\bibitem[{\citenamefont{Julio et~al.}(2022{\natexlab{b}})\citenamefont{Julio,
  Saad, and Thapa}}]{Julio:2022ton}
\bibinfo{author}{\bibfnamefont{J.}~\bibnamefont{Julio}},
  \bibinfo{author}{\bibfnamefont{S.}~\bibnamefont{Saad}}, \bibnamefont{and}
  \bibinfo{author}{\bibfnamefont{A.}~\bibnamefont{Thapa}}
  (\bibinfo{year}{2022}{\natexlab{b}}), \bibinfo{note}{pre-print},
  \eprint{2202.10479}.

\bibitem[{\citenamefont{Crivellin
  et~al.}(2021{\natexlab{b}})\citenamefont{Crivellin, Greub, M\"uller, and
  Saturnino}}]{Crivellin:2020mjs}
\bibinfo{author}{\bibfnamefont{A.}~\bibnamefont{Crivellin}},
  \bibinfo{author}{\bibfnamefont{C.}~\bibnamefont{Greub}},
  \bibinfo{author}{\bibfnamefont{D.}~\bibnamefont{M\"uller}}, \bibnamefont{and}
  \bibinfo{author}{\bibfnamefont{F.}~\bibnamefont{Saturnino}},
  \bibinfo{journal}{JHEP} \textbf{\bibinfo{volume}{02}}, \bibinfo{pages}{182}
  (\bibinfo{year}{2021}{\natexlab{b}}), \eprint{2010.06593}.

\bibitem[{\citenamefont{Cai et~al.}(2017{\natexlab{b}})\citenamefont{Cai,
  Gargalionis, Schmidt, and Volkas}}]{Cai:2017wry}
\bibinfo{author}{\bibfnamefont{Y.}~\bibnamefont{Cai}},
  \bibinfo{author}{\bibfnamefont{J.}~\bibnamefont{Gargalionis}},
  \bibinfo{author}{\bibfnamefont{M.~A.} \bibnamefont{Schmidt}},
  \bibnamefont{and} \bibinfo{author}{\bibfnamefont{R.~R.}
  \bibnamefont{Volkas}}, \bibinfo{journal}{JHEP} \textbf{\bibinfo{volume}{10}},
  \bibinfo{pages}{047} (\bibinfo{year}{2017}{\natexlab{b}}),
  \eprint{1704.05849}.

\bibitem[{\citenamefont{Morais et~al.}(2020)\citenamefont{Morais, Pasechnik,
  and Porod}}]{Morais:2020ypd}
\bibinfo{author}{\bibfnamefont{A.~P.} \bibnamefont{Morais}},
  \bibinfo{author}{\bibfnamefont{R.}~\bibnamefont{Pasechnik}},
  \bibnamefont{and} \bibinfo{author}{\bibfnamefont{W.}~\bibnamefont{Porod}},
  \bibinfo{journal}{Eur. Phys. J. C} \textbf{\bibinfo{volume}{80}},
  \bibinfo{pages}{1162} (\bibinfo{year}{2020}), \eprint{2001.06383}.

\bibitem[{\citenamefont{Morais et~al.}(2021)\citenamefont{Morais, Pasechnik,
  and Porod}}]{Morais:2020odg}
\bibinfo{author}{\bibfnamefont{A.~P.} \bibnamefont{Morais}},
  \bibinfo{author}{\bibfnamefont{R.}~\bibnamefont{Pasechnik}},
  \bibnamefont{and} \bibinfo{author}{\bibfnamefont{W.}~\bibnamefont{Porod}},
  \bibinfo{journal}{Universe} \textbf{\bibinfo{volume}{7}},
  \bibinfo{pages}{461} (\bibinfo{year}{2021}), \eprint{2001.04804}.

\bibitem[{\citenamefont{Borsanyi et~al.}(2021)}]{Borsanyi:2020mff}
\bibinfo{author}{\bibfnamefont{S.}~\bibnamefont{Borsanyi}}
  \bibnamefont{et~al.}, \bibinfo{journal}{Nature}
  \textbf{\bibinfo{volume}{593}}, \bibinfo{pages}{51} (\bibinfo{year}{2021}),
  \eprint{2002.12347}.

\bibitem[{\citenamefont{Alexandrou et~al.}(2023)}]{ExtendedTwistedMass:2022jpw}
\bibinfo{author}{\bibfnamefont{C.}~\bibnamefont{Alexandrou}}
  \bibnamefont{et~al.} (\bibinfo{collaboration}{Extended Twisted Mass}),
  \bibinfo{journal}{Phys. Rev. D} \textbf{\bibinfo{volume}{107}},
  \bibinfo{pages}{074506} (\bibinfo{year}{2023}), \eprint{2206.15084}.

\bibitem[{\citenamefont{C\`e et~al.}(2022)}]{Ce:2022kxy}
\bibinfo{author}{\bibfnamefont{M.}~\bibnamefont{C\`e}} \bibnamefont{et~al.},
  \bibinfo{journal}{Phys. Rev. D} \textbf{\bibinfo{volume}{106}},
  \bibinfo{pages}{114502} (\bibinfo{year}{2022}), \eprint{2206.06582}.

\bibitem[{\citenamefont{Aoyama et~al.}(2012)\citenamefont{Aoyama, Hayakawa,
  Kinoshita, and Nio}}]{aoyama:2012wk}
\bibinfo{author}{\bibfnamefont{T.}~\bibnamefont{Aoyama}},
  \bibinfo{author}{\bibfnamefont{M.}~\bibnamefont{Hayakawa}},
  \bibinfo{author}{\bibfnamefont{T.}~\bibnamefont{Kinoshita}},
  \bibnamefont{and} \bibinfo{author}{\bibfnamefont{M.}~\bibnamefont{Nio}},
  \bibinfo{journal}{Phys. Rev. Lett.} \textbf{\bibinfo{volume}{109}},
  \bibinfo{pages}{111808} (\bibinfo{year}{2012}), \eprint{1205.5370}.

\bibitem[{\citenamefont{Aoyama et~al.}(2019)\citenamefont{Aoyama, Kinoshita,
  and Nio}}]{Aoyama:2019ryr}
\bibinfo{author}{\bibfnamefont{T.}~\bibnamefont{Aoyama}},
  \bibinfo{author}{\bibfnamefont{T.}~\bibnamefont{Kinoshita}},
  \bibnamefont{and} \bibinfo{author}{\bibfnamefont{M.}~\bibnamefont{Nio}},
  \bibinfo{journal}{Atoms} \textbf{\bibinfo{volume}{7}}, \bibinfo{pages}{28}
  (\bibinfo{year}{2019}).

\bibitem[{\citenamefont{Czarnecki et~al.}(2003)\citenamefont{Czarnecki,
  Marciano, and Vainshtein}}]{czarnecki:2002nt}
\bibinfo{author}{\bibfnamefont{A.}~\bibnamefont{Czarnecki}},
  \bibinfo{author}{\bibfnamefont{W.~J.} \bibnamefont{Marciano}},
  \bibnamefont{and}
  \bibinfo{author}{\bibfnamefont{A.}~\bibnamefont{Vainshtein}},
  \bibinfo{journal}{Phys. Rev.} \textbf{\bibinfo{volume}{D67}},
  \bibinfo{pages}{073006} (\bibinfo{year}{2003}), \bibinfo{note}{[Erratum:
  Phys. Rev. {\bf D73}, 119901 (2006)]}, \eprint{hep-ph/0212229}.

\bibitem[{\citenamefont{Gnendiger et~al.}(2013)\citenamefont{Gnendiger,
  St{\"o}ckinger, and St{\"o}ckinger-Kim}}]{gnendiger:2013pva}
\bibinfo{author}{\bibfnamefont{C.}~\bibnamefont{Gnendiger}},
  \bibinfo{author}{\bibfnamefont{D.}~\bibnamefont{St{\"o}ckinger}},
  \bibnamefont{and}
  \bibinfo{author}{\bibfnamefont{H.}~\bibnamefont{St{\"o}ckinger-Kim}},
  \bibinfo{journal}{Phys. Rev.} \textbf{\bibinfo{volume}{D88}},
  \bibinfo{pages}{053005} (\bibinfo{year}{2013}), \eprint{1306.5546}.

\bibitem[{\citenamefont{Davier et~al.}(2017)\citenamefont{Davier, Hoecker,
  Malaescu, and Zhang}}]{davier:2017zfy}
\bibinfo{author}{\bibfnamefont{M.}~\bibnamefont{Davier}},
  \bibinfo{author}{\bibfnamefont{A.}~\bibnamefont{Hoecker}},
  \bibinfo{author}{\bibfnamefont{B.}~\bibnamefont{Malaescu}}, \bibnamefont{and}
  \bibinfo{author}{\bibfnamefont{Z.}~\bibnamefont{Zhang}},
  \bibinfo{journal}{Eur. Phys. J.} \textbf{\bibinfo{volume}{C77}},
  \bibinfo{pages}{827} (\bibinfo{year}{2017}), \eprint{1706.09436}.

\bibitem[{\citenamefont{Keshavarzi et~al.}(2018)\citenamefont{Keshavarzi,
  Nomura, and Teubner}}]{keshavarzi:2018mgv}
\bibinfo{author}{\bibfnamefont{A.}~\bibnamefont{Keshavarzi}},
  \bibinfo{author}{\bibfnamefont{D.}~\bibnamefont{Nomura}}, \bibnamefont{and}
  \bibinfo{author}{\bibfnamefont{T.}~\bibnamefont{Teubner}},
  \bibinfo{journal}{Phys. Rev.} \textbf{\bibinfo{volume}{D97}},
  \bibinfo{pages}{114025} (\bibinfo{year}{2018}), \eprint{1802.02995}.

\bibitem[{\citenamefont{Colangelo et~al.}(2019)\citenamefont{Colangelo,
  Hoferichter, and Stoffer}}]{colangelo:2018mtw}
\bibinfo{author}{\bibfnamefont{G.}~\bibnamefont{Colangelo}},
  \bibinfo{author}{\bibfnamefont{M.}~\bibnamefont{Hoferichter}},
  \bibnamefont{and} \bibinfo{author}{\bibfnamefont{P.}~\bibnamefont{Stoffer}},
  \bibinfo{journal}{JHEP} \textbf{\bibinfo{volume}{02}}, \bibinfo{pages}{006}
  (\bibinfo{year}{2019}), \eprint{1810.00007}.

\bibitem[{\citenamefont{Hoferichter et~al.}(2019)\citenamefont{Hoferichter,
  Hoid, and Kubis}}]{hoferichter:2019gzf}
\bibinfo{author}{\bibfnamefont{M.}~\bibnamefont{Hoferichter}},
  \bibinfo{author}{\bibfnamefont{B.-L.} \bibnamefont{Hoid}}, \bibnamefont{and}
  \bibinfo{author}{\bibfnamefont{B.}~\bibnamefont{Kubis}},
  \bibinfo{journal}{JHEP} \textbf{\bibinfo{volume}{08}}, \bibinfo{pages}{137}
  (\bibinfo{year}{2019}), \eprint{1907.01556}.

\bibitem[{\citenamefont{Davier et~al.}(2020)\citenamefont{Davier, Hoecker,
  Malaescu, and Zhang}}]{davier:2019can}
\bibinfo{author}{\bibfnamefont{M.}~\bibnamefont{Davier}},
  \bibinfo{author}{\bibfnamefont{A.}~\bibnamefont{Hoecker}},
  \bibinfo{author}{\bibfnamefont{B.}~\bibnamefont{Malaescu}}, \bibnamefont{and}
  \bibinfo{author}{\bibfnamefont{Z.}~\bibnamefont{Zhang}},
  \bibinfo{journal}{Eur. Phys. J. C} \textbf{\bibinfo{volume}{80}},
  \bibinfo{pages}{241} (\bibinfo{year}{2020}), \bibinfo{note}{[Erratum:
  Eur.Phys.J.C 80, 410 (2020)]}, \eprint{1908.00921}.

\bibitem[{\citenamefont{Keshavarzi et~al.}(2020)\citenamefont{Keshavarzi,
  Nomura, and Teubner}}]{keshavarzi:2019abf}
\bibinfo{author}{\bibfnamefont{A.}~\bibnamefont{Keshavarzi}},
  \bibinfo{author}{\bibfnamefont{D.}~\bibnamefont{Nomura}}, \bibnamefont{and}
  \bibinfo{author}{\bibfnamefont{T.}~\bibnamefont{Teubner}},
  \bibinfo{journal}{Phys. Rev. D} \textbf{\bibinfo{volume}{101}},
  \bibinfo{pages}{014029} (\bibinfo{year}{2020}), \eprint{1911.00367}.

\bibitem[{\citenamefont{Kurz et~al.}(2014)\citenamefont{Kurz, Liu, Marquard,
  and Steinhauser}}]{kurz:2014wya}
\bibinfo{author}{\bibfnamefont{A.}~\bibnamefont{Kurz}},
  \bibinfo{author}{\bibfnamefont{T.}~\bibnamefont{Liu}},
  \bibinfo{author}{\bibfnamefont{P.}~\bibnamefont{Marquard}}, \bibnamefont{and}
  \bibinfo{author}{\bibfnamefont{M.}~\bibnamefont{Steinhauser}},
  \bibinfo{journal}{Phys. Lett.} \textbf{\bibinfo{volume}{B734}},
  \bibinfo{pages}{144} (\bibinfo{year}{2014}), \eprint{1403.6400}.

\bibitem[{\citenamefont{Melnikov and Vainshtein}(2004)}]{melnikov:2003xd}
\bibinfo{author}{\bibfnamefont{K.}~\bibnamefont{Melnikov}} \bibnamefont{and}
  \bibinfo{author}{\bibfnamefont{A.}~\bibnamefont{Vainshtein}},
  \bibinfo{journal}{Phys. Rev.} \textbf{\bibinfo{volume}{D70}},
  \bibinfo{pages}{113006} (\bibinfo{year}{2004}), \eprint{hep-ph/0312226}.

\bibitem[{\citenamefont{Masjuan and
  S{\'a}nchez-Puertas}(2017)}]{masjuan:2017tvw}
\bibinfo{author}{\bibfnamefont{P.}~\bibnamefont{Masjuan}} \bibnamefont{and}
  \bibinfo{author}{\bibfnamefont{P.}~\bibnamefont{S{\'a}nchez-Puertas}},
  \bibinfo{journal}{Phys. Rev.} \textbf{\bibinfo{volume}{D95}},
  \bibinfo{pages}{054026} (\bibinfo{year}{2017}), \eprint{1701.05829}.

\bibitem[{\citenamefont{Colangelo et~al.}(2017)\citenamefont{Colangelo,
  Hoferichter, Procura, and Stoffer}}]{Colangelo:2017fiz}
\bibinfo{author}{\bibfnamefont{G.}~\bibnamefont{Colangelo}},
  \bibinfo{author}{\bibfnamefont{M.}~\bibnamefont{Hoferichter}},
  \bibinfo{author}{\bibfnamefont{M.}~\bibnamefont{Procura}}, \bibnamefont{and}
  \bibinfo{author}{\bibfnamefont{P.}~\bibnamefont{Stoffer}},
  \bibinfo{journal}{JHEP} \textbf{\bibinfo{volume}{04}}, \bibinfo{pages}{161}
  (\bibinfo{year}{2017}), \eprint{1702.07347}.

\bibitem[{\citenamefont{Hoferichter et~al.}(2018)\citenamefont{Hoferichter,
  Hoid, Kubis, Leupold, and Schneider}}]{hoferichter:2018kwz}
\bibinfo{author}{\bibfnamefont{M.}~\bibnamefont{Hoferichter}},
  \bibinfo{author}{\bibfnamefont{B.-L.} \bibnamefont{Hoid}},
  \bibinfo{author}{\bibfnamefont{B.}~\bibnamefont{Kubis}},
  \bibinfo{author}{\bibfnamefont{S.}~\bibnamefont{Leupold}}, \bibnamefont{and}
  \bibinfo{author}{\bibfnamefont{S.~P.} \bibnamefont{Schneider}},
  \bibinfo{journal}{JHEP} \textbf{\bibinfo{volume}{10}}, \bibinfo{pages}{141}
  (\bibinfo{year}{2018}), \eprint{1808.04823}.

\bibitem[{\citenamefont{G{\'e}rardin et~al.}(2019)\citenamefont{G{\'e}rardin,
  Meyer, and Nyffeler}}]{gerardin:2019vio}
\bibinfo{author}{\bibfnamefont{A.}~\bibnamefont{G{\'e}rardin}},
  \bibinfo{author}{\bibfnamefont{H.~B.} \bibnamefont{Meyer}}, \bibnamefont{and}
  \bibinfo{author}{\bibfnamefont{A.}~\bibnamefont{Nyffeler}},
  \bibinfo{journal}{Phys. Rev.} \textbf{\bibinfo{volume}{D100}},
  \bibinfo{pages}{034520} (\bibinfo{year}{2019}), \eprint{1903.09471}.

\bibitem[{\citenamefont{Bijnens et~al.}(2019)\citenamefont{Bijnens,
  Hermansson-Truedsson, and Rodr{\'i}guez-S{\'a}nchez}}]{bijnens:2019ghy}
\bibinfo{author}{\bibfnamefont{J.}~\bibnamefont{Bijnens}},
  \bibinfo{author}{\bibfnamefont{N.}~\bibnamefont{Hermansson-Truedsson}},
  \bibnamefont{and}
  \bibinfo{author}{\bibfnamefont{A.}~\bibnamefont{Rodr{\'i}guez-S{\'a}nchez}},
  \bibinfo{journal}{Phys. Lett.} \textbf{\bibinfo{volume}{B798}},
  \bibinfo{pages}{134994} (\bibinfo{year}{2019}), \eprint{1908.03331}.

\bibitem[{\citenamefont{Colangelo et~al.}(2020)\citenamefont{Colangelo,
  Hagelstein, Hoferichter, Laub, and Stoffer}}]{colangelo:2019uex}
\bibinfo{author}{\bibfnamefont{G.}~\bibnamefont{Colangelo}},
  \bibinfo{author}{\bibfnamefont{F.}~\bibnamefont{Hagelstein}},
  \bibinfo{author}{\bibfnamefont{M.}~\bibnamefont{Hoferichter}},
  \bibinfo{author}{\bibfnamefont{L.}~\bibnamefont{Laub}}, \bibnamefont{and}
  \bibinfo{author}{\bibfnamefont{P.}~\bibnamefont{Stoffer}},
  \bibinfo{journal}{JHEP} \textbf{\bibinfo{volume}{03}}, \bibinfo{pages}{101}
  (\bibinfo{year}{2020}), \eprint{1910.13432}.

\bibitem[{\citenamefont{Blum et~al.}(2020)\citenamefont{Blum, Christ, Hayakawa,
  Izubuchi, Jin, Jung, and Lehner}}]{Blum:2019ugy}
\bibinfo{author}{\bibfnamefont{T.}~\bibnamefont{Blum}},
  \bibinfo{author}{\bibfnamefont{N.}~\bibnamefont{Christ}},
  \bibinfo{author}{\bibfnamefont{M.}~\bibnamefont{Hayakawa}},
  \bibinfo{author}{\bibfnamefont{T.}~\bibnamefont{Izubuchi}},
  \bibinfo{author}{\bibfnamefont{L.}~\bibnamefont{Jin}},
  \bibinfo{author}{\bibfnamefont{C.}~\bibnamefont{Jung}}, \bibnamefont{and}
  \bibinfo{author}{\bibfnamefont{C.}~\bibnamefont{Lehner}},
  \bibinfo{journal}{Phys. Rev. Lett.} \textbf{\bibinfo{volume}{124}},
  \bibinfo{pages}{132002} (\bibinfo{year}{2020}), \eprint{1911.08123}.

\bibitem[{\citenamefont{Colangelo et~al.}(2014)\citenamefont{Colangelo,
  Hoferichter, Nyffeler, Passera, and Stoffer}}]{colangelo:2014qya}
\bibinfo{author}{\bibfnamefont{G.}~\bibnamefont{Colangelo}},
  \bibinfo{author}{\bibfnamefont{M.}~\bibnamefont{Hoferichter}},
  \bibinfo{author}{\bibfnamefont{A.}~\bibnamefont{Nyffeler}},
  \bibinfo{author}{\bibfnamefont{M.}~\bibnamefont{Passera}}, \bibnamefont{and}
  \bibinfo{author}{\bibfnamefont{P.}~\bibnamefont{Stoffer}},
  \bibinfo{journal}{Phys. Lett.} \textbf{\bibinfo{volume}{B735}},
  \bibinfo{pages}{90} (\bibinfo{year}{2014}), \eprint{1403.7512}.

\bibitem[{\citenamefont{Aoyama et~al.}(2020)}]{Aoyama:2020ynm}
\bibinfo{author}{\bibfnamefont{T.}~\bibnamefont{Aoyama}} \bibnamefont{et~al.},
  \bibinfo{journal}{Phys. Rept.} \textbf{\bibinfo{volume}{887}},
  \bibinfo{pages}{1} (\bibinfo{year}{2020}), \eprint{2006.04822}.

\bibitem[{\citenamefont{Bennett et~al.}(2006)}]{Muong-2:2006rrc}
\bibinfo{author}{\bibfnamefont{G.~W.} \bibnamefont{Bennett}}
  \bibnamefont{et~al.} (\bibinfo{collaboration}{Muon g-2}),
  \bibinfo{journal}{Phys. Rev. D} \textbf{\bibinfo{volume}{73}},
  \bibinfo{pages}{072003} (\bibinfo{year}{2006}), \eprint{hep-ex/0602035}.

\bibitem[{\citenamefont{Dor\v{s}ner et~al.}(2020)\citenamefont{Dor\v{s}ner,
  Fajfer, and Sumensari}}]{Dorsner:2019itg}
\bibinfo{author}{\bibfnamefont{I.}~\bibnamefont{Dor\v{s}ner}},
  \bibinfo{author}{\bibfnamefont{S.}~\bibnamefont{Fajfer}}, \bibnamefont{and}
  \bibinfo{author}{\bibfnamefont{O.}~\bibnamefont{Sumensari}},
  \bibinfo{journal}{JHEP} \textbf{\bibinfo{volume}{06}}, \bibinfo{pages}{089}
  (\bibinfo{year}{2020}), \eprint{1910.03877}.

\bibitem[{\citenamefont{Arcadi et~al.}(2021)\citenamefont{Arcadi, Calibbi,
  Fedele, and Mescia}}]{Arcadi:2021cwg}
\bibinfo{author}{\bibfnamefont{G.}~\bibnamefont{Arcadi}},
  \bibinfo{author}{\bibfnamefont{L.}~\bibnamefont{Calibbi}},
  \bibinfo{author}{\bibfnamefont{M.}~\bibnamefont{Fedele}}, \bibnamefont{and}
  \bibinfo{author}{\bibfnamefont{F.}~\bibnamefont{Mescia}},
  \bibinfo{journal}{Phys. Rev. Lett.} \textbf{\bibinfo{volume}{127}},
  \bibinfo{pages}{061802} (\bibinfo{year}{2021}), \eprint{2104.03228}.

\bibitem[{\citenamefont{Perez et~al.}(2021)\citenamefont{Perez, Murgui, and
  Plascencia}}]{Perez:2021ddi}
\bibinfo{author}{\bibfnamefont{P.~F.} \bibnamefont{Perez}},
  \bibinfo{author}{\bibfnamefont{C.}~\bibnamefont{Murgui}}, \bibnamefont{and}
  \bibinfo{author}{\bibfnamefont{A.~D.} \bibnamefont{Plascencia}},
  \bibinfo{journal}{Phys. Rev. D} \textbf{\bibinfo{volume}{104}},
  \bibinfo{pages}{035041} (\bibinfo{year}{2021}), \eprint{2104.11229}.

\bibitem[{\citenamefont{Amhis
  et~al.}(2023)}]{HeavyFlavorAveragingGroup:2022wzx}
\bibinfo{author}{\bibfnamefont{Y.~S.} \bibnamefont{Amhis}} \bibnamefont{et~al.}
  (\bibinfo{collaboration}{Heavy Flavor Averaging Group, HFLAV}),
  \bibinfo{journal}{Phys. Rev. D} \textbf{\bibinfo{volume}{107}},
  \bibinfo{pages}{052008} (\bibinfo{year}{2023}), \eprint{2206.07501}.

\bibitem[{\citenamefont{Straub}(2022)}]{Straub:2018kue}
\bibinfo{author}{\bibfnamefont{D.~M.} \bibnamefont{Straub}}
  (\bibinfo{year}{2022}), \bibinfo{note}{pre-print},
  \eprint{hep-ph/1810.08132}.

\bibitem[{\citenamefont{Bailey et~al.}(2015)}]{MILC:2015uhg}
\bibinfo{author}{\bibfnamefont{J.~A.} \bibnamefont{Bailey}}
  \bibnamefont{et~al.} (\bibinfo{collaboration}{MILC}), \bibinfo{journal}{Phys.
  Rev. D} \textbf{\bibinfo{volume}{92}}, \bibinfo{pages}{034506}
  (\bibinfo{year}{2015}), \eprint{1503.07237}.

\bibitem[{\citenamefont{Bigi and Gambino}(2016)}]{Bigi:2016mdz}
\bibinfo{author}{\bibfnamefont{D.}~\bibnamefont{Bigi}} \bibnamefont{and}
  \bibinfo{author}{\bibfnamefont{P.}~\bibnamefont{Gambino}},
  \bibinfo{journal}{Phys. Rev. D} \textbf{\bibinfo{volume}{94}},
  \bibinfo{pages}{094008} (\bibinfo{year}{2016}), \eprint{1606.08030}.

\bibitem[{\citenamefont{Bardhan et~al.}(2017)\citenamefont{Bardhan, Byakti, and
  Ghosh}}]{Bardhan:2016uhr}
\bibinfo{author}{\bibfnamefont{D.}~\bibnamefont{Bardhan}},
  \bibinfo{author}{\bibfnamefont{P.}~\bibnamefont{Byakti}}, \bibnamefont{and}
  \bibinfo{author}{\bibfnamefont{D.}~\bibnamefont{Ghosh}},
  \bibinfo{journal}{JHEP} \textbf{\bibinfo{volume}{01}}, \bibinfo{pages}{125}
  (\bibinfo{year}{2017}), \eprint{1610.03038}.

\bibitem[{\citenamefont{Ganiev}(2023)}]{BelleII_Knunu}
\bibinfo{author}{\bibfnamefont{E.}~\bibnamefont{Ganiev}},
  \emph{\bibinfo{title}{On radiative and electroweak penguin decays}},
  \bibinfo{howpublished}{\url{https://indico.desy.de/event/34916/contributions/146877/attachments/84380/111798/EWP@Belle2_EPS.pdf}}
  (\bibinfo{year}{2023}).

\bibitem[{\citenamefont{Aaboud et~al.}(2018)}]{ATLAS:2017rzl}
\bibinfo{author}{\bibfnamefont{M.}~\bibnamefont{Aaboud}} \bibnamefont{et~al.}
  (\bibinfo{collaboration}{ATLAS}), \bibinfo{journal}{Eur. Phys. J. C}
  \textbf{\bibinfo{volume}{78}}, \bibinfo{pages}{110} (\bibinfo{year}{2018}),
  \bibinfo{note}{[Erratum: Eur.Phys.J.C 78, 898 (2018)]}, \eprint{1701.07240}.

\bibitem[{\citenamefont{Aaij et~al.}(2022{\natexlab{b}})}]{LHCb:2021bjt}
\bibinfo{author}{\bibfnamefont{R.}~\bibnamefont{Aaij}} \bibnamefont{et~al.}
  (\bibinfo{collaboration}{LHCb}), \bibinfo{journal}{JHEP}
  \textbf{\bibinfo{volume}{01}}, \bibinfo{pages}{036}
  (\bibinfo{year}{2022}{\natexlab{b}}), \eprint{2109.01113}.

\bibitem[{ATL(2023)}]{ATLAS:2023fsi}
 (\bibinfo{year}{2023}).

\bibitem[{\citenamefont{de~Blas et~al.}(2022)\citenamefont{de~Blas, Pierini,
  Reina, and Silvestrini}}]{deBlas:2022hdk}
\bibinfo{author}{\bibfnamefont{J.}~\bibnamefont{de~Blas}},
  \bibinfo{author}{\bibfnamefont{M.}~\bibnamefont{Pierini}},
  \bibinfo{author}{\bibfnamefont{L.}~\bibnamefont{Reina}}, \bibnamefont{and}
  \bibinfo{author}{\bibfnamefont{L.}~\bibnamefont{Silvestrini}},
  \bibinfo{journal}{Phys. Rev. Lett.} \textbf{\bibinfo{volume}{129}},
  \bibinfo{pages}{271801} (\bibinfo{year}{2022}), \eprint{2204.04204}.

\bibitem[{\citenamefont{Crivellin
  et~al.}(2020{\natexlab{b}})\citenamefont{Crivellin, M\"uller, and
  Saturnino}}]{Crivellin:2020ukd}
\bibinfo{author}{\bibfnamefont{A.}~\bibnamefont{Crivellin}},
  \bibinfo{author}{\bibfnamefont{D.}~\bibnamefont{M\"uller}}, \bibnamefont{and}
  \bibinfo{author}{\bibfnamefont{F.}~\bibnamefont{Saturnino}},
  \bibinfo{journal}{JHEP} \textbf{\bibinfo{volume}{11}}, \bibinfo{pages}{094}
  (\bibinfo{year}{2020}{\natexlab{b}}), \eprint{2006.10758}.

\bibitem[{\citenamefont{Schael et~al.}(2006)}]{ALEPH:2005ab}
\bibinfo{author}{\bibfnamefont{S.}~\bibnamefont{Schael}} \bibnamefont{et~al.}
  (\bibinfo{collaboration}{ALEPH, DELPHI, L3, OPAL, SLD, LEP Electroweak
  Working Group, SLD Electroweak Group, SLD Heavy Flavour Group}),
  \bibinfo{journal}{Phys. Rept.} \textbf{\bibinfo{volume}{427}},
  \bibinfo{pages}{257} (\bibinfo{year}{2006}), \eprint{hep-ex/0509008}.

\bibitem[{\citenamefont{Arnan et~al.}(2019)\citenamefont{Arnan, Becirevic,
  Mescia, and Sumensari}}]{Arnan:2019olv}
\bibinfo{author}{\bibfnamefont{P.}~\bibnamefont{Arnan}},
  \bibinfo{author}{\bibfnamefont{D.}~\bibnamefont{Becirevic}},
  \bibinfo{author}{\bibfnamefont{F.}~\bibnamefont{Mescia}}, \bibnamefont{and}
  \bibinfo{author}{\bibfnamefont{O.}~\bibnamefont{Sumensari}},
  \bibinfo{journal}{JHEP} \textbf{\bibinfo{volume}{02}}, \bibinfo{pages}{109}
  (\bibinfo{year}{2019}), \eprint{1901.06315}.

\bibitem[{\citenamefont{Bobeth and Buras}(2018)}]{Bobeth:2017ecx}
\bibinfo{author}{\bibfnamefont{C.}~\bibnamefont{Bobeth}} \bibnamefont{and}
  \bibinfo{author}{\bibfnamefont{A.~J.} \bibnamefont{Buras}},
  \bibinfo{journal}{JHEP} \textbf{\bibinfo{volume}{02}}, \bibinfo{pages}{101}
  (\bibinfo{year}{2018}), \eprint{1712.01295}.

\bibitem[{\citenamefont{Dor\v{s}ner et~al.}(2016)\citenamefont{Dor\v{s}ner,
  Fajfer, Greljo, Kamenik, and Ko\v{s}nik}}]{Dorsner:2016wpm}
\bibinfo{author}{\bibfnamefont{I.}~\bibnamefont{Dor\v{s}ner}},
  \bibinfo{author}{\bibfnamefont{S.}~\bibnamefont{Fajfer}},
  \bibinfo{author}{\bibfnamefont{A.}~\bibnamefont{Greljo}},
  \bibinfo{author}{\bibfnamefont{J.~F.} \bibnamefont{Kamenik}},
  \bibnamefont{and}
  \bibinfo{author}{\bibfnamefont{N.}~\bibnamefont{Ko\v{s}nik}},
  \bibinfo{journal}{Phys. Rept.} \textbf{\bibinfo{volume}{641}},
  \bibinfo{pages}{1} (\bibinfo{year}{2016}), \eprint{1603.04993}.

\bibitem[{\citenamefont{Crivellin
  et~al.}(2021{\natexlab{c}})\citenamefont{Crivellin, M\"uller, and
  Schnell}}]{Crivellin:2021egp}
\bibinfo{author}{\bibfnamefont{A.}~\bibnamefont{Crivellin}},
  \bibinfo{author}{\bibfnamefont{D.}~\bibnamefont{M\"uller}}, \bibnamefont{and}
  \bibinfo{author}{\bibfnamefont{L.}~\bibnamefont{Schnell}},
  \bibinfo{journal}{Phys. Rev. D} \textbf{\bibinfo{volume}{103}},
  \bibinfo{pages}{115023} (\bibinfo{year}{2021}{\natexlab{c}}),
  \eprint{2104.06417}.

\bibitem[{\citenamefont{Aad et~al.}(2021{\natexlab{a}})}]{ATLAS:2021jyv}
\bibinfo{author}{\bibfnamefont{G.}~\bibnamefont{Aad}} \bibnamefont{et~al.}
  (\bibinfo{collaboration}{ATLAS}), \bibinfo{journal}{Phys. Rev. D}
  \textbf{\bibinfo{volume}{104}}, \bibinfo{pages}{112005}
  (\bibinfo{year}{2021}{\natexlab{a}}), \eprint{2108.07665}.

\bibitem[{\citenamefont{Aad et~al.}(2021{\natexlab{b}})}]{ATLAS:2021oiz}
\bibinfo{author}{\bibfnamefont{G.}~\bibnamefont{Aad}} \bibnamefont{et~al.}
  (\bibinfo{collaboration}{ATLAS}), \bibinfo{journal}{JHEP}
  \textbf{\bibinfo{volume}{06}}, \bibinfo{pages}{179}
  (\bibinfo{year}{2021}{\natexlab{b}}), \eprint{2101.11582}.

\bibitem[{\citenamefont{Sirunyan et~al.}(2019)}]{CMS:2018iye}
\bibinfo{author}{\bibfnamefont{A.~M.} \bibnamefont{Sirunyan}}
  \bibnamefont{et~al.} (\bibinfo{collaboration}{CMS}), \bibinfo{journal}{JHEP}
  \textbf{\bibinfo{volume}{03}}, \bibinfo{pages}{170} (\bibinfo{year}{2019}),
  \eprint{1811.00806}.

\bibitem[{\citenamefont{Sirunyan et~al.}(2018)}]{CMS:2018oaj}
\bibinfo{author}{\bibfnamefont{A.~M.} \bibnamefont{Sirunyan}}
  \bibnamefont{et~al.} (\bibinfo{collaboration}{CMS}), \bibinfo{journal}{Phys.
  Rev. Lett.} \textbf{\bibinfo{volume}{121}}, \bibinfo{pages}{241802}
  (\bibinfo{year}{2018}), \eprint{1809.05558}.

\bibitem[{\citenamefont{Workman et~al.}(2022)}]{ParticleDataGroup:2022pth}
\bibinfo{author}{\bibfnamefont{R.~L.} \bibnamefont{Workman}}
  \bibnamefont{et~al.} (\bibinfo{collaboration}{Particle Data Group}),
  \bibinfo{journal}{PTEP} \textbf{\bibinfo{volume}{2022}},
  \bibinfo{pages}{083C01} (\bibinfo{year}{2022}).

\bibitem[{\citenamefont{Staub}(2014)}]{Staub:2013tta}
\bibinfo{author}{\bibfnamefont{F.}~\bibnamefont{Staub}},
  \bibinfo{journal}{Comput. Phys. Commun.} \textbf{\bibinfo{volume}{185}},
  \bibinfo{pages}{1773} (\bibinfo{year}{2014}), \eprint{1309.7223}.

\bibitem[{\citenamefont{Porod and Staub}(2012)}]{Porod:2011nf}
\bibinfo{author}{\bibfnamefont{W.}~\bibnamefont{Porod}} \bibnamefont{and}
  \bibinfo{author}{\bibfnamefont{F.}~\bibnamefont{Staub}},
  \bibinfo{journal}{Comput. Phys. Commun.} \textbf{\bibinfo{volume}{183}},
  \bibinfo{pages}{2458} (\bibinfo{year}{2012}), \eprint{1104.1573}.

\bibitem[{\citenamefont{Aebischer et~al.}(2018)\citenamefont{Aebischer, Kumar,
  and Straub}}]{Aebischer:2018bkb}
\bibinfo{author}{\bibfnamefont{J.}~\bibnamefont{Aebischer}},
  \bibinfo{author}{\bibfnamefont{J.}~\bibnamefont{Kumar}}, \bibnamefont{and}
  \bibinfo{author}{\bibfnamefont{D.~M.} \bibnamefont{Straub}},
  \bibinfo{journal}{Eur. Phys. J. C} \textbf{\bibinfo{volume}{78}},
  \bibinfo{pages}{1026} (\bibinfo{year}{2018}), \eprint{1804.05033}.

\bibitem[{\citenamefont{{W}es {M}c{K}inney}(2010)}]{mckinney-proc-scipy-2010}
\bibinfo{author}{\bibnamefont{{W}es {M}c{K}inney}}, in
  \emph{\bibinfo{booktitle}{{P}roceedings of the 9th {P}ython in {S}cience
  {C}onference}}, edited by \bibinfo{editor}{\bibnamefont{{S}t\'efan van~der
  {W}alt}} \bibnamefont{and} \bibinfo{editor}{\bibnamefont{{J}arrod {M}illman}}
  (\bibinfo{year}{2010}), pp. \bibinfo{pages}{56 -- 61}.

\bibitem[{\citenamefont{Mandal and Pich}(2019)}]{Mandal:2019gff}
\bibinfo{author}{\bibfnamefont{R.}~\bibnamefont{Mandal}} \bibnamefont{and}
  \bibinfo{author}{\bibfnamefont{A.}~\bibnamefont{Pich}},
  \bibinfo{journal}{JHEP} \textbf{\bibinfo{volume}{12}}, \bibinfo{pages}{089}
  (\bibinfo{year}{2019}), \eprint{1908.11155}.

\bibitem[{\citenamefont{Faroughy et~al.}(2017)\citenamefont{Faroughy, Greljo,
  and Kamenik}}]{Faroughy:2016osc}
\bibinfo{author}{\bibfnamefont{D.~A.} \bibnamefont{Faroughy}},
  \bibinfo{author}{\bibfnamefont{A.}~\bibnamefont{Greljo}}, \bibnamefont{and}
  \bibinfo{author}{\bibfnamefont{J.~F.} \bibnamefont{Kamenik}},
  \bibinfo{journal}{Phys. Lett. B} \textbf{\bibinfo{volume}{764}},
  \bibinfo{pages}{126} (\bibinfo{year}{2017}), \eprint{1609.07138}.

\bibitem[{\citenamefont{Greljo and Marzocca}(2017)}]{Greljo:2017vvb}
\bibinfo{author}{\bibfnamefont{A.}~\bibnamefont{Greljo}} \bibnamefont{and}
  \bibinfo{author}{\bibfnamefont{D.}~\bibnamefont{Marzocca}},
  \bibinfo{journal}{Eur. Phys. J. C} \textbf{\bibinfo{volume}{77}},
  \bibinfo{pages}{548} (\bibinfo{year}{2017}), \eprint{1704.09015}.

\bibitem[{\citenamefont{D'Amico et~al.}(2017)\citenamefont{D'Amico, Nardecchia,
  Panci, Sannino, Strumia, Torre, and Urbano}}]{DAmico:2017mtc}
\bibinfo{author}{\bibfnamefont{G.}~\bibnamefont{D'Amico}},
  \bibinfo{author}{\bibfnamefont{M.}~\bibnamefont{Nardecchia}},
  \bibinfo{author}{\bibfnamefont{P.}~\bibnamefont{Panci}},
  \bibinfo{author}{\bibfnamefont{F.}~\bibnamefont{Sannino}},
  \bibinfo{author}{\bibfnamefont{A.}~\bibnamefont{Strumia}},
  \bibinfo{author}{\bibfnamefont{R.}~\bibnamefont{Torre}}, \bibnamefont{and}
  \bibinfo{author}{\bibfnamefont{A.}~\bibnamefont{Urbano}},
  \bibinfo{journal}{JHEP} \textbf{\bibinfo{volume}{09}}, \bibinfo{pages}{010}
  (\bibinfo{year}{2017}), \eprint{1704.05438}.

\bibitem[{\citenamefont{Amhis et~al.}(2021)}]{HFLAV:2019otj}
\bibinfo{author}{\bibfnamefont{Y.~S.} \bibnamefont{Amhis}} \bibnamefont{et~al.}
  (\bibinfo{collaboration}{HFLAV}), \bibinfo{journal}{Eur. Phys. J. C}
  \textbf{\bibinfo{volume}{81}}, \bibinfo{pages}{226} (\bibinfo{year}{2021}),
  \eprint{1909.12524}.

\bibitem[{\citenamefont{Inami et~al.}(2003)}]{Belle:2002nla}
\bibinfo{author}{\bibfnamefont{K.}~\bibnamefont{Inami}} \bibnamefont{et~al.}
  (\bibinfo{collaboration}{Belle}), \bibinfo{journal}{Phys. Lett. B}
  \textbf{\bibinfo{volume}{551}}, \bibinfo{pages}{16} (\bibinfo{year}{2003}),
  \eprint{hep-ex/0210066}.

\bibitem[{\citenamefont{Grygier et~al.}(2017)}]{Belle:2017oht}
\bibinfo{author}{\bibfnamefont{J.}~\bibnamefont{Grygier}} \bibnamefont{et~al.}
  (\bibinfo{collaboration}{Belle}), \bibinfo{journal}{Phys. Rev. D}
  \textbf{\bibinfo{volume}{96}}, \bibinfo{pages}{091101}
  (\bibinfo{year}{2017}), \bibinfo{note}{[Addendum: Phys.Rev.D 97, 099902
  (2018)]}, \eprint{1702.03224}.

\bibitem[{\citenamefont{Aaij et~al.}(2014)}]{LHCb:2014mit}
\bibinfo{author}{\bibfnamefont{R.}~\bibnamefont{Aaij}} \bibnamefont{et~al.}
  (\bibinfo{collaboration}{LHCb}), \bibinfo{journal}{JHEP}
  \textbf{\bibinfo{volume}{09}}, \bibinfo{pages}{177} (\bibinfo{year}{2014}),
  \eprint{1408.0978}.

\bibitem[{\citenamefont{Aaij et~al.}(2021)}]{LHCb:2020gog}
\bibinfo{author}{\bibfnamefont{R.}~\bibnamefont{Aaij}} \bibnamefont{et~al.}
  (\bibinfo{collaboration}{LHCb}), \bibinfo{journal}{Phys. Rev. Lett.}
  \textbf{\bibinfo{volume}{126}}, \bibinfo{pages}{161802}
  (\bibinfo{year}{2021}), \eprint{2012.13241}.

\bibitem[{\citenamefont{Aaij et~al.}(2020)}]{LHCb:2020lmf}
\bibinfo{author}{\bibfnamefont{R.}~\bibnamefont{Aaij}} \bibnamefont{et~al.}
  (\bibinfo{collaboration}{LHCb}), \bibinfo{journal}{Phys. Rev. Lett.}
  \textbf{\bibinfo{volume}{125}}, \bibinfo{pages}{011802}
  (\bibinfo{year}{2020}), \eprint{2003.04831}.

\end{thebibliography}

\end{document}